\documentclass[fleqn,usenatbib]{mnras}

\usepackage{newtxtext,newtxmath}

\usepackage[T1]{fontenc}
\usepackage{ae,aecompl}
\usepackage{makecell}

\usepackage{graphicx}	%
\usepackage{amsmath}	%

\usepackage{amssymb}	%
\usepackage{hyperref}
\usepackage{xcolor}
\usepackage{orcidlink}

\newcommand{\comt}[1]{}

\newcommand{\HII}{\rm \ion{H}{ii}}

\newcommand{\MassIndependent}{\texttt{MassIndependent}}
\newcommand{\LowMassDriven}{\texttt{LowMassDriven}}
\newcommand{\HighMassDriven}{\texttt{HighMassDriven}}

\newcommand{\addorcid}[1]{\orcidlink{\csname orcidauthor#1\endcsname}}

\title[The phase gradient spectrum of the 21-cm signal during reionization]{The phase gradient spectrum: probing the squeezed bispectrum of the 21-cm signal during reionization}

\author[S. K. Giri et al.]{Sambit K. Giri\addorcid{SG},$^{1,2}$\thanks{E-mail: \href{mailto:sambit.giri@astro.su.se}{sambit.giri@astro.su.se}}, 
P. Daniel Meerburg\addorcid{DM},$^{2}$ and 
L\'eon V. E. Koopmans\addorcid{LVEK}$^{3}$
\\
$^{1}$Department of Astronomy and Oskar Klein Center, Stockholm University, AlbaNova, SE-106 91 Stockholm, Sweden\\
$^{2}$Van Swinderen Institute for Particle Physics and Gravity, University of Groningen, Nijenborgh 3, 9747 AG Groningen, The Netherlands\\
$^{3}$Kapteyn Astronomical Institute, University of Groningen, PO Box 800, NL-9700 AV Groningen, the Netherlands
}

\date{Accepted XXX. Received YYY; in original form ZZZ}

\pubyear{2026}

\begin{document}
\label{firstpage}
\pagerange{\pageref{firstpage}--\pageref{lastpage}}
\maketitle

\begin{abstract}
The 21-cm signal from the epoch of reionization is highly non-Gaussian. Standard analyses rely on the power spectrum, which is blind to the Fourier phases, while higher-order statistics such as the bispectrum are computationally expensive and spread information across triangle configurations. As an efficient alternative, we develop the \textit{phase gradient spectrum}, which measures the circular coherence of phase differences between Fourier modes separated by a long-wavelength mode. We show analytically that this statistic measures the angle-averaged normalised squeezed bispectrum, which describes the response of small-scale power to long-wavelength fluctuations.
Applied to reionization simulations with different ionizing source properties, the phase gradient spectrum of the 21-cm signal shows a characteristic non-monotonic evolution. At early times, the spectrum tracks the non-Gaussianity of the matter density field. Once the mean neutral fraction $\langle x_\mathrm{HI}\rangle \lesssim 0.9$, the spectrum follows the ionization topology. The phase coherence reaches a minimum midway through reionization and rises again as the remaining neutral regions shrink into isolated islands. Statistics that keep the sign of the squeezed bispectrum show that this minimum marks its sign change. This evolution holds for different reionization models and simulation frameworks.
A Fisher forecast for a 1000-h Square Kilometre Array (SKA)-Low observation, combining five redshifts across reionization, shows that the phase gradient spectrum alone constrains the astrophysical parameters, and these constraints tighten by a factor of $\gtrsim 2$ for a 10,000-h observation. Combining the phase gradient spectrum with the power spectrum tightens the marginalised constraints by up to $\sim$20 per cent relative to the power spectrum alone.
\end{abstract}


\section{Introduction}\label{sec:intro}

The Epoch of Reionization (EoR) represents the last major phase transition of hydrogen gas in the intergalactic medium (IGM), marking the transition of the Universe from a dark, neutral state to a fully ionized one \citep[e.g.][]{barkana2001beginning,furlanetto2006cosmology}. Triggered by the birth of the first stars, galaxies, and early active galactic nuclei, this process was driven by the release of ionizing ultraviolet radiation from early luminous sources that escaped into the IGM, ionizing the surrounding gas to form growing bubbles (\HII{} regions) \citep[e.g.][]{furlanetto2004growth,zaroubi2012epoch}. Over hundreds of millions of years ($z \sim 15$--6), these bubbles expanded, overlapped with neighbouring ionized regions, and eventually percolated on cosmic scales \citep{furlanetto2016reionization,giri2018bubble}.

The physical processes governing this cosmic milestone are exceptionally complex, dictated by multi-scale astrophysical phenomena such as the transfer of ionizing radiation through an inhomogeneous medium, feedback-regulated star formation, and the growth of dark matter halos hosting the ionizing sources \citep[e.g.][]{gnedin2000cosmological,iliev2007self,ciardi2005first,dixon2016large}. This non-linear coupling imprints a highly non-Gaussian topology on the IGM \citep[e.g.][]{iliev2006simulating,mcquinn2007morphology,friedrich2011topology,giri2021measuring}, which is vastly more non-linear than the underlying, gravitationally evolved matter density field at these redshifts. In the observed 21-cm signal, this non-Gaussianity manifests itself through a combination of the ionization field and the spin-temperature fluctuations set by the inhomogeneous heating of the IGM \citep[e.g.][]{pritchard2007inhomogeneous,watkinson2019bispectrum}. The present work focuses on the contribution of the ionization field.

Reionization has been probed indirectly through Lyman-$\alpha$ absorption in high-redshift quasar and galaxy spectra, which constrains the neutral fraction at individual redshifts \citep[e.g.][]{fan2006constraining,mason2018universe}, and through the integrated optical depth to the cosmic microwave background (CMB) \citep[e.g.][]{komatsu2011seven,planck2020cosmology}. These indirect probes, however, lack sensitivity to the spatial morphology and astrophysical drivers of the reionization process. The redshifted 21-cm hyperfine transition of neutral hydrogen represents the premier observational probe to trace this evolution directly in three dimensions \citep[e.g.][]{pritchard201221}. This has motivated the development of a suite of current and next-generation low-frequency radio interferometers dedicated to mapping the 21-cm signal, including the Low Frequency Array \citep[LOFAR; e.g.,][]{mertens2020improved,mertens2025deeper}, the Murchison Widefield Array \citep[MWA; e.g.,][]{trott2025improved}, the Hydrogen Epoch of Reionization Array \citep[HERA; e.g.,][]{hera2022phaseI,hera2026phaseII}, and the upcoming Square Kilometre Array \citep[SKA-Low; e.g.,][]{koopmans2015cosmic}. As a direct tracer of neutral hydrogen in the IGM, the 21-cm signal probes the geometry of the distribution of \HII{} regions and thereby encodes the highly non-Gaussian morphology of the reionization field \citep[e.g.][]{mellema2006simulating,giri2019position}. Standard analyses have historically relied on the power spectrum to extract physical parameters from this signal. However, the power spectrum only captures two-point amplitude correlations and is entirely blind to the spatial phase correlations. It is therefore fundamentally incapable of capturing this non-Gaussian structure, leaving the rich phase information of the Fourier modes unexploited.

To fully characterise this non-Gaussianity, one must access higher-order statistical correlations. In Fourier space, this hierarchy is represented by the polyspectra, starting with the three-point correlation function or bispectrum. However, the bispectrum has two key limitations. First, it is computationally expensive, requiring the evaluation of many Fourier-space triangles across large volumes \citep[e.g.][]{watkinson2017fast}. Second, its information is highly distributed across diverse triangle configurations, making physical interpretation non-trivial \citep[e.g.][]{majumdar2018quantifying,watkinson2019bispectrum}. Alternatively, simulation-based inference (SBI) utilising neural posterior estimators trained on forward-modelled data has emerged as a powerful technique to extract non-Gaussian summaries directly from the raw observations \citep[e.g.][]{alsing2018massive,zhao2022implicit}. However, SBI requires exceptionally accurate forward models of instrumental systematics and bright astrophysical foregrounds, which remain notoriously difficult to model to the precision required for 21-cm experiments \citep[e.g.][]{patil2016systematic,mevius2022numerical,kern2025differentiable,giri2026foregrounds}.

Several alternative summary statistics have been proposed to probe non-Gaussian structure in 21-cm observations efficiently. In the image domain, morphological descriptors such as size distributions of ionized or neutral regions \citep[e.g.][]{kakiichi2017granulometry,giri2018bubble,giri2019neutral}, Euler characteristics \citep[e.g.][]{friedrich2011topology,giri2019neutral}, and Betti numbers from persistent homology \citep{elbers2019persistent,giri2021measuring} have shown immense promise. However, image-domain metrics are highly sensitive to reconstruction artefacts, antenna calibration errors, and incomplete foreground subtraction \citep[e.g.,][]{gagnon2021recovering,bianco2024deep}.

The position-dependent power spectrum \citep{giri2019position} is a computationally cheap alternative in Fourier space. It measures how small-scale power is modulated by large-scale overdensities and acts as an indirect probe of the squeezed-limit bispectrum. However, the position-dependent power spectrum requires the survey volume to be subdivided into subvolumes. The mean overdensity of each subvolume is set by the long-wavelength modes, which spectrally smooth foregrounds contaminate most strongly \citep[e.g.,][]{datta2012light}. Each subvolume also holds only a fraction of the short-wavelength modes, which makes the local power spectra noisy. This motivates exploring whether squeezed-limit information can be accessed through the phases of the short-wavelength modes alone, using the full survey volume.

An often overlooked but powerful diagnostic of non-Gaussianity lies in the Fourier phase information of a field. Radio interferometers measure visibilities, which are proportional to the Fourier transform of the brightness temperature field and carry both the amplitude and the phase of each Fourier mode. This phase information is already present in every visibility dataset recorded by current and future 21-cm experiments. No hardware modifications, dedicated observing modes, or additional telescope time are required to access it. Standard analyses, however, reduce the complex visibilities to the power spectrum by taking the squared modulus of each Fourier mode, discarding the phase content in the process.

For a purely Gaussian random field (GRF), the Fourier phases are uniformly distributed over the interval $[-\pi, \pi)$ and are statistically independent. In such cases, the power spectrum contains all the statistical information about the field. As the field becomes non-Gaussian, the distribution of individual phases can remain consistent with uniform \citep{Suginohara1991Statistics,Scherrer1991AQuantitative}, and the non-Gaussian information instead appears as correlations between the phases of different modes \citep[e.g.,][]{watts2003universal,coles2004phase}. These deviations encode information that is not captured by the power spectrum, %
making phase statistics a valuable complementary tool for probing non-Gaussian structure \citep[e.g.,][]{chiang2000phase,chiang2002return,Matsubara2003PhaseFields,matsubara2007statistics}. Existing phase-based estimators, such as the triangle correlation function of normalised phases \citep{gorce2019studying} and the closure phase measured on triangles of interferometer baselines \citep{thyagarajan2018detecting,thyagarajan2020detectionI}, are built from closed triangles of Fourier modes. Covering many triangle configurations with them faces the same computational bottlenecks as the standard bispectrum.

In this work, we present the \textit{phase gradient spectrum}, a computationally efficient Fourier-space statistic, built from pairs of Fourier modes, that directly captures this phase coherence. The statistic builds on the phase gradient, the phase difference between two Fourier modes separated by a fixed wavenumber increment \citep{chiang2000phase}. Unlike the individual phases, the phase gradients of a non-Gaussian field deviate from a uniform distribution. The phase gradient spectrum measures the circular coherence of the phase gradients between Fourier modes separated by a long-wavelength mode. Circular coherence quantifies how strongly the distribution of these phase differences on the circle $[-\pi, \pi)$ concentrates around a preferred direction. A uniform distribution, as expected for a GRF, has zero coherence, and the coherence increases as the distribution becomes more peaked. By working with phase differences of individual modes, the phase gradient spectrum avoids expensive triangle configurations and operates entirely in Fourier space, circumventing the image-quality sensitivities of morphological measures. We show analytically that the phase gradient spectrum measures the angle-averaged normalised squeezed bispectrum, linking it to the response of small-scale structure to long-wavelength modes.

The paper is organised as follows. Sec.~\ref{sec:phase_info} introduces the phase gradient spectrum and validates its connection to the squeezed bispectrum on log-normal fields. Sec.~\ref{sec:reion_model} describes the reionization simulations, which span a range of source models and two simulation frameworks. Sec.~\ref{sec:21cm_phases} applies the phase gradient spectrum to these simulations, follows its evolution through reionization, and interprets this evolution with statistics that keep the sign of the squeezed bispectrum. Sec.~\ref{sec:forecast_study} forecasts parameter constraints for a mock SKA-Low observation using the Fisher matrix formalism. We summarise our conclusions in Sec.~\ref{sec:summary}. Appendix~\ref{app:bispectrum_connection} derives the connection between the phase gradient spectrum and the squeezed bispectrum, Appendix~\ref{app:fisher} describes the forecast methodology, and Appendix~\ref{app:phase_info_tests} tests the information content of the phase gradient spectrum. Throughout this work, we assume a flat $\Lambda$CDM cosmology consistent with \citet{planck2020cosmology}: $\Omega_m = 0.31$, $\Omega_\Lambda = 0.69$, $\Omega_b = 0.049$, $h = 0.67$, $\sigma_8 = 0.83$, and $n_s = 0.96$.

\section{Phase information}\label{sec:phase_info}

In this section, we develop a summary statistic that captures the non-Gaussian information encoded in Fourier phases. We first introduce log-normal fields (Sec.~\ref{sec:phases_log_normal}) as a controlled testbed where the level of non-Gaussianity can be tuned continuously. In Sec.~\ref{sec:phases_phase_grad}, we define phase gradients, motivate the use of circular coherence as a sufficient summary of their distribution, and introduce the phase gradient spectrum in Sec.~\ref{sec:phases_phase_grad_hist}. There, we relate it to the squeezed bispectrum (Sec.~\ref{sec:formalism}) and test it on log-normal fields (Sec.~\ref{sec:lognormal_spectrum}).

\subsection{Fourier phases of log-normal models}\label{sec:phases_log_normal}
To validate our phase-based summary statistics, we first consider log-normal fields \citep[e.g.,][]{coles1991lognormal}, defined as
\begin{equation}
    f(\pmb{x}) = \exp\left(\gamma\, g(\pmb{x}) - \frac{\gamma^2}{2}\right) - 1 \, ,
    \label{eq:lognormal_field}
\end{equation}
where $\gamma$ is a free parameter controlling the level of non-Gaussianity, $\pmb{x}$ is the position vector, and $g(\pmb{x})$ is a GRF with zero mean, unit variance and a power-law power spectrum $P_g(k) \propto k^{-2}$. For small $\gamma$, $f(\pmb{x})$ approximates a Gaussian field. We generate discrete realisations of such fields on a cubic mesh of $128^3$ cells spanning a volume that is 300 Mpc in each direction.

Any real-space field $\delta(\pmb{x})$ can be decomposed into a sum of plane waves through its Fourier transform:
\begin{equation}
    \delta(\pmb{x}) = \sum_{\pmb{k}} \Tilde{\delta}(\pmb{k})\, \exp{(\mathrm{i}\,\pmb{k} \cdot \pmb{x})} \, ,
\end{equation}
where $\Tilde{\delta}(\pmb{k})$ denotes the Fourier coefficients and $\pmb{k}$ is the wavevector. These coefficients are complex numbers that can be expressed as
\begin{equation}
    \Tilde{\delta}(\pmb{k}) = A(\pmb{k})\, \exp{(\mathrm{i}\, \Phi_{\pmb{k}})} \, ,
\end{equation}
where $A(\pmb{k})$ is the Fourier amplitude and $\Phi_{\pmb{k}}$ is the Fourier phase. The power spectrum is proportional to the mean of $A^2(\pmb{k})$ over modes of the same $k$ (see Appendix~\ref{app:conventions} for the normalisation), encapsulating the variance of the amplitudes but not the phase information. In this work, we explore the information contained in the phases $\Phi_{\pmb{k}}$ by developing a summary statistic based on these phases, which is described in the following subsections.

\subsection{Phase gradient coherence}\label{sec:phases_phase_grad}

\begin{figure}
    \centering
    \includegraphics[width=0.96\linewidth]{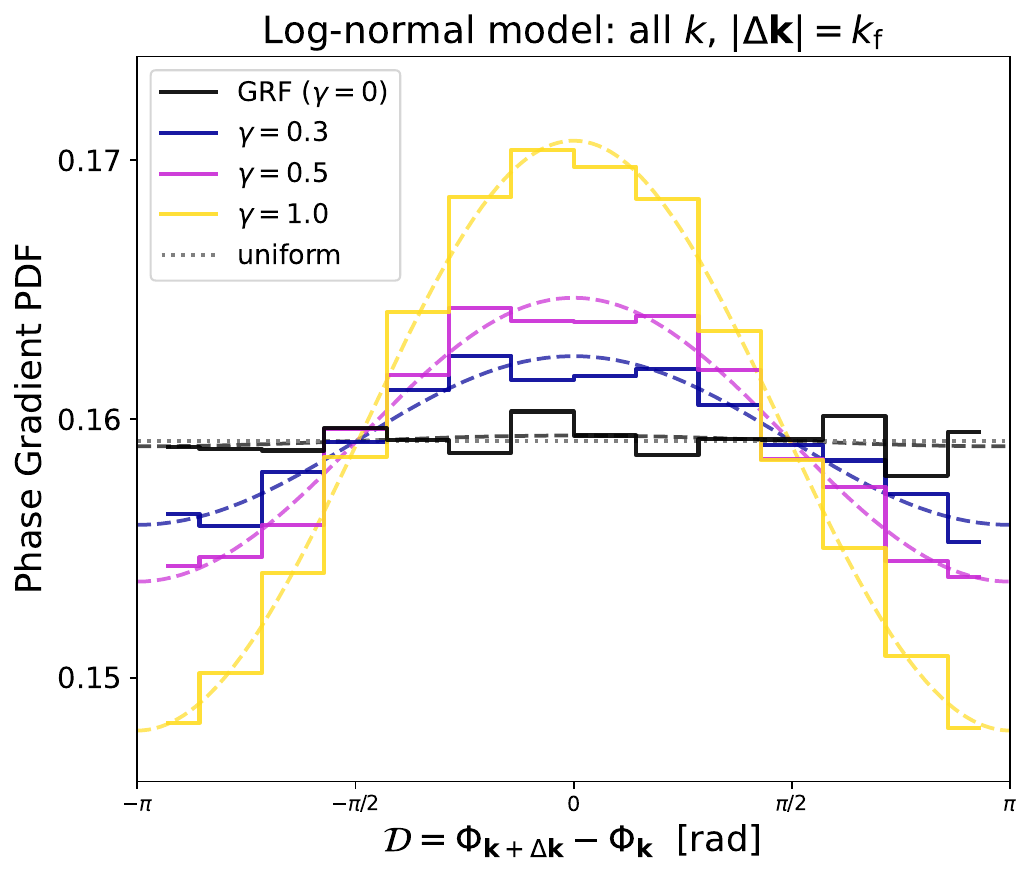}
    \\[-0.05cm]
    \includegraphics[width=0.96\linewidth]{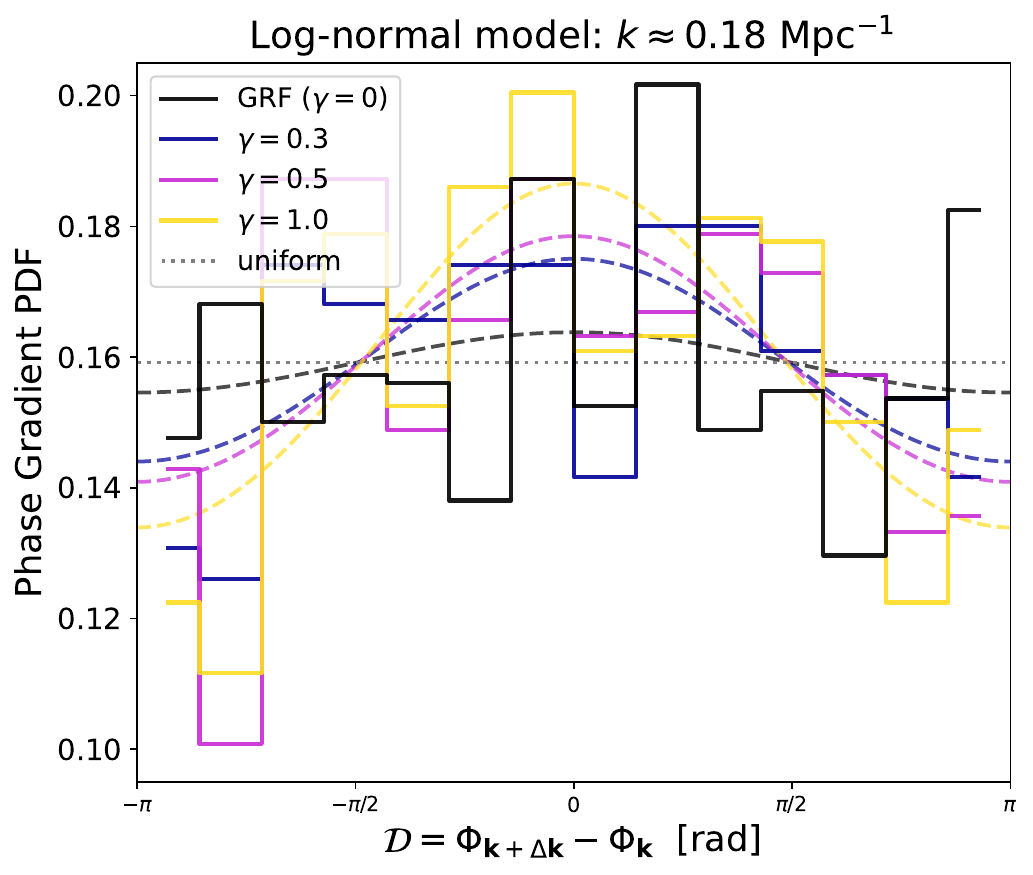}
    \\[-0.05cm]
    \includegraphics[width=0.96\linewidth]{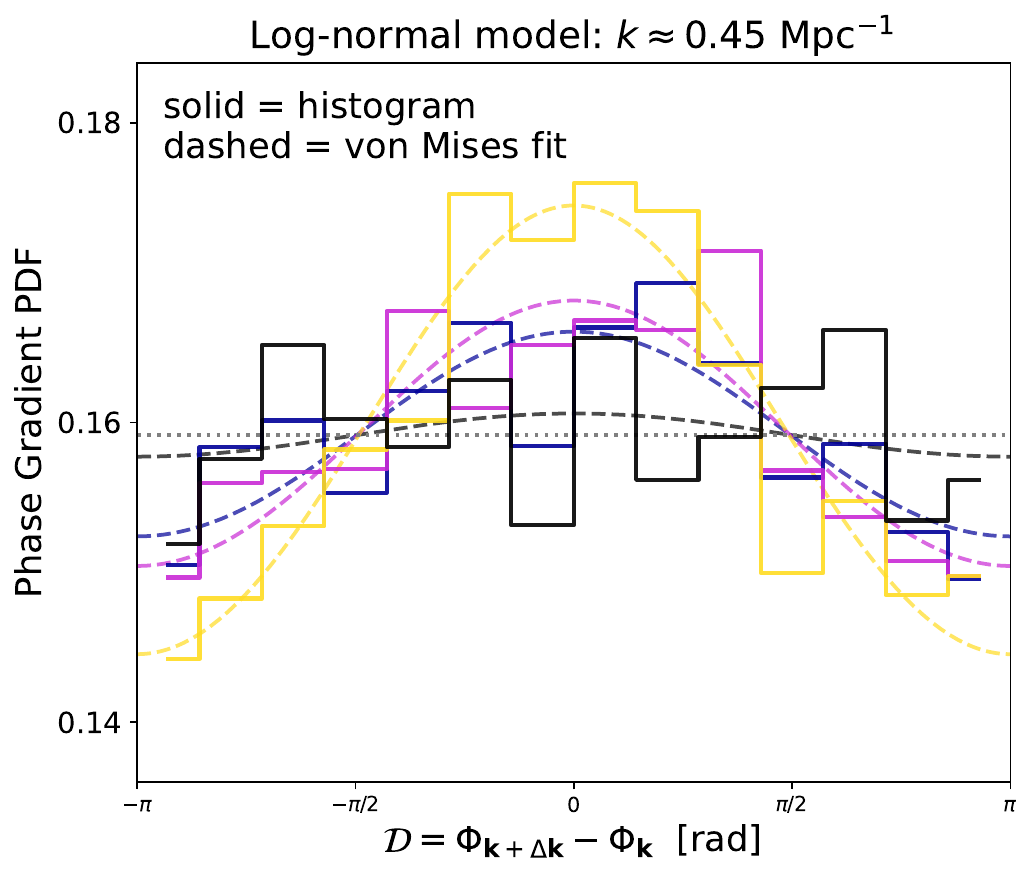}
    \vspace{-0.3cm}
    \caption{
    Phase gradient PDFs of log-normal fields with different non-Gaussianity parameters $\gamma$, for one realisation and a single long-wavelength mode with $|\Delta\pmb{k}| = k_\mathrm{f}$. The top panel uses all modes in the box, and the middle and bottom panels use the modes in $k$-bins at $k \approx 0.18$ and $0.45\,\mathrm{Mpc}^{-1}$. Each PDF is centred on its circular mean, and the vertical ranges differ between panels. Solid curves show the measured PDFs, dashed curves of the same colour are von Mises fits, and the dotted horizontal line represents a uniform distribution. The PDFs narrow as non-Gaussianity ($\gamma$) increases.
    }
    \label{fig:phase_grad_pdf_lognormal}
\end{figure}

In a GRF, Fourier phases are independently and uniformly distributed over $[-\pi, \pi)$\footnote{Some studies use the interval $[0, 2\pi)$ instead. Both describe the same distribution on the circle, as phases are defined modulo $2\pi$.} \citep[e.g.,][]{watts2003universal}. The power spectrum therefore fully characterises a GRF. Moreover, even in mildly non-Gaussian fields, individual phases may still appear consistent with a uniform distribution \citep{Suginohara1991Statistics}. The non-Gaussian information is therefore encoded not in the absolute phase values but in the correlations between phases of neighbouring Fourier modes \citep[e.g.,][]{Scherrer1991AQuantitative,chiang2000phase}. We quantify this through the \textit{phase gradient}, the difference between the phases of two modes separated by a wavenumber increment $\Delta\pmb{k}$, defined as
\begin{equation}
    \mathcal{D}(\pmb{k}, \Delta\pmb{k}) = \Phi_{\pmb{k}+\Delta\pmb{k}} - \Phi_{\pmb{k}} \, ,
\end{equation}
where $\mathcal{D}$ is wrapped onto the interval $[-\pi, \pi)$. The increment $\Delta\pmb{k}$ is a long-wavelength mode of the discrete Fourier grid. Each increment is an integer combination of the fundamental mode $k_\mathrm{f} = 2\pi/L_B$ of a box of size $L_B$ along the three grid axes, and $k_\mathrm{f}$ is the smallest increment on the grid. For adjacent modes ($|\Delta\pmb{k}| = k_\mathrm{f}$), $\mathcal{D}$ is a finite-difference phase gradient, which motivates the name used by \citet{chiang2000phase}. We keep this name for the larger increments used below. We do not divide $\mathcal{D}$ by $|\Delta\pmb{k}|$, which keeps it an angle on the circle, as the circular statistics below require.

The distribution of phase gradients is well approximated by a von Mises distribution \citep[e.g.,][]{mardia1975statistics,jammalamadaka2001topics}, the circular analogue of the Gaussian, characterised by two parameters: a mean direction $\mu$ and a concentration $\kappa$, which sets the width of the distribution \citep{watts2003universal}. The mean direction $\mu$ is set by the positions of the dominant structures in the field, which differ between realisations of a statistically homogeneous field \citep{chiang2000phase,watts2003universal}. We therefore use the width of the distribution, which does not depend on these positions. The top panel of Fig.~\ref{fig:phase_grad_pdf_lognormal} shows the phase gradient probability distribution functions (PDFs) of log-normal fields for a single long-wavelength mode with $|\Delta\pmb{k}| = k_\mathrm{f}$, using all modes in the box, as in \citet{watts2003universal}. The phase gradients are rotated so that their mean direction $\mu$ lies at $\mathcal{D} = 0$, which allows the widths to be compared directly. The von Mises distribution (dashed curves) describes the measured PDFs well. The PDF of the GRF is consistent with a uniform distribution, and the distribution narrows as the non-Gaussianity parameter $\gamma$ increases, reflecting increasingly coherent phase gradients.

Using all modes, however, mixes all scales. To follow the scale dependence of the phase coherence, we group the modes into $k$-bins. The middle and bottom panels of Fig.~\ref{fig:phase_grad_pdf_lognormal} show the PDFs in two such bins, at $k \approx 0.18$ and $0.45\,\mathrm{Mpc}^{-1}$, respectively. The GRF now shows a weak peak at $\mathcal{D} = 0$. This peak is a finite-sample effect, not phase coherence. The unit phase vectors of the modes in a bin never sum exactly to zero, and the rotation aligns this residual with $\mathcal{D} = 0$. The higher $k$-bin contains more modes and shows a weaker GRF peak. The von Mises fits show that the PDFs of the non-Gaussian fields remain narrower than that of the GRF in both bins.

We measure the width of the PDF at each wavenumber with the mean resultant length \citep{jammalamadaka2001topics}. Each phase gradient defines a unit vector $\exp(\mathrm{i}\,\mathcal{D})$ in the complex plane. The mean resultant length is the length of the average of these unit vectors over the modes in the $k$-bin,
\begin{eqnarray}
    \bar{R}(k, \Delta\pmb{k}) = \left| \frac{1}{N_k} \sum_{|\pmb{k}| \in \mathrm{bin}} \exp\!\left(\mathrm{i}\, \mathcal{D}(\pmb{k}, \Delta\pmb{k})\right) \right| \, ,
\end{eqnarray}
with $N_k$ the number of modes in the $k$-bin. The mean resultant length is 0 for uniformly distributed phase gradients. A GRF with many modes approaches this limit. The mean resultant length reaches 1 when all phase gradients are identical. The complement $\sigma^2_\mathrm{circ} = 1 - \bar{R}$ is the circular variance. For a von Mises distribution, $\bar{R} = I_1(\kappa)/I_0(\kappa)$, with $I_n$ the modified Bessel functions. A single increment gives a noisy estimate, because $\bar{R}$ scales with the amplitude of the one long-wavelength mode $\Delta\pmb{k}$, which varies between realisations (Sec.~\ref{sec:lognormal_spectrum} and Appendix~\ref{app:bispectrum_connection}). We therefore evaluate $\bar{R}$ for every long-wavelength mode inside a sphere of radius $3\,k_\mathrm{f}$ ($0 < |\Delta\pmb{k}| \leq 3\,k_\mathrm{f}$) and average the resultant lengths, $\bar{R}(k) = N_L^{-1} \sum_{\Delta\pmb{k}} \bar{R}(k, \Delta\pmb{k})$. This gives $N_L = 61$ long-wavelength modes.\footnote{The long-wavelength modes are the grid vectors $\Delta\pmb{k} = k_\mathrm{f}\,\pmb{n}$, where $\pmb{n}$ is a vector of integers. There are 122 such vectors with $0 < |\pmb{n}| \leq 3$. As the field is real, $\Delta\pmb{k}$ and $-\Delta\pmb{k}$ give the same $\bar{R}$, and each pair is therefore counted once.} We motivate the radius of $3\,k_\mathrm{f}$ in Sec.~\ref{sec:lognormal_spectrum}. The narrowing of the PDFs with $\gamma$ in Fig.~\ref{fig:phase_grad_pdf_lognormal} corresponds to a rising $\bar{R}$. We refer to $\bar{R}$ as the \textit{circular coherence} and use it throughout this work.

\subsection{Phase gradient spectrum}\label{sec:phases_phase_grad_hist}

We develop the formalism for the phase gradient spectrum in Sec.~\ref{sec:formalism} and apply it to log-normal fields in Sec.~\ref{sec:lognormal_spectrum}.

\subsubsection{Formalism}\label{sec:formalism}

The collection of circular coherence values $\bar{R}(k)$ across multiple $k$-bins constitutes the \textit{phase gradient spectrum}, providing a phase-based analogue of the power spectrum that encodes how phase coherence, and hence non-Gaussianity, varies with spatial scale. In this work, we use logarithmic $k$-bins with 10 bins per decade ($\Delta\log_{10}k = 0.1$) from $k = 0.1$ to $1.2\,\mathrm{Mpc}^{-1}$, just below the Nyquist wavenumber of our grids. In Appendix~\ref{app:bispectrum_connection}, we derive the expected value of the phase gradient spectrum for a weakly non-Gaussian field. To leading order in the normalised bispectrum and in the limit $\Delta k \ll k$,
\begin{eqnarray}
    \langle\bar{R}(k, \Delta k)\rangle^2 \approx \frac{\pi}{4\,N_\mathrm{indep}} + \left(\frac{\pi^{3/2}}{8}\right)^2 \left|\bar{p}^{(3)}(k, \Delta k)\right|^2 \, ,
    \label{eq:Rbar_full}
\end{eqnarray}
where $N_\mathrm{indep}$ is the number of independent phase vectors in the $k$-bin, and $\bar{p}^{(3)}(k, \Delta k) \equiv B_\mathrm{sq}(\Delta k,\,k)/\sqrt{V P(\Delta k)\,P(k)^2}$ is the normalised squeezed bispectrum \citep{Matsubara2003PhaseFields}. Here $V$ is the survey (box) volume, $\Delta k = |\Delta\pmb{k}|$ is the magnitude of the long-wavelength mode, on which the ensemble mean depends by statistical isotropy, and $k$ sets the short-wavelength modes. The first term in Eq.~\eqref{eq:Rbar_full} is the mode-counting noise of the estimator. A finite number of randomly oriented unit phase vectors never sums exactly to zero, and this term sets a floor that vanishes only as $N_\mathrm{indep} \to \infty$. The second term is the non-Gaussian signal. This term measures how strongly the long-wavelength mode at $\Delta k$ coherently modulates the phases of the neighbouring short-wavelength modes at scale $k$, which is the squeezed-limit bispectrum. The factor $V^{-1/2}$ in $\bar{p}^{(3)}$ is required dimensionally and reflects the fact that individual phase correlations are diluted in larger volumes.

Eq.~\eqref{eq:Rbar_full} has two useful limits. For a GRF, $\bar{p}^{(3)} = 0$ and only the first term remains,
\begin{eqnarray}
    \langle\bar{R}\rangle_\mathrm{GRF} = \sqrt{\frac{\pi}{4\, N_\mathrm{indep}}} \, ,
    \label{eq:grf_noise_limit}
\end{eqnarray}
which is the null expectation of the Rayleigh test of circular uniformity \citep[e.g.,][]{mardia2009directional,jammalamadaka2001topics}. We refer to this as the GRF noise limit. It depends on $k$ only through the number of modes in the bin. With the logarithmic binning used here, $N_k$ grows as $\sim$$k^3$, and the GRF noise limit therefore falls as $\sim$$k^{-3/2}$. When the non-Gaussian signal dominates, the second term gives
\begin{eqnarray}
    \langle\bar{R}(k)\rangle \approx \frac{\pi^{3/2}}{8}\,|\bar{p}^{(3)}(k)| \, .
    \label{eq:variance_spectrum_main}
\end{eqnarray}
Since the first term is additive in Eq.~\eqref{eq:Rbar_full}, the mode-counting noise can be removed in the same way as the noise bias of the 21-cm power spectrum. For each long-wavelength mode, the squared estimator satisfies $\langle\bar{R}^2\rangle = (4/\pi)\,\langle\bar{R}\rangle^2$, and $\bar{R}^2 - 1/N_\mathrm{indep}$ is therefore an unbiased estimator of the non-Gaussian signal. The correction must be applied to each mode before averaging over the $N_L$ modes. As $N_\mathrm{indep}$ depends only on the box size, the $k$-binning and $\Delta\pmb{k}$, this correction uses no information beyond the measured resultant lengths. We do not apply it in this work, and the forecast of Sec.~\ref{sec:forecast_study} uses the uncorrected $\bar{R}(k)$ consistently.

The phase gradient spectrum is also cheap to compute. For a field on $N$ grid cells, the phase gradient spectrum requires a single fast Fourier transform (FFT), with a cost of $\mathcal{O}(N \log N)$, followed by $\mathcal{O}(N_L N)$ operations to form and bin the phase differences of the $N_L$ long-wavelength modes. As for the power spectrum, only one FFT is needed. The fastest bispectrum estimators also rely on FFTs but require several inverse transforms of shell-filtered fields for every triangle configuration \citep{scoccimarro2015fast,watkinson2017fast}, and direct triangle-counting estimators are far slower. Even when restricted to the squeezed configurations probed by the phase gradient spectrum, these FFT-based estimators still require one inverse FFT per $k$-bin in addition to that of the long-wavelength shell, compared with a single FFT for the phase gradient spectrum. The speed gain of the phase gradient spectrum therefore grows with the number of $k$-bins.

\subsubsection{Phase gradient spectrum of log-normal fields}\label{sec:lognormal_spectrum}

\begin{figure}
    \centering
    \includegraphics[width=0.96\linewidth]{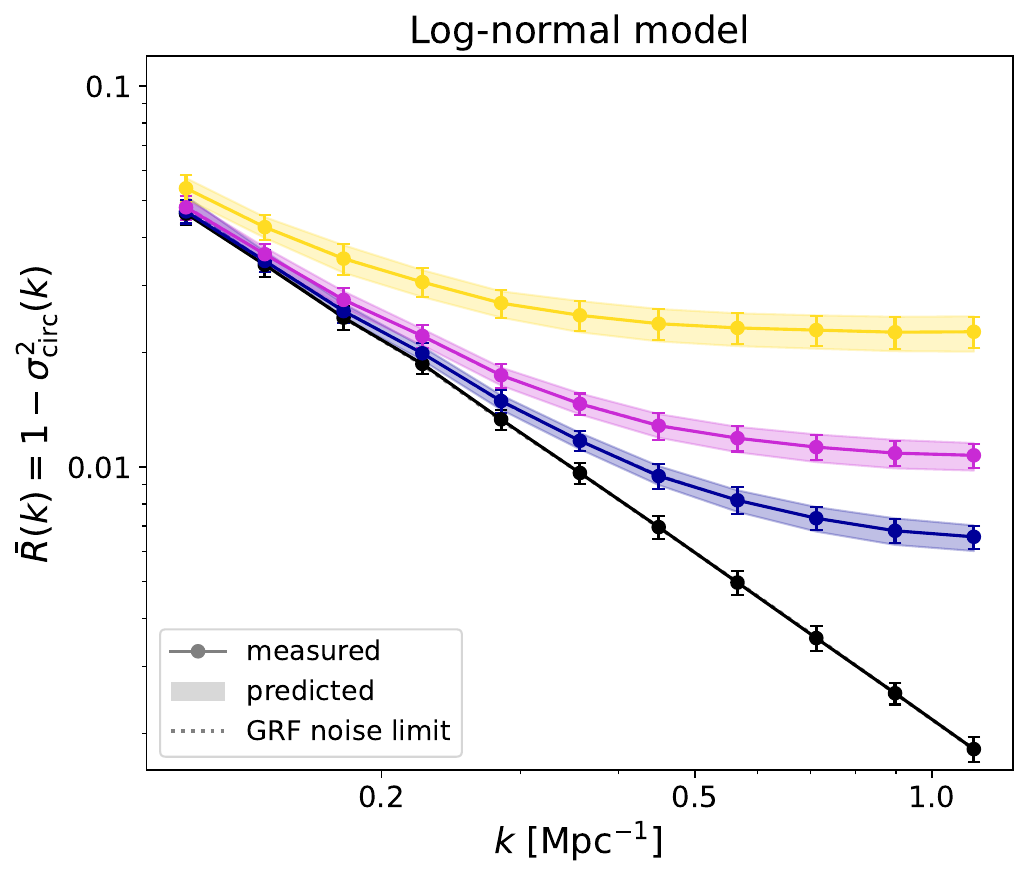}
    \\[-0.05cm]
    \includegraphics[width=0.96\linewidth,trim={-9 0 0 25},clip]{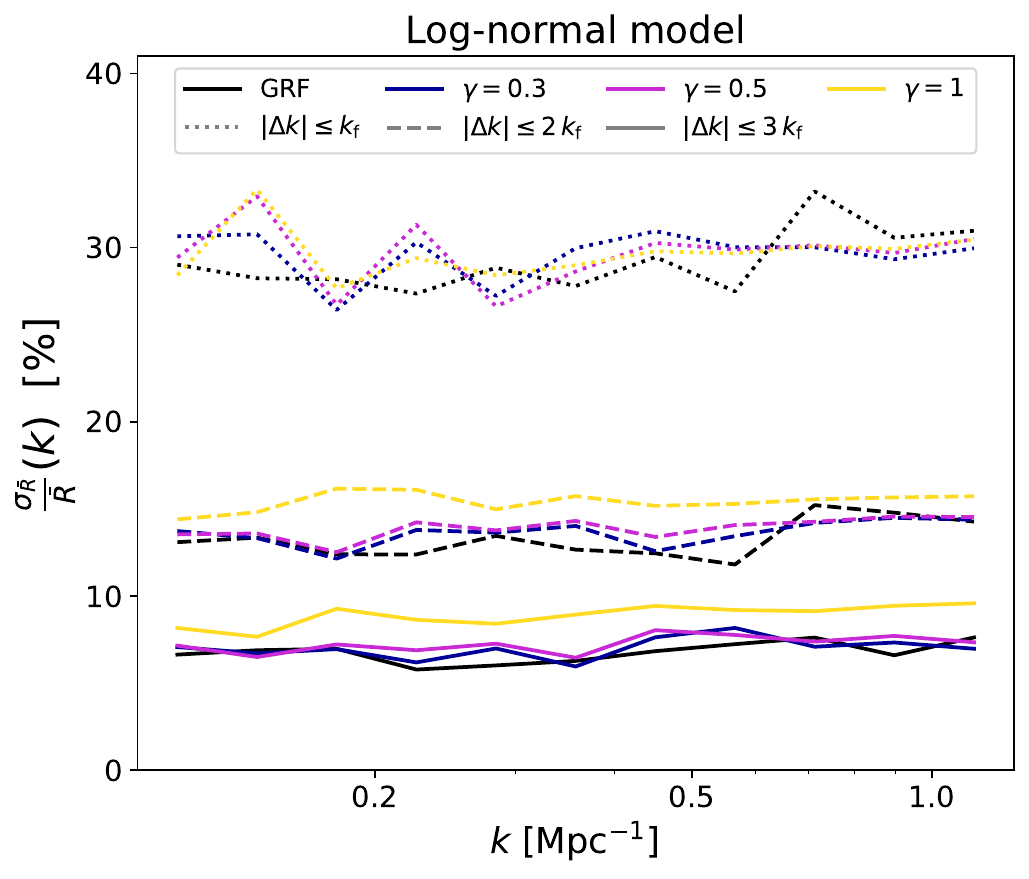}
    \vspace{-0.3cm}
    \caption{
    \textit{Top panel}: Phase gradient spectrum $\bar{R}(k)$ of the log-normal fields of Fig.~\ref{fig:phase_grad_pdf_lognormal}, averaged over the $N_L = 61$ long-wavelength modes with $|\Delta\pmb{k}| \leq 3\,k_\mathrm{f}$. Circles with error bars show the mean and standard deviation over 100 realisations. The mean of the bispectrum-based prediction (Eq.~\ref{eq:Rbar_full}) falls on the same solid curves as the measured mean. Shaded regions show the scatter of the prediction between realisations. For the GRF (black circles), the prediction reduces to the GRF noise limit (Eq.~\ref{eq:grf_noise_limit}). \textit{Bottom panel}: Relative scatter of $\bar{R}(k)$ between the 100 realisations, for the long-wavelength modes within $k_\mathrm{f}$ (dotted), $2\,k_\mathrm{f}$ (dashed) and $3\,k_\mathrm{f}$ (solid).
    }
    \label{fig:phase_grad_spectrum_lognormal}
\end{figure}

The top panel of Fig.~\ref{fig:phase_grad_spectrum_lognormal} shows $\bar{R}(k)$ for log-normal fields with increasing $\gamma$. The measured values (circles) rise monotonically with $\gamma$ at all wavenumbers, confirming that the phase gradient spectrum is a sensitive and scale-resolved tracer of non-Gaussianity. For the GRF (black circles), the measured spectrum closely follows the GRF noise limit of Eq.~\eqref{eq:grf_noise_limit}, confirming that any excess of $\bar{R}(k)$ above this limit is a genuine non-Gaussian signal. The circles and error bars give the mean and standard deviation over 100 realisations with different initial seeds. The prediction of Eq.~\eqref{eq:Rbar_full} uses the squared normalised bispectrum averaged over the same realisations, and the shaded regions show the scatter of this prediction between realisations. For $\gamma = 0.3$, 0.5 and 1.0, the prediction agrees with the measured spectrum to within about 1 per cent in all $k$-bins, and the two means fall on the same solid curves. This agreement confirms the analytic connection.

At fixed $\gamma$, the measured spectrum is also nearly independent of $k$ wherever it lies well above the GRF noise limit. This $k$-independence is a property of the log-normal transformation. The Gaussian field can be split into long- and short-wavelength parts, $g = g_L + g_S$, which are independent because they are built from disjoint sets of Fourier modes, with variances $\sigma_L^2 + \sigma_S^2 = 1$. Eq.~\eqref{eq:lognormal_field} then factorises exactly as
\begin{eqnarray}
    1 + f = \underbrace{e^{\gamma g_L - \gamma^2\sigma_L^2/2}}_{1 + f_L}\;\underbrace{e^{\gamma g_S - \gamma^2\sigma_S^2/2}}_{1 + f_S} \, .
\end{eqnarray}
A long-wavelength mode therefore rescales the local short-wavelength fluctuations by the same factor $(1 + f_L)$ at all $k$, and the local power spectrum by $(1 + f_L)^2$. The modulation of short-wavelength modes by the long-wavelength mode, and hence $\bar{p}^{(3)}(k)$ in Eq.~\eqref{eq:Rbar_full}, is then the same at every $k$.

The bottom panel of Fig.~\ref{fig:phase_grad_spectrum_lognormal} shows how the average over long-wavelength modes reduces the scatter of $\bar{R}(k)$ between realisations. With the 3 modes of $|\Delta\pmb{k}| = k_\mathrm{f}$, the relative scatter is about 30 per cent. It falls to about 14 per cent for the 16 modes within $2\,k_\mathrm{f}$ and to about 7 per cent for the 61 modes within $3\,k_\mathrm{f}$. The scatter is nearly the same for all $k$ and for $\gamma \leq 0.5$, and slightly larger for $\gamma = 1$ (about 9 per cent with 61 modes). The scatter falls as $N_L^{-1/2}$. This scatter is the sample variance of the long-wavelength modes. The resultant length of each mode follows a Rayleigh distribution (Appendix~\ref{app:bispectrum_connection}), whose relative scatter is $\sqrt{4/\pi - 1} \approx 0.52$. Averaging $N_L$ independent modes gives $0.52/\sqrt{N_L}$, which is 30, 13 and 7 per cent for the three sets. These fields are noiseless. In an observation, instrumental noise adds to this scatter, which we include in the forecast of Sec.~\ref{sec:forecast_study}.

The scatter would keep falling if more long-wavelength modes were included, but each mode probes a different bispectrum configuration. The resultant length for a given $\Delta\pmb{k}$ measures the triangles $(\pmb{k}, -\pmb{k}-\Delta\pmb{k}, \Delta\pmb{k})$ with $|\Delta\pmb{k}| \ll k$. In this squeezed limit, $\bar{p}^{(3)}$ is nearly independent of $\Delta k$, and the long-wavelength modes are independent measurements of the same quantity. Averaging them then only reduces the scatter. As $\Delta k$ approaches $k$, the triangles are no longer squeezed and Eq.~\eqref{eq:Rbar_full} no longer applies. The average would then mix different triangle shapes into one number, and the dependence on $\Delta k$ would be lost. Instead of averaging these larger increments, one can measure $\bar{R}(k, \Delta k)$ separately in bins of $\Delta k$. This two-dimensional statistic would trace how the phase coherence changes from squeezed to less squeezed triangle shapes, and we leave it for future work. With $|\Delta\pmb{k}| \leq 3\,k_\mathrm{f} \approx 0.06\,\mathrm{Mpc}^{-1}$, the long-wavelength modes have wavelengths of at least 100\,Mpc and lie below the smallest $k$ of $0.1\,\mathrm{Mpc}^{-1}$ used in this work. In the lowest $k$-bin, $\Delta k/k$ reaches about 0.6 for the longest of these modes, and the squeezed limit therefore holds only approximately there. For the log-normal fields, the factorisation above makes $\bar{p}^{(3)}$ independent of $\Delta k$ as well as of $k$, and indeed the mean of $\bar{R}(k)$ changes by less than 2 per cent between the three sets of long-wavelength modes, including the lowest $k$-bin.

\section{Modelling the 21-cm signal}
\label{sec:reion_model}

The observable quantity in 21-cm observations is the differential brightness temperature \citep[e.g.,][]{pritchard201221},
\begin{align}
\delta T_{\mathrm{b}} (\pmb{x}, z) &= T_{0} (z) x_{\rm HI}(\pmb{x},z) (1+\delta_{b}(\pmb{x},z)) \left( \frac{T_{\mathrm{s}}(\pmb{x},z) -T_{\mathrm{CMB}}(z) }{T_{\mathrm{s}}(\pmb{x},z)} \right) , \label{eq:dTb} \\
T_0 (z) &\approx 27 \left(\frac{1+z}{10}\right)^{\frac{1}{2}} \left( \frac{\Omega_{\mathrm{b}}}{0.044} \frac{h}{0.7} \right) \left( \frac{\Omega_{\mathrm{m}}}{0.27} \right)^{-\frac{1}{2}} \left( \frac{1-Y_{\mathrm{p}}}{1-0.248} \right) ~\mathrm{mK},
\end{align}
where $\delta_b$ is the matter overdensity, $x_{\rm HI}$ is the neutral fraction, $Y_\mathrm{p}$ is the primordial helium mass fraction, and $T_{\mathrm{s}}$ and $T_{\mathrm{CMB}}$ are the spin and CMB temperatures, respectively. Throughout this work, we assume the high spin temperature limit ($T_{\mathrm{s}} \gg T_{\mathrm{CMB}}$), which is a fair approximation during the EoR \citep[e.g.,][]{ross2017simulating}. Sec.~\ref{sec:model_inputs} presents our models for the cosmological structure formation and the ionizing sources during the EoR, and Sec.~\ref{sec:sim_framework} describes the reionization simulations used to compute the 21-cm signal.

\subsection{Cosmological and astrophysical model}\label{sec:model_inputs}

Both reionization codes of Sec.~\ref{sec:sim_framework} share the same cosmological and astrophysical inputs. Sec.~\ref{sec:structure_formation} describes the evolution of the matter density field and the treatment of redshift-space distortions. Sec.~\ref{sec:ionising_source} describes the model of the ionizing sources hosted by dark matter haloes.

\subsubsection{Cosmological structure formation}\label{sec:structure_formation}

We generate the initial conditions as a GRF normalised to the linear matter power spectrum. We evolve them to the redshifts studied in this work with second-order Lagrangian perturbation theory \citep[2LPT; e.g.,][]{bouchet1995perturbative,crocce2006renormalized} in \textsc{21cmFAST} \citep{mesinger201121cmfast}. We simulate boxes of 300\,Mpc on a side. The initial conditions and their 2LPT displacements are computed on a $384^3$ grid with cells of 0.78\,Mpc, as the perturbative evolution underestimates the density fluctuations unless the initial grid resolves scales below about 1\,Mpc \citep{mesinger201121cmfast}. The evolved density, ionization and 21-cm fields are then computed on a coarser $128^3$ grid. This configuration matches the log-normal tests in Sec.~\ref{sec:phases_log_normal}. Both reionization codes of Sec.~\ref{sec:sim_framework} use the same evolved density fields.
Peculiar velocities along the line of sight introduce redshift-space distortions in the 21-cm signal \citep[e.g.,][]{mao2012redshift}. All simulated signals in this work include these distortions through the line-of-sight gradient of the linear peculiar velocity field. We defer the impact of other line-of-sight anisotropies, such as the lightcone effect \citep[e.g.,][]{datta2012light,giri2018bubble}, and of non-linear redshift-space distortions \citep[e.g.,][]{mao2012redshift,ross2021redshift} to future work. %

\subsubsection{Ionizing source model}\label{sec:ionising_source}

Following \citet{park2019inferring}, each dark matter halo of mass $M_h$ hosts a galaxy with stellar mass
\begin{eqnarray}
    M_\star = f_\star(M_h) \left(\frac{\Omega_b}{\Omega_m}\right) M_h \ ,
    \label{eq:Mstar}
\end{eqnarray}
where $f_\star(M_h)$ is the fraction of the halo's baryonic content converted into stars. The stars form over a characteristic fraction $t_\star$ of the Hubble time, giving a star formation rate
\begin{eqnarray}
    \Dot{M}_\star = \frac{M_\star}{t_\star H(z)^{-1}} \ ,
    \label{eq:sfr}
\end{eqnarray}
with $H(z)$ the Hubble parameter and $t_\star$ a dimensionless parameter that sets the star formation time-scale. The rate at which ionizing photons escape a halo into the IGM is then
\begin{eqnarray}
    \Dot{N}_\gamma (M_h) = f_\mathrm{duty}(M_h)\, f_\mathrm{esc}(M_h)\, N_\mathrm{ion,b}\, \frac{\Dot{M}_\star}{m_b} \ ,
    \label{eq:Ndot_gamma}
\end{eqnarray}
where $N_\mathrm{ion,b}$ is the number of ionizing photons produced per stellar baryon, $m_b$ is the mean baryon mass, $f_\mathrm{esc}(M_h)$ is the ionizing photon escape fraction, and $f_\mathrm{duty}(M_h)$ is the duty cycle defined below. Both $f_\star$ and $f_\mathrm{esc}$ are power-law functions of halo mass,
\begin{eqnarray}
    f_i(M_h) = f_{i,10} \left(\frac{M_h}{10^{10} M_\odot}\right)^{\alpha_i} , \quad i \in \{\star, \mathrm{esc}\} \ ,
\end{eqnarray}
with both fractions capped at unity.

We fix the escape fraction parameters to $\log_{10} f_{\mathrm{esc},10} = -1$ and $\alpha_\mathrm{esc} = -0.5$, and treat $\log_{10} f_{\star,10}$ and $\alpha_\star$ as free parameters. Because $f_\star$ and $f_\mathrm{esc}$ enter Eq.~\eqref{eq:Ndot_gamma} only through their product, $\log_{10} f_{\star,10}$ is degenerate with $\log_{10} f_{\mathrm{esc},10}$ and $\alpha_\star$ is degenerate with $\alpha_\mathrm{esc}$. The net mass dependence of the ionizing efficiency is therefore governed by $\alpha_\star + \alpha_\mathrm{esc}$: sources are mass-independent when $\alpha_\star + \alpha_\mathrm{esc} = 0$, low-mass dominated when $\alpha_\star + \alpha_\mathrm{esc} < 0$, and high-mass dominated when $\alpha_\star + \alpha_\mathrm{esc} > 0$. We also fix $N_\mathrm{ion,b} = 5000$ ionizing photons per stellar baryon, a value motivated by a Salpeter initial mass function \citep{park2019inferring}. As $N_\mathrm{ion,b}$ enters Eq.~\eqref{eq:Ndot_gamma} only through its product with $f_\star$, it is degenerate with $f_{\star,10}$, and fixing it does not reduce the range of models explored.
Low-mass haloes are suppressed by inefficient gas cooling and radiative feedback \citep[e.g.,][]{nebrin2023starbursts}. We model this with a turnover mass $M_\mathrm{turn}$ parameter through the mass-dependent duty cycle entering Eq.~\eqref{eq:Ndot_gamma},
\begin{eqnarray}
    f_\mathrm{duty} (M_h) = \exp \left(-\frac{M_\mathrm{turn}}{M_h}\right) \ ,
    \label{eq:fduty}
\end{eqnarray}
which smoothly suppresses the contribution of low-mass haloes while leaving high-mass haloes unaffected. Thus, the ionizing source models are explored in this work with three free parameters, which are $\log_{10}(f_{\star,10})$, $\alpha_\star$ and $\log_{10}(M_\mathrm{turn}/M_\odot)$.

\subsection{Reionization simulations}\label{sec:sim_framework}

\begin{figure}
    \centering
    \includegraphics[width=0.96\linewidth,trim={0.0cm 0.0cm 0.1cm 0.2cm},clip]{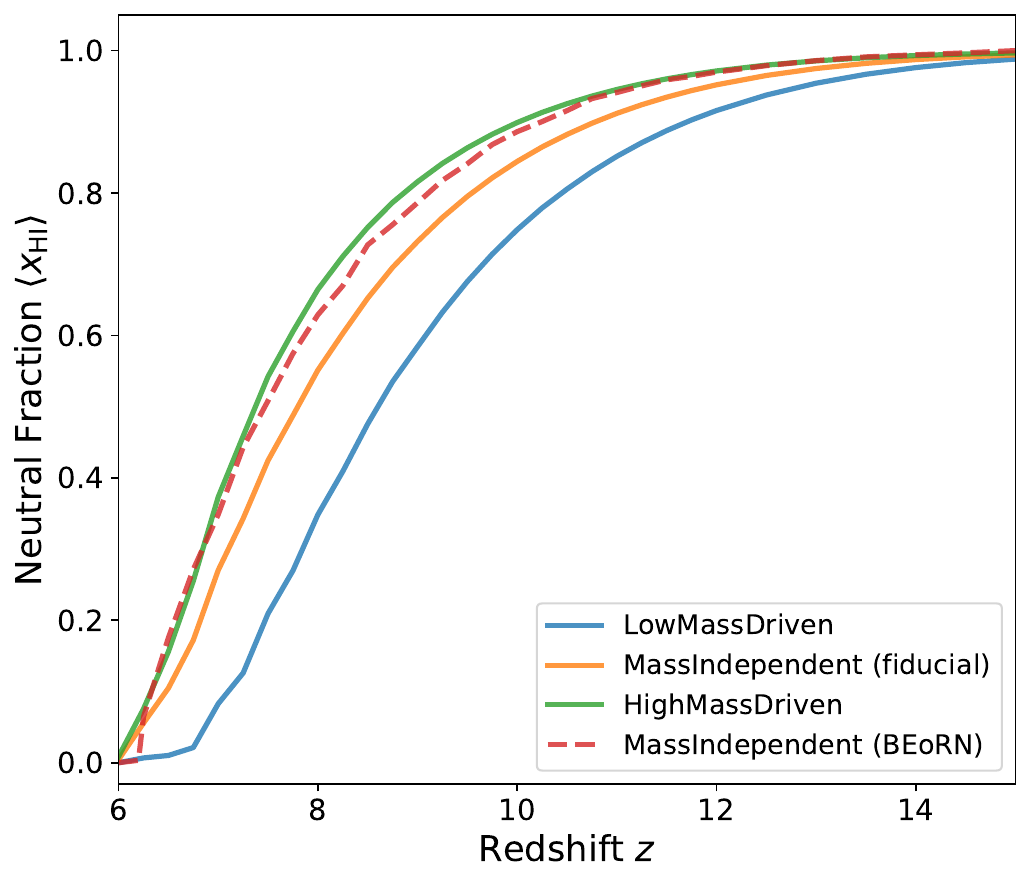}\\
    \includegraphics[width=0.96\linewidth,trim={0.0cm 0.0cm 0.1cm 0.2cm},clip]{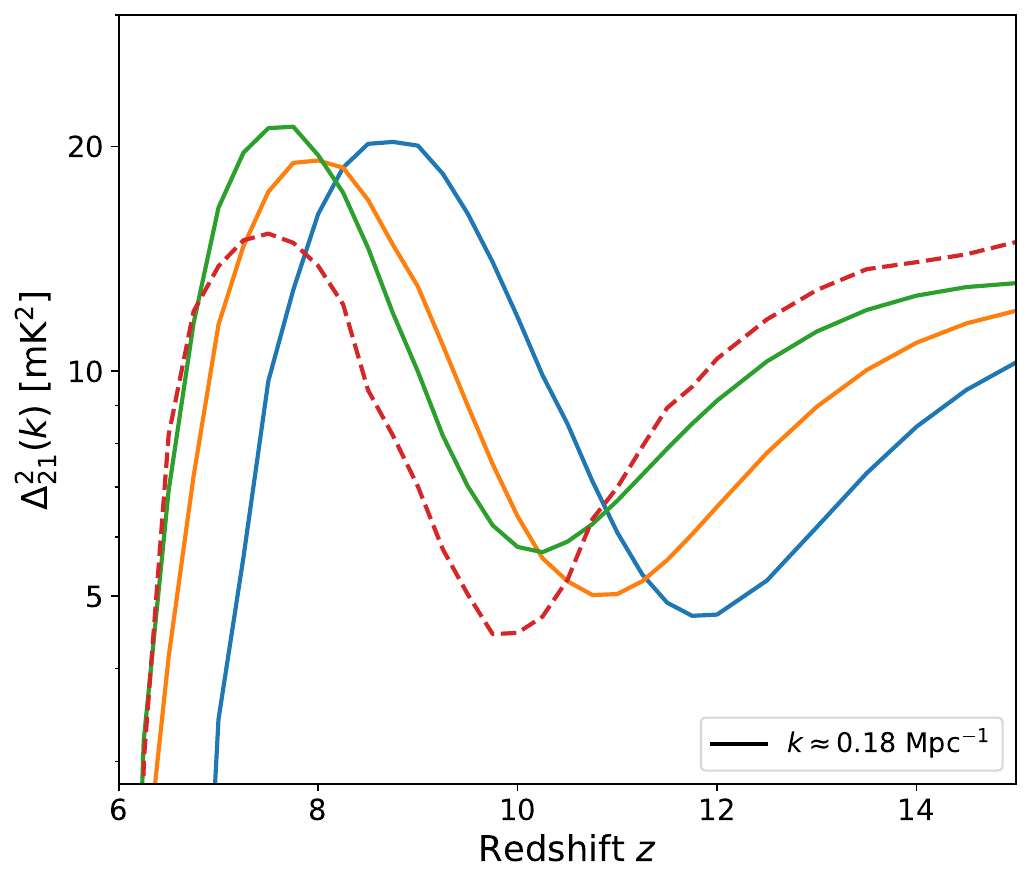}
    \vspace{-0.3cm}
    \caption{
    \textit{Top panel}: Reionization history for the three \textsc{21cmFAST} models (solid curves), differing in the mass dependence of the ionizing photon production efficiency, and for the \textsc{BEoRN} model (dashed curve).
    \textit{Bottom panel}: Evolution of the dimensionless 21-cm power spectrum $\Delta^2_{21}(k)$ with redshift at $k = 0.18\,\mathrm{Mpc}^{-1}$ for the same four models. This power spectrum evolution differs between models, reflecting their different ionization morphologies and reionization time-scales.
    }
    \label{fig:reion_hist}
\end{figure}

We simulate the 21-cm signal with two codes that share the same initial conditions. The main analysis uses the semi-numerical code \textsc{21cmFAST} (Sec.~\ref{sec:21cmfast_models}). To test whether our conclusions depend on the reionization algorithm, we also simulate reionization with the one-dimensional radiative transfer code \textsc{BEoRN} (Sec.~\ref{sec:beorn_model}). Table~\ref{tab:models} summarises the four models.

\begin{table*}
    \centering
    \caption{List of reionization models used in this work. All models share $\log_{10} f_{\mathrm{esc},10} = -1$, $\alpha_\mathrm{esc} = -0.5$, $N_\mathrm{ion,b} = 5000$ and $\log_{10}(M_\mathrm{turn}/M_\odot) = 8.7$.}
    \label{tab:models}
    \begin{tabular}{llccccl}
        \hline
        Model & Code & $\alpha_\star$ & $\alpha_\star + \alpha_\mathrm{esc}$ & $\log_{10} f_{\star,10}$ & $t_\star$ & Unresolved recombinations \\
        \hline
        \LowMassDriven{} & \textsc{21cmFAST} & 0.25 & $-0.25$ & $-1.3$ & 0.5 & \citet{sobacchi2014inhomogeneous} \\
        \MassIndependent{} (fiducial) & \textsc{21cmFAST} & 0.5 & 0 & $-1.3$ & 0.5 & \citet{sobacchi2014inhomogeneous} \\
        \HighMassDriven{} & \textsc{21cmFAST} & 0.75 & 0.25 & $-1.3$ & 0.5 & \citet{sobacchi2014inhomogeneous} \\
        \MassIndependent{} (\textsc{BEoRN}) & \textsc{BEoRN} & 0.5 & 0 & $-0.92$ & $[0.79(1+z)]^{-1}$ & Clumping factor in Eq.~\eqref{eq:clumping} \\
        \hline
    \end{tabular}
\end{table*}

\subsubsection{\textsc{21cmFAST} models}\label{sec:21cmfast_models}

\textsc{21cmFAST} simulates the reionization of the IGM using an excursion-set-based formalism developed by \citet{furlanetto2004growth}, applied to gridded cosmological fields \citep{mesinger2007efficient}. This approach allows for efficient modelling of large-scale ionization patterns driven by early astrophysical sources. In this framework, the ionization field is derived from the evolved matter density field of Sec.~\ref{sec:structure_formation} by comparing, within spheres of decreasing radii, the cumulative number of ionizing photons per baryon to the fraction of atoms that remains to be ionized, boosted by the cumulative recombinations \citep[e.g.,][]{mesinger201121cmfast}. Specifically, a voxel at coordinates $(\pmb{x}, z)$ is flagged as fully ionized if it satisfies the condition
\begin{eqnarray}
    n_\mathrm{ion}(\pmb{x},z|\mathcal{R},\delta_\mathcal{R}) \geq (1+\Bar{n}_\mathrm{rec})\,(1 - \Bar{x}_e)
    \label{eq:ion_criterion}
\end{eqnarray}
for any smoothing radius $\mathcal{R}$. Here $n_\mathrm{ion}$ is the cumulative number of ionizing photons per baryon produced within the sphere of radius $\mathcal{R}$ \citep{park2019inferring},
\begin{eqnarray}
    n_\mathrm{ion}(\pmb{x},z|\mathcal{R},\delta_\mathcal{R}) &=& \frac{N_\mathrm{ion,b}}{\Bar{\rho}_b} \int \mathrm{d}M_h \, \frac{\mathrm{d}n(M_h,z|\mathcal{R},\delta_\mathcal{R})}{\mathrm{d}M_h} \nonumber \\
    && \times\, f_\mathrm{duty}(M_h)\, f_\mathrm{esc}(M_h)\, M_\star(M_h) \ ,
    \label{eq:n_ion}
\end{eqnarray}
where $\mathrm{d}n/\mathrm{d}M_h$ is the conditional halo mass function within the sphere, $\Bar{\rho}_b$ is the mean baryon density, $\delta_\mathcal{R}$ is the matter overdensity smoothed with a spherical top-hat filter of the same radius, $\Bar{n}_\mathrm{rec}$ is the sphere-averaged cumulative number of recombinations per baryon due to unresolved sinks\footnote{Density fluctuations below the resolution of the simulations consume ionizing photons through recombinations and therefore affect the reionization topology \citep[e.g.,][]{mao2020clumping,bianco2021clumping}.}, modelled using the formalism of \citet{sobacchi2014inhomogeneous}, and $\Bar{x}_e$ is the sphere-averaged electron fraction produced by X-ray pre-ionization. As we assume a fully heated IGM ($T_\mathrm{s} \gg T_\mathrm{CMB}$), we do not follow the spin temperature and X-ray sources, and $\Bar{x}_e = 0$. In the model of \citet{sobacchi2014inhomogeneous}, the sinks are self-shielded gas clouds whose recombination rate grows with the local density and decreases with the local photoionization rate. Dense regions far from the sources therefore recombine faster and are ionized later, which slows the growth of the ionized regions as they become large.

We consider three models that differ in the mass-dependence of the ionizing photon production efficiency. We construct these models by varying $\alpha_\star$, with all other parameters held fixed at $\log_{10} f_{\star,10}=-1.3$ and $\log_{10}(M_\mathrm{turn}/M_\odot)=8.7$. In \textsc{21cmFAST}, the star formation time-scale enters only through the recombinations, and we fix it to $t_\star = 0.5$ in all models. The fiducial model (\MassIndependent{}) adopts $\alpha_\star = 0.5$, which gives $\alpha_\star + \alpha_\mathrm{esc} = 0$ and thus sources that produce the same number of ionizing photons per unit halo mass above the turnover mass. \LowMassDriven{} uses $\alpha_\star = 0.25$, giving low-mass haloes a larger relative contribution to the ionizing budget, while \HighMassDriven{} uses $\alpha_\star = 0.75$, making high-mass haloes the dominant ionizing sources. 

The top panel of Fig.~\ref{fig:reion_hist} shows the reionization history of these three models. The \LowMassDriven{} model begins and completes reionization earlier, while \HighMassDriven{} does so later, compared to the fiducial model. The fiducial parameters are those of \citet{park2019inferring}, which are consistent with the observed ultraviolet luminosity functions of high-redshift galaxies. The other two models are not calibrated against current observations, such as the timing of the end of reionization or the galaxy luminosity functions, and serve only to illustrate the impact of the source properties on the phase gradient spectrum.

The bottom panel of Fig.~\ref{fig:reion_hist} shows the evolution of the 21-cm power spectrum at $k = 0.18\,\mathrm{Mpc}^{-1}$ for all the models of Table~\ref{tab:models}. The power spectrum initially decreases as the densest peaks are ionized first, since they host the most ionizing sources \citep{georgiev2022large}. It then rises as the growing bubble network increasingly dominates the large-scale structure of the 21-cm field, reaching a peak when the typical ionized bubble size matches the chosen scale $k^{-1}$, before declining again once the bubbles exceed this scale \citep{lidz2008detecting}. The four models peak at different redshifts, reflecting differences in their typical bubble sizes at equivalent stages of reionization. The results in Sec.~\ref{sec:21cm_phases} draw on these four models to characterise the phase gradient spectrum of the 21-cm signal, while the full three-parameter space of the source model is explored in the Fisher forecast study presented in Sec.~\ref{sec:forecast_study}.

\subsubsection{\textsc{BEoRN} model}\label{sec:beorn_model}

\begin{figure*}
    \centering
    \includegraphics[width=1.0\linewidth,trim={-0.05cm 0.05cm 0.10cm 0.0cm},clip]{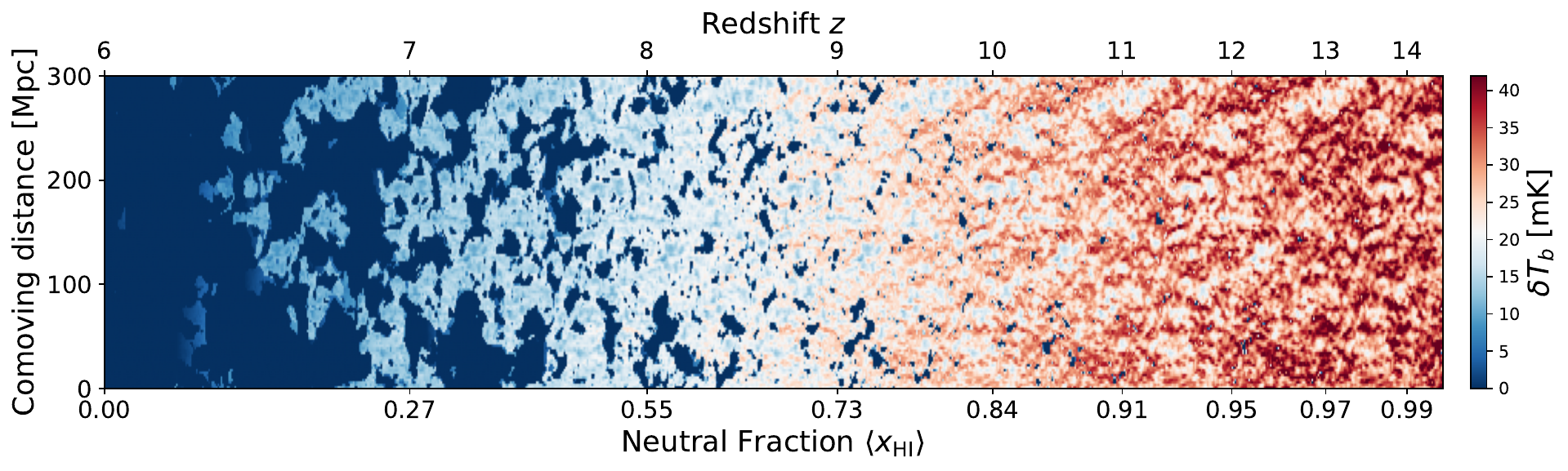}\\
    \begin{minipage}{0.49\linewidth}
        \centering
        \text{\large Matter density}\\
        \includegraphics[width=0.96\linewidth,trim={0.25cm 0.25cm 0.0cm 0.18cm},clip]{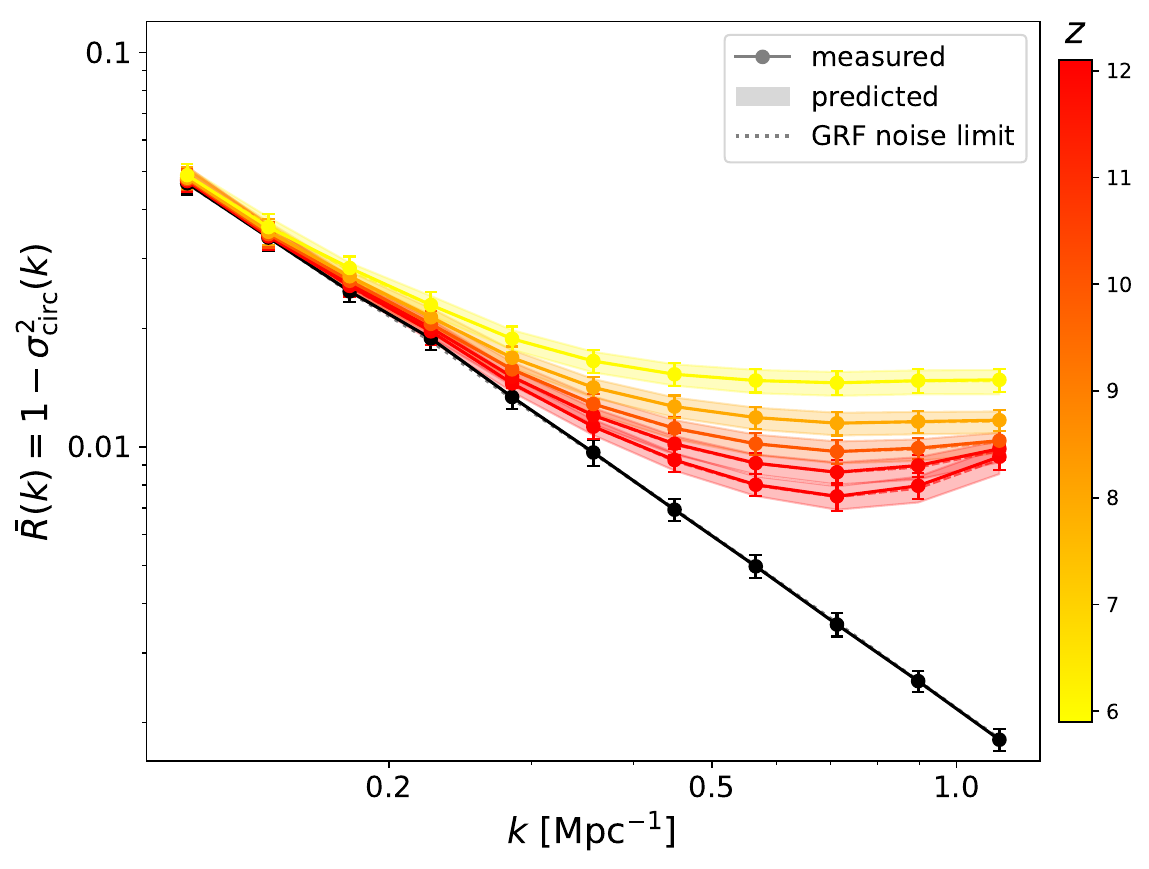}
    \end{minipage}
    \hfill
    \begin{minipage}{0.49\linewidth}
        \centering
        \text{\large 21-cm signal}\\
        \includegraphics[width=0.96\linewidth,trim={0.25cm 0.25cm 0.0cm 0.18cm},clip]{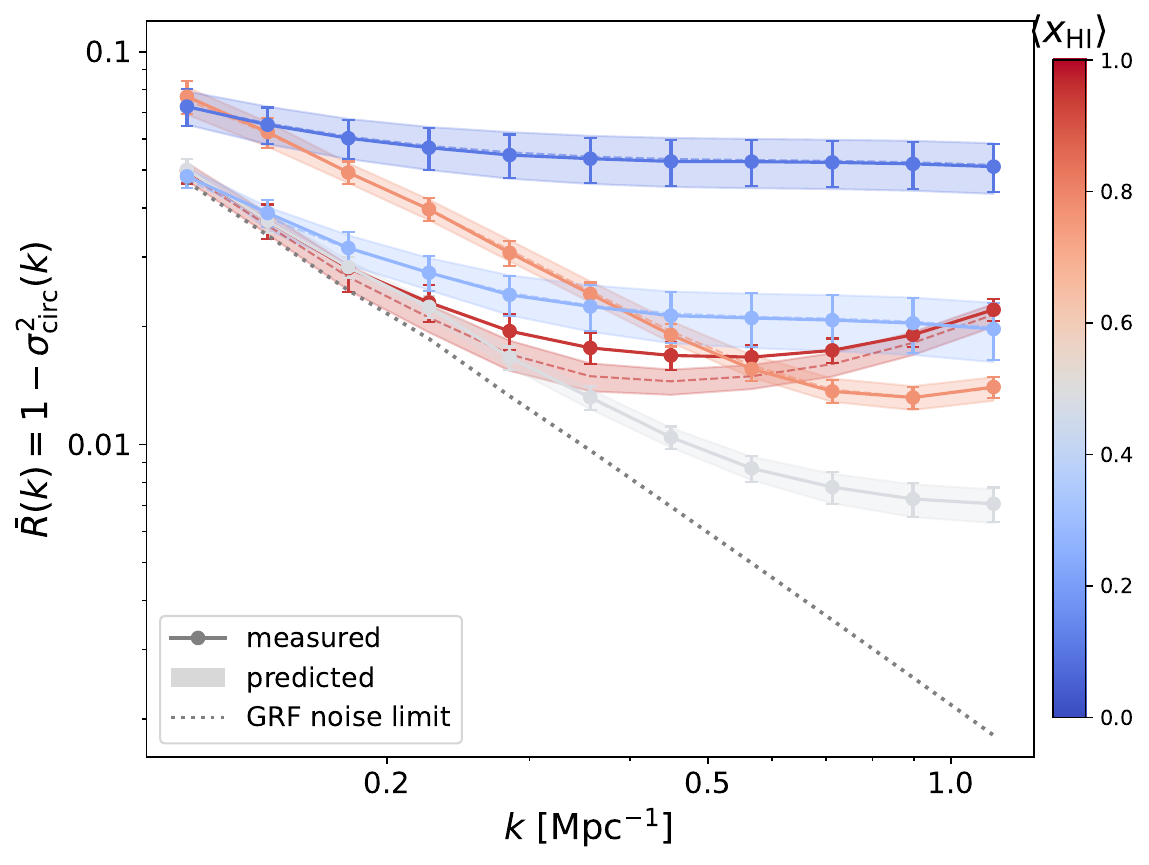}
    \end{minipage}
    \caption{
    All panels show the fiducial \MassIndependent{} model of \textsc{21cmFAST}.
    \textit{Top panel}: Lightcone slice of the 21-cm brightness temperature field from one of the 100 realisations. Redshift decreases to the left, and we see ionized bubbles (blue regions) forming and growing over time.
    \textit{Bottom left panel}: Phase gradient spectrum $\bar{R}(k)$ of the matter overdensity field $\delta_b$ at several redshifts. The black circles show the initial conditions of the simulation (a GRF), which agree with the analytic GRF noise limit of Eq.~\eqref{eq:grf_noise_limit} and hide it in this panel. Coloured circles show measured values and the dashed curves with shaded regions show the analytic prediction from Eq.~\eqref{eq:Rbar_full} with its $1\sigma$ uncertainty.
    \textit{Bottom right panel}: Phase gradient spectrum of the 21-cm brightness temperature field $\delta T_\mathrm{b}$, coloured by mean neutral fraction $\langle x_\mathrm{HI} \rangle$ (decreasing from red to blue). Dashed curves with shaded regions show the analytic prediction, and the grey dotted curve marks the GRF noise limit. Both bottom panels are computed from coeval simulation boxes of the model shown in the top panel, and the mean and standard deviation values are estimated over 100 realisations of varying initial cosmic seeds.
    }
    \label{fig:power_phase_spect_21cmfast_fiducial}
\end{figure*}

\textsc{BEoRN} \citep{schaeffer2023beorn} models reionization around individual dark matter haloes. For each halo, it follows the growth of the surrounding ionized bubble and paints the resulting spherical ionization profile on the grid. We run \textsc{BEoRN} on the evolved density fields of Sec.~\ref{sec:structure_formation}. Haloes with $M_h \geq 3\times10^{8}\,M_\odot$ are drawn from these fields with the stochastic halo sampler of \textsc{21cmFAST} presented in \citet{davies2025efficient}. The \textsc{BEoRN} model therefore shares the large-scale structure of the \textsc{21cmFAST} models and differs only in the reionization algorithm.

We use the source model of Sec.~\ref{sec:ionising_source} with $\alpha_\star = 0.5$, $\log_{10} f_{\mathrm{esc},10} = -1$, $\alpha_\mathrm{esc} = -0.5$ and $N_\mathrm{ion,b} = 5000$, as in \MassIndependent{}, together with a turnover function fitted to Eq.~\eqref{eq:fduty} with the same $M_\mathrm{turn}$. Stars form as the haloes grow, $\dot{M}_\star = f_\star (\Omega_b/\Omega_m) \dot{M}_h$, with an exponential mass accretion history $\dot{M}_h = \alpha_h (1+z) H(z) M_h$ and $\alpha_h = 0.79$ \citep{schneider2021halo,schaeffer2023beorn}. This star formation rate corresponds to Eq.~\eqref{eq:sfr} with a redshift-dependent $t_\star = [\alpha_h (1+z)]^{-1}$, which decreases from 0.18 at $z = 6$ to 0.08 at $z = 15$. \textsc{BEoRN} obtains the ionizing photons of each halo by integrating Eq.~\eqref{eq:Ndot_gamma} along its accretion history, with $f_\mathrm{duty}$ evaluated at each past halo mass. The turnover therefore also suppresses the star formation of the low-mass progenitors of each halo, whereas \textsc{21cmFAST} applies $f_\mathrm{duty}$ only at the current halo mass in Eq.~\eqref{eq:n_ion}. A \textsc{BEoRN} halo therefore produces fewer ionizing photons than a \textsc{21cmFAST} halo of the same mass and parameters, by about a factor of two at $z \approx 10$. We compensate for this deficit, and for the stronger recombinations described below, with $\log_{10} f_{\star,10} = -0.92$. We refer to this model as \MassIndependent{} (\textsc{BEoRN}), because its ionizing photon production per unit halo mass is independent of halo mass above the turnover mass, as in the fiducial model.

\textsc{BEoRN} accounts for recombinations in the growth of each ionized bubble through a clumping factor $C$. We adopt the redshift-dependent clumping factor of the intergalactic gas from \citet{mellema2006simulating},
\begin{eqnarray}
    C(z) = 27.466\, \exp\left(-0.114\,z + 0.001328\,z^2\right) \ ,
    \label{eq:clumping}
\end{eqnarray}
which increases from about 7 at $z = 15$ to about 15 at $z = 6$. Where ionized bubbles overlap, \textsc{BEoRN} redistributes the excess ionizing photons to nearby neutral cells. The \textsc{BEoRN} model therefore conserves ionizing photons, unlike the excursion-set algorithm of \textsc{21cmFAST}\footnote{Note that \textsc{21cmFAST} has implemented a photon-conserving method \citep{park2022calibrating}. However, we do not use that feature for the simulations produced for this work.}.
The top panel of Fig.~\ref{fig:reion_hist} shows that the reionization history of \MassIndependent{} (\textsc{BEoRN}) is close to that of \HighMassDriven{}, with reionization completing at $z \approx 6$. The bottom panel shows that the 21-cm power spectrum follows the same evolution as in the \textsc{21cmFAST} models. The initial decrease is more pronounced, with a minimum at $z \approx 10$, and the subsequent peak at $z \approx 7.5$ is lower. These differences reflect the different treatments of the ionizing sources and sinks in the two codes described above. We use this model in Sec.~\ref{sec:reion_models} to test whether the evolution of the phase gradient spectrum depends on the reionization algorithm.

\section{Results}
\label{sec:21cm_phases}

We investigate the information encoded in the Fourier phases of the 21-cm signal from reionization using the simulations presented in Fig.~\ref{fig:reion_hist}. In Sec.~\ref{sec:reion_bispectrum}, we study the non-Gaussian information in the matter distribution and the 21-cm signal probed by the phase gradient spectrum. Sec.~\ref{sec:reion_fields} presents the evolution of phase gradient spectrum during reionization epochs and isolates the physical origin of this non-Gaussian information by decomposing the 21-cm signal into its constituent fields. Finally, Sec.~\ref{sec:reion_models} demonstrates the capability of the phase gradient spectrum to distinguish between models.

\subsection{Phase gradient spectrum of the matter and 21-cm fields}\label{sec:reion_bispectrum}

The top panel of Fig.~\ref{fig:power_phase_spect_21cmfast_fiducial} shows a lightcone slice of the 21-cm brightness temperature field from the fiducial \MassIndependent{} model, which we use throughout this section. The signal transitions from density-modulated fluctuations at high redshift to a pattern increasingly dominated by the growing ionized bubble network, which is the primary driver of non-Gaussian structure during reionization. In Sec.~\ref{sec:reion_bispectrum_delta}, we first study the phase gradient spectrum of the matter overdensity field as a clean baseline, before examining the 21-cm signal in Sec.~\ref{sec:reion_bispectrum_21cm}.

\subsubsection{Matter overdensity field}\label{sec:reion_bispectrum_delta}

The matter overdensity field $\delta_b$ provides a clean baseline, as it is identical across all four models and evolves purely through non-linear gravitational structure formation. The bottom-left panel of Fig.~\ref{fig:power_phase_spect_21cmfast_fiducial} shows $\bar{R}(k)$ for the matter field, with curves coloured by redshift. The black reference points show $\bar{R}(k)$ measured from the initial conditions of the simulation, which are a GRF: they sit on the analytic GRF noise limit of Eq.~\eqref{eq:grf_noise_limit} (grey dotted curve in the bottom-right panel, hidden behind the black circles in the bottom-left panel), confirming that the estimator registers no spurious coherence for Gaussian input. At all scales, $\bar{R}(k)$ increases monotonically with time as structure forms non-linearly, as expected even though the density fields are evolved with 2LPT, an approximation to the full non-linear gravitational evolution. At $k \gtrsim 0.3\,\mathrm{Mpc}^{-1}$, the curves lie clearly above the GRF noise limit. At larger scales, the matter field is close to Gaussian, and the curves approach the GRF noise limit. The leading-order bispectrum-based prediction from Eq.~\eqref{eq:Rbar_full} (dashed curves with shaded 1$\sigma$ bands) agrees well with the measured spectra across all epochs and scales, validating the analytic connection derived in Appendix~\ref{app:bispectrum_connection}.

\subsubsection{21-cm brightness temperature}\label{sec:reion_bispectrum_21cm}

Next, we study the 21-cm brightness temperature ($\delta T_\mathrm{b}$), which is illustrated in the top panel of Fig.~\ref{fig:power_phase_spect_21cmfast_fiducial}. Before reionization begins (high $z$), $\delta T_\mathrm{b}$ traces the underlying matter density field. As the first ionizing sources switch on, small dark bubbles of ionized hydrogen appear against the bright neutral background, growing and merging as reionization progresses. At mid-reionization, the field is dominated by a complex, percolating network of ionized and neutral regions. By the final stages, only isolated compact neutral islands remain, before the signal vanishes entirely.

The bottom-right panel of Fig.~\ref{fig:power_phase_spect_21cmfast_fiducial} shows $\bar{R}(k)$ for the $\delta T_\mathrm{b}$, with curves coloured by $\langle x_\mathrm{HI} \rangle$. In contrast to the matter density case, $\bar{R}(k)$ does not increase monotonically with cosmic time. Instead, the phase gradient spectrum of the 21-cm signal rises and falls as the ionized bubble population grows, percolates, and eventually leaves only isolated neutral islands. This evolution directly reflects the stages visible in the lightcone above. Crucially, the leading-order bispectrum-based prediction (dashed curves with shaded bands) also reproduces this non-monotonic evolution at all epochs, and at $k \gtrsim 0.3\,\mathrm{Mpc}^{-1}$ the measured spectra lie clearly above the GRF noise limit (grey dotted curve) at all stages, confirming that the behaviour is a genuine signature of the non-Gaussian ionization structure. In the log-normal tests, the measurement and the prediction agree to within about 1 per cent (Sec.~\ref{sec:lognormal_spectrum}). For the 21-cm field, the measurement and the prediction agree within the scatter at most stages. The exception is the earliest stage shown ($\langle x_\mathrm{HI} \rangle \approx 0.94$), where the measured spectrum lies above the prediction by about $1\sigma$ at $0.25 \lesssim k \lesssim 0.6\,\mathrm{Mpc}^{-1}$. The physical origin of this evolution is examined in Sec.~\ref{sec:reion_fields}.

\subsection{Sources of non-Gaussian information}\label{sec:reion_fields}

\begin{figure*}
    \centering
    \includegraphics[width=0.95\linewidth,trim={0.0cm 0.0cm 0.0cm 0.0cm},clip]{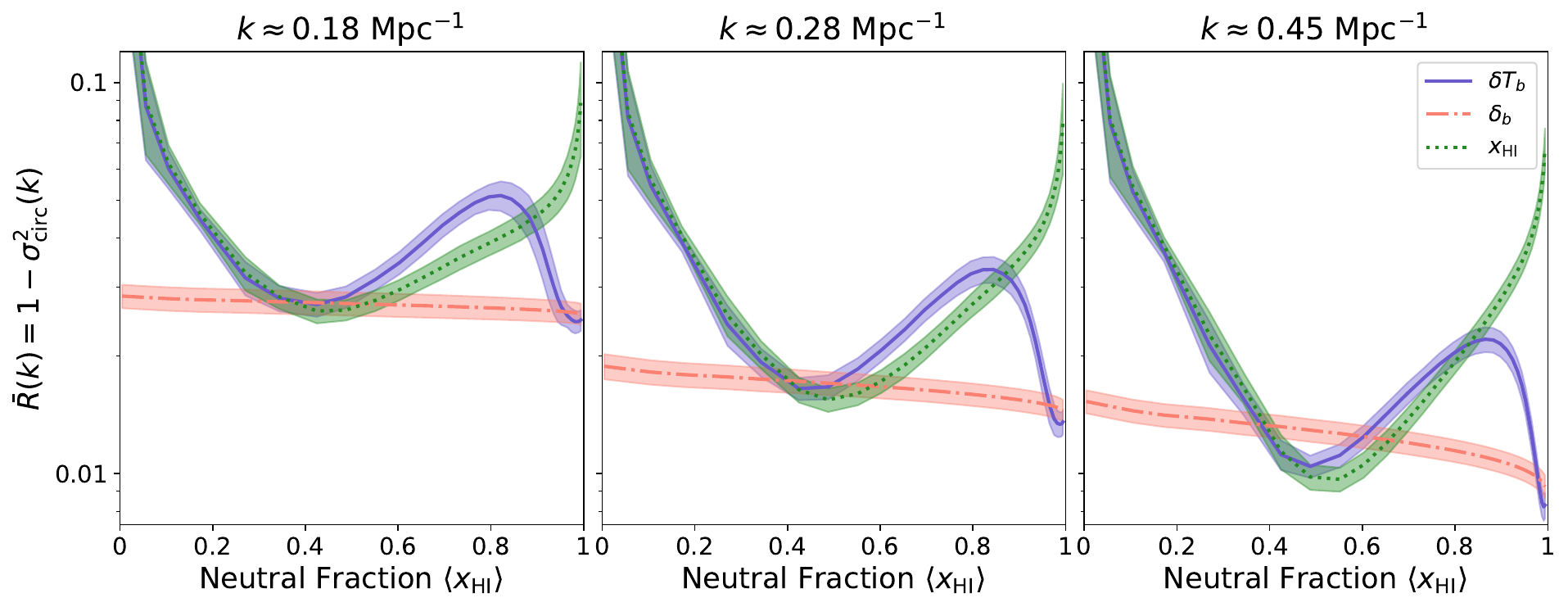}\\
    \includegraphics[width=0.97\linewidth,trim={0.1cm 0.0cm 0.0cm -0.05cm},clip]{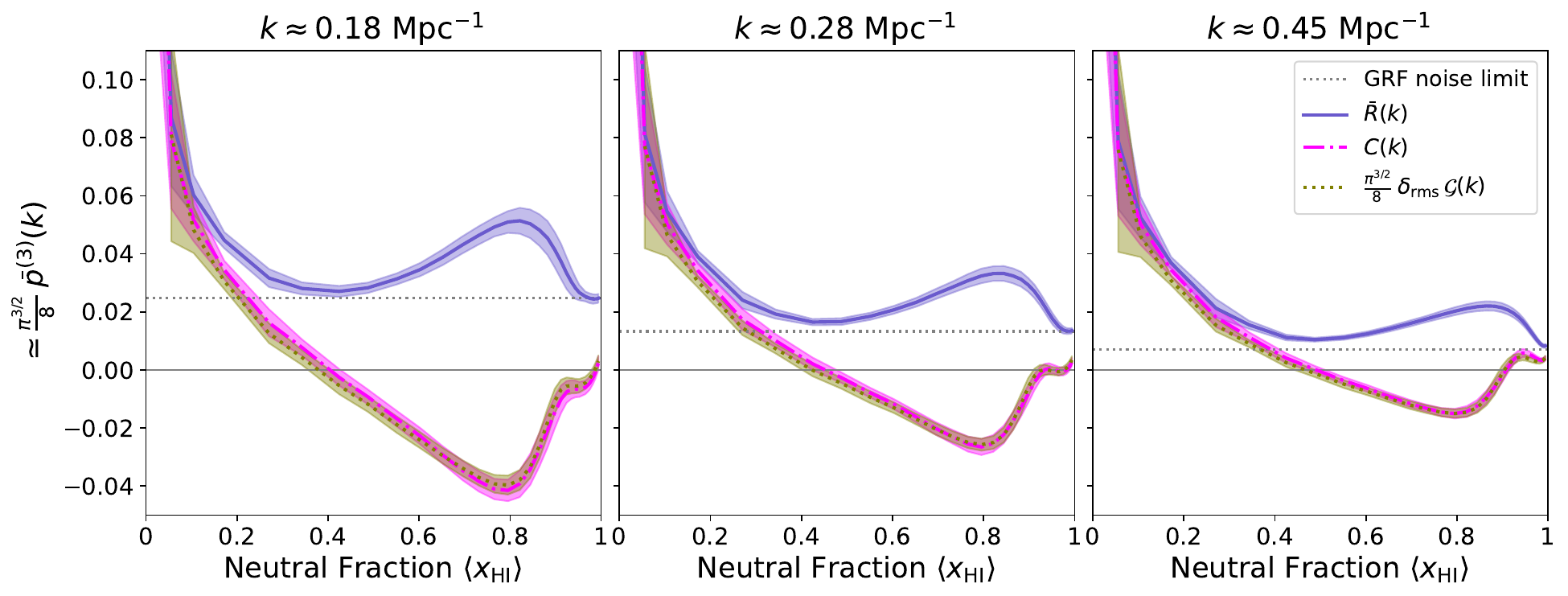}
    \caption{
    Evolution of the phase coherence of the fiducial \MassIndependent{} model with the mean neutral fraction $\langle x_\mathrm{HI} \rangle$, for three representative wavenumber bins (columns). \textit{Top row}: $\bar{R}(k)$ of the matter overdensity $\delta_b$ (dash-dotted), the neutral hydrogen fraction $x_\mathrm{HI}$ (dotted), and the 21-cm brightness temperature $\delta T_\mathrm{b}$ (solid). \textit{Bottom row}: $\bar{R}(k)$ (solid), the closure-phase spectrum $C(k)$ (dash-dotted) and the prediction $\frac{\pi^{3/2}}{8}\,\delta_\mathrm{rms}\,\mathcal{G}(k)$ from the position-dependent power spectrum (dotted) of $\delta T_\mathrm{b}$, all on the scale of $\frac{\pi^{3/2}}{8}\,\bar{p}^{(3)}(k)$. The grey dotted line marks the GRF noise limit. In all panels, the curves and shaded regions show the mean and standard deviation over 100 realisations of varying initial cosmic seeds.
    }
    \label{fig:phase_grad_field_decomp}
\end{figure*}

In this section, we identify what drives the evolution of the phase gradient spectrum of the 21-cm signal. We first compare this evolution with that of the constituent fields of the signal (Sec.~\ref{sec:reion_fields_constituents}). We then interpret the evolution through the connection between $\bar{R}(k)$ and the squeezed bispectrum (Sec.~\ref{sec:reion_fields_bispectrum}).

\subsubsection{Constituent fields of the 21-cm signal}\label{sec:reion_fields_constituents}

We now ask which physical component of the 21-cm signal drives the non-monotonic evolution seen in Sec.~\ref{sec:reion_bispectrum}. The top row of Fig.~\ref{fig:phase_grad_field_decomp} shows $\bar{R}(k)$ as a function of $\langle x_\mathrm{HI} \rangle$ for each constituent field of the fiducial model. The three columns show three representative wavenumbers, $k \approx 0.18$, 0.28 and $0.45\,\mathrm{Mpc}^{-1}$.
The matter overdensity $\delta_b$ (dash-dotted curves) shows a steady, monotonically increasing $\bar{R}(k)$ throughout reionization, which is consistent with what we found in Sec.~\ref{sec:reion_bispectrum_delta}. The neutral hydrogen fraction $x_\mathrm{HI}$ (dotted curves), by contrast, exhibits a non-monotonic evolution: at the earliest stages, the field is nearly uniform with only a few isolated ionized cells, and this sharp binary contrast between the almost entirely neutral background and sparsely ionized regions drives a high initial phase coherence. As reionization progresses and the bubble network grows and percolates, the morphological complexity increases and the coherence decreases, reaching a minimum around mid-reionization. Thereafter, $\bar{R}$ rises again as the remaining neutral regions contract into compact, isolated islands with a strongly non-Gaussian morphology.

At the earliest stages, the 21-cm brightness temperature $\delta T_\mathrm{b}$ is dominated by density fluctuations and its phase gradient spectrum closely tracks that of $\delta_b$. As reionization advances, the 21-cm field is increasingly modulated by the ionization structure. Its phase gradient spectrum rises above those of both $\delta_b$ and $x_\mathrm{HI}$, peaks at $\langle x_\mathrm{HI} \rangle \approx 0.8$, and converges onto that of $x_\mathrm{HI}$ at $\langle x_\mathrm{HI} \rangle \approx 0.5$, following it for the remainder of reionization. The early peak has no counterpart in either constituent field, and we explain it in Sec.~\ref{sec:reion_fields_bispectrum}. The non-Gaussian information in the phase gradient spectrum of the 21-cm signal is overwhelmingly sourced by the ionization field topology, not by the matter density.

\subsubsection{Connection to the squeezed bispectrum}\label{sec:reion_fields_bispectrum}

After mid-reionization, the phase coherence of the 21-cm signal follows that of $x_\mathrm{HI}$. The evolution during the early stages, however, cannot be understood by comparing with the constituent fields alone. To understand the full evolution, we relate $\bar{R}(k)$ to the physical response of the small-scale fluctuations to long-wavelength modes and to two statistics that keep the sign of the squeezed bispectrum.

The squeezed bispectrum has a simple physical interpretation. A long-wavelength mode raises the local density in some parts of the volume and lowers it in others, and the short-wavelength modes respond to this. To first order, the local power spectrum is $P(k\,|\,\delta_L) \approx P(k)\left[1 + \mathcal{G}(k)\,\delta_L\right]$, where $\delta_L$ is the long-wavelength overdensity and $\mathcal{G}(k) = \partial \ln P(k\,|\,\delta_L)/\partial \delta_L$ is the response function. The squeezed bispectrum is then $B_\mathrm{sq}(\Delta k, k) \approx P(\Delta k)\,P(k)\,\mathcal{G}(k)$ \citep[e.g.,][]{bernardeau2002large,chiang2014position}, and Eq.~\eqref{eq:variance_spectrum_main} becomes (Appendix~\ref{app:interpretation})
\begin{eqnarray}
    \langle \bar{R}(k, \Delta k) \rangle \approx \frac{\pi^{3/2}}{8}\, \delta_\mathrm{rms}(\Delta k)\, \left|\mathcal{G}(k)\right| \, ,
    \label{eq:variance_response}
\end{eqnarray}
where $\delta_\mathrm{rms}(\Delta k) = \sqrt{P(\Delta k)/V}$ is the root-mean-square (rms) amplitude of the long-wavelength mode in the volume. The phase gradient spectrum therefore measures the magnitude of the response of the small-scale power to the long-wavelength modes, weighted by their typical amplitude.

The position-dependent power spectrum measures the same response in configuration space \citep{chiang2014position,giri2019position}. The survey volume is divided into subvolumes, and the power spectrum $P(k, \pmb{r}_L)$ of the subvolume centred at $\pmb{r}_L$ is correlated with its mean overdensity $\bar{\delta}(\pmb{r}_L)$. In the squeezed limit, this correlation gives the response directly,
\begin{eqnarray}
    \frac{\langle P(k, \pmb{r}_L)\,\bar{\delta}(\pmb{r}_L)\rangle}{\sigma_L^2\,P(k)} \approx \mathcal{G}(k) \, ,
    \label{eq:position_dependent_ps}
\end{eqnarray}
where $\sigma_L^2$ is the variance of $\bar{\delta}$ between subvolumes. Unlike $\bar{R}(k)$, this estimate keeps the sign of the response.

The phase gradient is also related to the closure phase, the phase of the product of three Fourier modes whose wavevectors form a closed triangle, which has been proposed as a probe of the 21-cm signal \citep{thyagarajan2018detecting,thyagarajan2020detectionI}. For the triangle $(\Delta\pmb{k},\, \pmb{k},\, -\pmb{k}-\Delta\pmb{k})$, the closure phase is $\psi = \Phi_{\Delta\pmb{k}} - \mathcal{D}$ (Appendix~\ref{app:conditional}). Averaging the closure phases over the $k$-bin gives
\begin{eqnarray}
    Z_\psi(k, \Delta\pmb{k}) \equiv \frac{1}{N_k}\sum_{|\pmb{k}| \in \mathrm{bin}} e^{\mathrm{i}\psi} = e^{\mathrm{i}\Phi_{\Delta\pmb{k}}}\, Z^*(k, \Delta\pmb{k}) \, ,
    \label{eq:closure_phase_average}
\end{eqnarray}
where $Z(k, \Delta\pmb{k}) = N_k^{-1}\sum e^{\mathrm{i}\mathcal{D}}$ is the mean phase vector, whose modulus is $\bar{R}(k, \Delta\pmb{k})$. For a given long-wavelength mode, the ensemble mean of $Z_\psi$ is real and proportional to $\bar{p}^{(3)}(k, \Delta k)$ (Appendix~\ref{app:estimator}). The real part of $Z_\psi$ therefore keeps the sign of the bispectrum. The mode-counting noise also averages to zero in this real part, which leaves no GRF noise limit. Averaging the real part over the long-wavelength modes defines the closure-phase spectrum,
\begin{eqnarray}
    C(k) \equiv \frac{1}{N_L}\sum_{\Delta\pmb{k}} \mathrm{Re}\, Z_\psi(k, \Delta\pmb{k}) \, , \qquad \langle C(k) \rangle = \frac{\pi^{3/2}}{8}\, \bar{p}^{(3)}(k) \, ,
    \label{eq:closure_phase_spectrum}
\end{eqnarray}
which has the same scale as $\bar{R}(k)$ in Eq.~\eqref{eq:variance_spectrum_main} but keeps the sign of $\bar{p}^{(3)}$. Dividing by the amplitude of the long-wavelength mode further removes the scatter of this amplitude between realisations, and the result is closely related to the direct estimate of the normalised squeezed bispectrum (defined in Appendix~\ref{app:bispectrum_comparison}).

The simulations contain the long-wavelength modes, which allows us to measure these signed quantities and to test the interpretation of $\bar{R}(k)$ directly. The bottom row of Fig.~\ref{fig:phase_grad_field_decomp} shows the closure-phase spectrum $C(k)$ (Eq.~\ref{eq:closure_phase_spectrum}) of the 21-cm signal, computed from the long-wavelength modes of $\bar{R}(k)$ except the three along the line of sight, which interferometers do not measure. The bottom row also shows the prediction of Eq.~\eqref{eq:variance_response} with the sign of the response kept. Here $\mathcal{G}(k)$ is the response of the small-scale 21-cm power to the long-wavelength 21-cm mode, measured with the position-dependent power spectrum (Eq.~\ref{eq:position_dependent_ps}) in $4^3$ subvolumes of 75\,Mpc, and $\delta_\mathrm{rms}$ is the rms amplitude of the long-wavelength modes used for $\bar{R}(k)$. The two agree closely at all three wavenumbers and at all stages, although the long-wavelength mode of the position-dependent power spectrum is set by the subvolume size and differs from those of $C(k)$ and $\bar{R}(k)$. This agreement confirms the response-function form of Eq.~\eqref{eq:variance_response} for the 21-cm signal. $\bar{R}(k)$ follows $|C(k)|$ where the signal is strong and approaches the GRF noise limit where $C(k)$ crosses zero, as expected from Eq.~\eqref{eq:Rbar_full}.

$C(k)$ also shows the sign of the response, which $\bar{R}(k)$ discards. At the earliest stages, $C(k)$ is small and positive, as the 21-cm signal traces the matter density. $C(k)$ turns negative at $\langle x_\mathrm{HI} \rangle \approx 0.95$, reaches its most negative value at $\langle x_\mathrm{HI} \rangle \approx 0.8$, where $\bar{R}(k)$ peaks, and crosses zero again at $\langle x_\mathrm{HI} \rangle \approx 0.4$, where $\bar{R}(k)$ has its trough. During the late stages, $C(k)$ is positive and grows steeply. The rise–fall–rise pattern of $\bar{R}(k)$ in Fig.~\ref{fig:power_phase_spect_21cmfast_fiducial} is therefore the modulus of a response that changes sign. The early peak of $\bar{R}(k)$ above both constituent fields corresponds to a strong negative response of the small-scale 21-cm power to the long-wavelength 21-cm mode, and the trough marks the subsequent sign change. In a linear bias model of the neutral fraction, the sign of $C(k)$ is that of $(1+b_1)\,\mathcal{G}_b(k)$, where $b_1$ is the large-scale bias of the neutral fraction and $\mathcal{G}_b$ is the response to the long-wavelength matter density (Appendix~\ref{app:interpretation}). A sign change of $C(k)$ can therefore arise from the response to the matter density or from the sign change of the large-scale 21-cm bias at $b_1 = -1$ \citep{giri2019position}. Only the former also produces a trough in $\bar{R}(k)$, as the factor $(1+b_1)$ cancels in $\bar{R}(k)$.

\begin{figure*}
    \centering
    \includegraphics[width=0.95\linewidth,trim={0.0cm 0.0cm 0.0cm 0.0cm},clip]{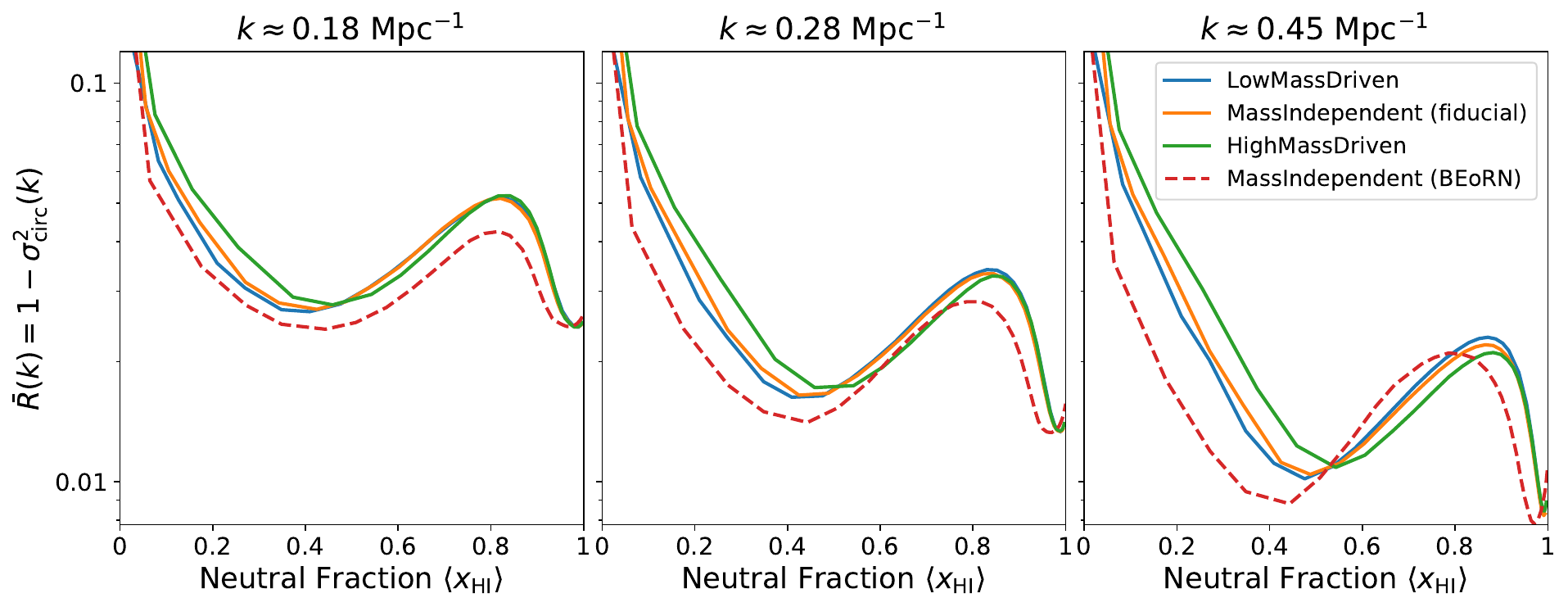}\\
    \includegraphics[width=0.95\linewidth,trim={0.0cm 0.0cm 0.0cm 0.95cm},clip]{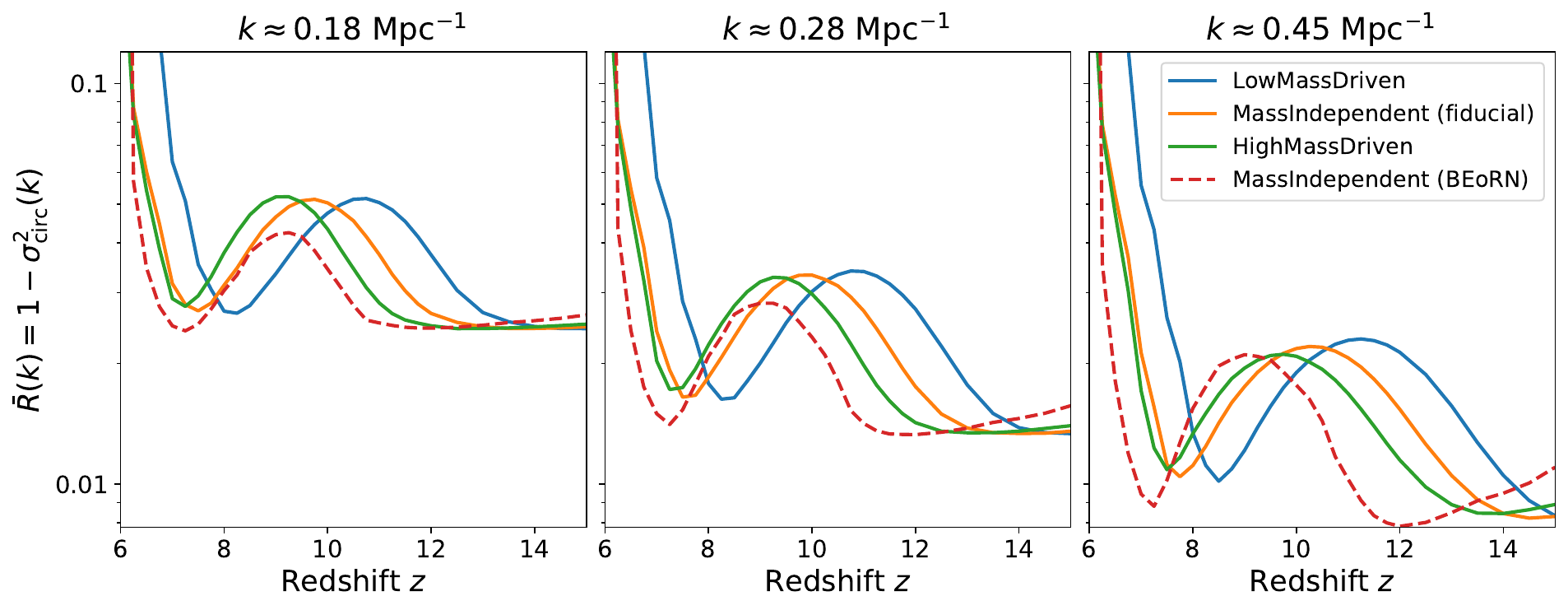}
    \caption{
    Phase gradient spectrum $\bar{R}(k)$ of $\delta T_\mathrm{b}$ for the three \textsc{21cmFAST} models (solid) and the \textsc{BEoRN} model (dashed), for three representative wavenumber bins (columns), as a function of the mean neutral fraction $\langle x_\mathrm{HI} \rangle$ (\textit{top row}) and of redshift (\textit{bottom row}). All curves are means over 100 realisations.
    }
    \label{fig:phase_grad_spect_model_dependence}
\end{figure*}

This trough is closely related to the peak of the 21-cm power spectrum as a function of redshift at fixed $k$ (bottom panel of Fig.~\ref{fig:reion_hist}). Both features occur around the stage at which the typical ionized bubble size becomes $\sim k^{-1}$ \citep{lidz2008detecting}. Before this stage, a long-wavelength matter overdensity enhances the small-scale 21-cm fluctuations. After this stage, the same overdensity is largely ionized and suppresses them \citep{giri2019position}. The response $\mathcal{G}_b$ to the long-wavelength matter density therefore changes sign. The two features do not coincide exactly. At $k \approx 0.18\,\mathrm{Mpc}^{-1}$, the trough in $\bar{R}(k)$ (bottom left panel of Fig.~\ref{fig:phase_grad_spect_model_dependence}) occurs at a redshift lower by $\Delta z \approx 0.25$--$0.75$ than the peak of $\Delta^2_{21}(k)$ for all four models. The sign change of the response therefore follows shortly after the power spectrum turns over. In our simulations, the subsequent rise at $\langle x_\mathrm{HI} \rangle \lesssim 0.4$ shows that the magnitude of the response grows again after the sign change. At these stages, the 21-cm signal is confined to the remaining neutral islands, which lie preferentially in long-wavelength underdensities. The growing response is consistent with the small-scale fluctuations of these islands being increasingly modulated by the long-wavelength modes. The high $\bar{R}$ of the $x_\mathrm{HI}$ field in Fig.~\ref{fig:phase_grad_field_decomp} supports this picture.

$C(k)$ and the position-dependent power spectrum thus give a more complete picture than $\bar{R}(k)$, but both require the long-wavelength modes to be measured. $C(k)$ needs the phase of each long-wavelength mode, and the position-dependent power spectrum needs the mean overdensity of each subvolume, which is set by the same modes. With $|\Delta\pmb{k}| \leq 3\,k_\mathrm{f} \approx 0.06\,\mathrm{Mpc}^{-1}$, these modes have small line-of-sight wavenumbers, where the spectrally smooth foregrounds dominate \citep[e.g.,][]{datta2012light}. Foreground removal can recover part of these modes but leaves residuals that are largest there \citep[e.g.,][]{mertens2020improved,giri2026foregrounds}. The modulus in $\bar{R} = |Z| = |Z_\psi|$ removes the phase of the long-wavelength mode, and the phase gradient spectrum therefore needs only the short-wavelength modes. Discarding this phase has three costs: (i) the sign of the response is lost, (ii) the mode-counting noise no longer averages to zero and sets the GRF noise limit, and (iii) the amplitude of each long-wavelength mode scatters the signal between realisations. The position-dependent power spectrum has the further cost that each subvolume contains only a fraction of the short-wavelength modes, while the phase gradient spectrum uses the full volume. We compare the constraining power of $\bar{R}(k)$ with that of a direct estimate of the normalised squeezed bispectrum, which keeps the sign, in Sec.~\ref{sec:constraints}.

\subsection{Reionization model comparison}\label{sec:reion_models}

A key question is whether the non-monotonic evolution of the phase gradient spectrum is a generic property of reionization or an artefact of the particular simulation. Fig.~\ref{fig:phase_grad_spect_model_dependence} compares $\bar{R}(k)$ across the three \textsc{21cmFAST} models and the \textsc{BEoRN} model. The top row shows $\bar{R}(k)$ as a function of $\langle x_\mathrm{HI} \rangle$, which removes differences in reionization timing and compares the models at equivalent stages of reionization. The bottom row shows $\bar{R}(k)$ as a function of redshift, which is what an observer would measure directly.

The top row shows that during the early stages of reionization the phase gradient spectra of the three \textsc{21cmFAST} models are nearly identical at all three wavenumbers. At a given neutral fraction, the early ionization morphology of these models is therefore set mainly by the shared density field and the excursion-set algorithm, and only weakly by the mass dependence of the sources. Below $\langle x_\mathrm{HI} \rangle \approx 0.9$, the curves diverge, first at small scales, reflecting the different bubble morphologies produced by each source prescription. All three models nonetheless exhibit the same non-monotonic rise--fall--rise behaviour seen in the fiducial case, indicating that this behaviour is a signature of reionization morphology rather than of a particular source model.

The \MassIndependent{} (\textsc{BEoRN}) model (dashed curves) tests whether this behaviour depends on the reionization algorithm. Its phase gradient spectrum shows the same rise--fall--rise evolution at all three wavenumbers. For $\langle x_\mathrm{HI} \rangle \gtrsim 0.5$, it follows the \textsc{21cmFAST} models, with a peak about 20 per cent lower at $k \lesssim 0.3\,\mathrm{Mpc}^{-1}$, and its minimum occurs at a similar neutral fraction, $\langle x_\mathrm{HI} \rangle \approx 0.45$. During the later stages, $\bar{R}(k)$ rises again but remains below the \textsc{21cmFAST} models for $0.05 \lesssim \langle x_\mathrm{HI} \rangle \lesssim 0.4$, by up to a factor of two at $k \approx 0.45\,\mathrm{Mpc}^{-1}$, and it approaches the \textsc{21cmFAST} values at the end of reionization. These differences arise even though both codes use the same density field. The ionizing sources of \textsc{BEoRN} are the haloes drawn with the stochastic halo sampler, which do not follow exactly the mean collapsed fraction of each smoothed region used by the excursion-set algorithm of \textsc{21cmFAST}. \textsc{BEoRN} also grows a spherical bubble around each halo and treats recombinations with a global clumping factor (Sec.~\ref{sec:beorn_model}). The ionized regions, and hence $\bar{R}(k)$, therefore differ between the two codes at the same neutral fraction. However, the key result of this comparison is that a different reionization algorithm also shows the same non-monotonic evolution. This evolution of the phase gradient spectrum is therefore not specific to the excursion-set algorithm of \textsc{21cmFAST}.

In the bottom row, where the same phase gradient spectra are shown as a function of observed redshift, all four models are distinguishable once reionization is under way. As the models reionize at different times, the near-identical early evolution of the \textsc{21cmFAST} models in the top row appears here as a shift in redshift. The redshifts of the peak and trough of $\bar{R}(k)$ follow the reionization histories in Fig.~\ref{fig:reion_hist}. \LowMassDriven{} reaches these features at the earliest redshifts, and \HighMassDriven{} and \MassIndependent{} (\textsc{BEoRN}) at the latest. In every model, the trough marks the sign change of the response of the small-scale 21-cm power to the long-wavelength modes (Sec.~\ref{sec:reion_fields_bispectrum}). This sensitivity of the phase gradient spectrum to the timing and morphology of reionization motivates its use in astrophysical parameter inference, which we explore in Sec.~\ref{sec:forecast_study}.

\section{Fisher forecast study}\label{sec:forecast_study}

Section~\ref{sec:reion_models} showed that the phase gradient spectrum is sensitive to the reionization morphology and can distinguish between source prescriptions. We now quantify this sensitivity through a Fisher forecast, assessing the information content of the phase gradient spectrum for constraining the astrophysical parameters of reionization. The detailed description of the methodology is given in Appendix~\ref{app:fisher}. In Sec.~\ref{sec:mock_obs}, we introduce the instrumental simulation used to model a mock SKA-Low observation, which is required to estimate the covariance matrix. In Sec.~\ref{sec:constraints}, we present the expected constraints on the astrophysical parameters.

\subsection{Mock observation}\label{sec:mock_obs}

We model a mock \textsc{SKA-Low} observation using the \textsc{AA*} configuration \citep[e.g.,][]{bag2026imaging,acharya2026inferring}, which comprises 307 stations. The system temperature is modelled as
\begin{eqnarray}
    T_\mathrm{sys}(\nu) = \Big[ 140 + 60 \left(\frac{\nu}{150\,\mathrm{MHz}}\right)^{-2.55} \Big] \,\mathrm{K} \ ,
    \label{eq:tsys}
\end{eqnarray}
where the constant term represents the receiver temperature and the power law represents the sky temperature. The rms thermal noise of a visibility measured by one baseline is
\begin{eqnarray}
    \sigma_\mathrm{N} = \frac{\sqrt{2}\,k_\mathrm{B}\,T_\mathrm{sys}}{A_\mathrm{eff}\,\sqrt{\Delta\nu\,t_\mathrm{int}}} \ ,
    \label{eq:noise_rms}
\end{eqnarray}
where $A_\mathrm{eff}$ is the effective collecting area per station, $\Delta\nu$ is the channel bandwidth, and $t_\mathrm{int}$ is the total integration time \citep{giri2018optimal}. The $uv$ coverage is computed for 6\,h of observation per day using the \textsc{Tools21cm} package \citep{giri2020tools21cm}, and thermal noise realisations consistent with this baseline distribution are generated and added directly to the simulated 21-cm brightness temperature cubes. In Sec.~\ref{sec:constraints}, we present results for total observation times of $t_\mathrm{obs} = 1000$ and $10\,000\,\mathrm{h}$, combining measurements at $z = 10$, 9, 8, 7 and 6.5.

\begin{figure}
    \centering
    \includegraphics[width=0.96\linewidth]{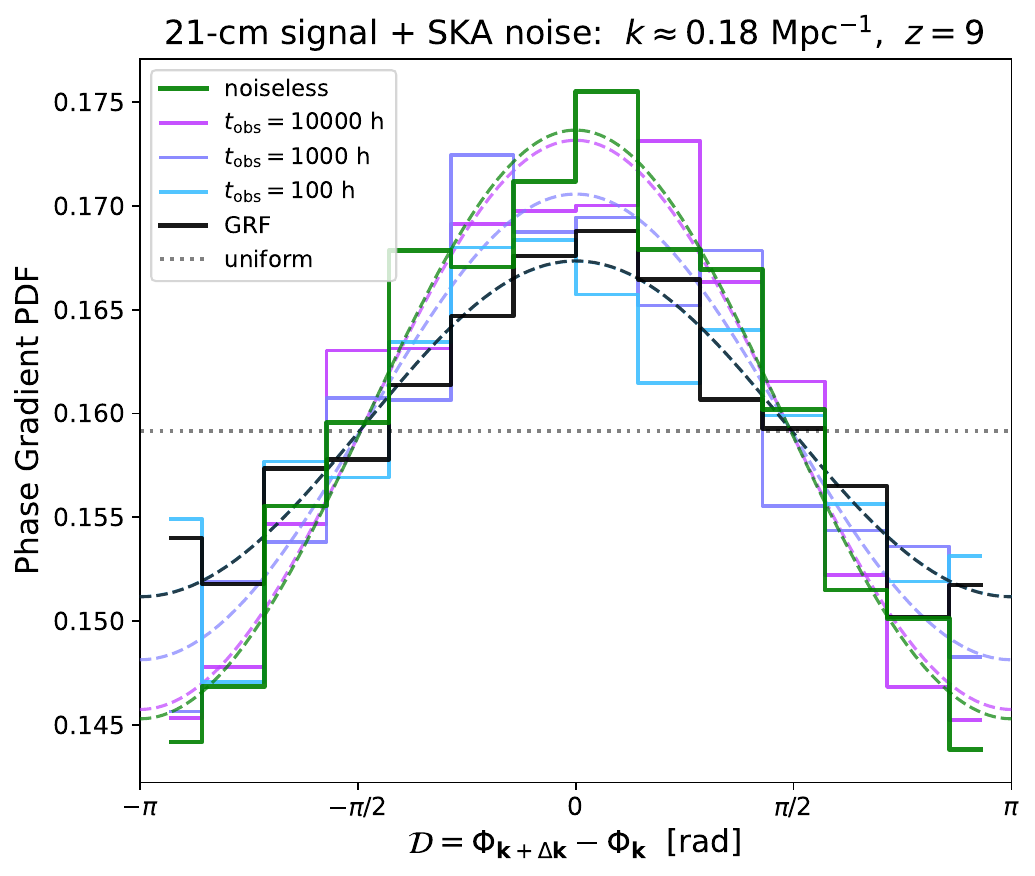}\\[-0.05cm]
    \includegraphics[width=0.96\linewidth]{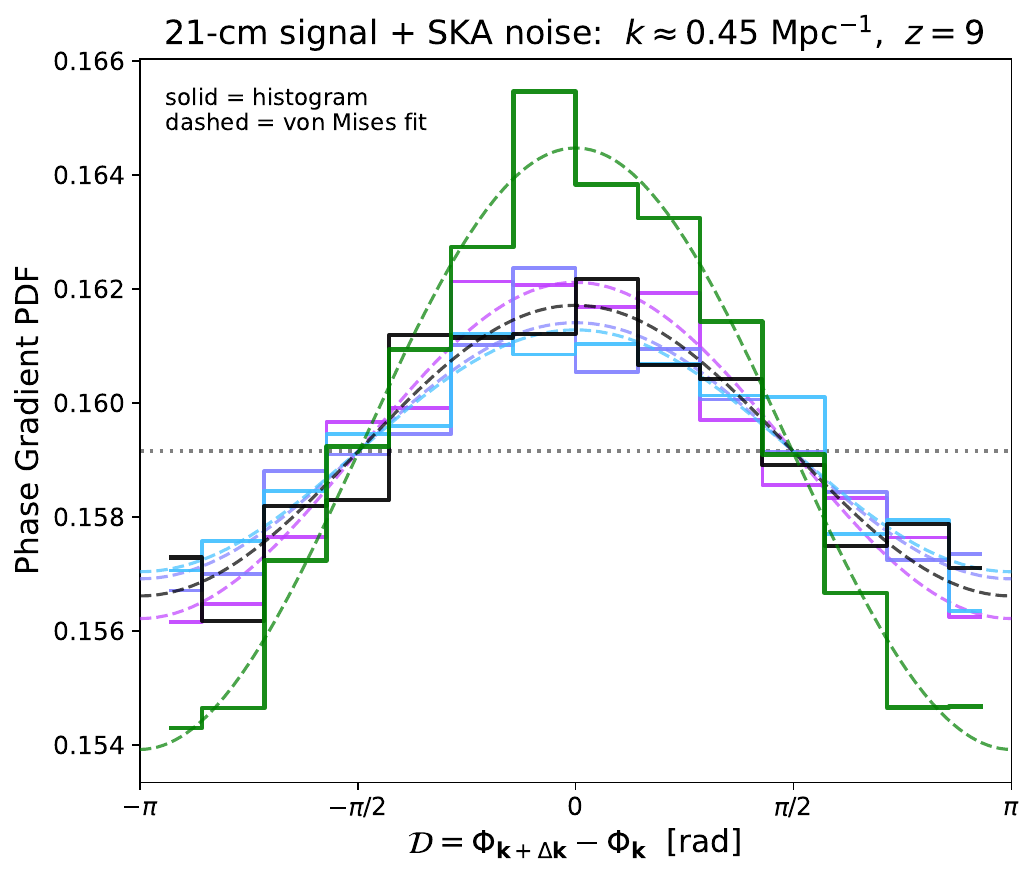}
    \vspace{-0.3cm}
    \caption{
    Phase gradient PDFs of the 21-cm signal of the \MassIndependent{} model at $z = 9$, for $k \approx 0.18\,\mathrm{Mpc}^{-1}$ (top panel) and $0.45\,\mathrm{Mpc}^{-1}$ (bottom panel), stacked over the $N_L$ long-wavelength modes of $\bar{R}(k)$ rather than a single mode as in Fig.~\ref{fig:phase_grad_pdf_lognormal}. Each mode is centred on its circular mean before stacking. Solid curves are measured PDFs, dashed curves of the same colour are von Mises fits, and the dotted horizontal line represents a uniform distribution. Green shows the noiseless signal, and purple, lavender and light blue show it with SKA-Low thermal noise for $t_\mathrm{obs} = 10\,000$, 1000 and $100\,\mathrm{h}$. Black shows a GRF on the same grid.
    }
    \label{fig:phase_grad_spectrum_fiducial}
\end{figure}

Fig.~\ref{fig:phase_grad_spectrum_fiducial} shows how thermal noise affects the phase gradient PDFs at $z = 9$. The same noise realisation is rescaled to $t_\mathrm{obs} = 10\,000$, 1000 and $100\,\mathrm{h}$ and added to one realisation of the 21-cm signal. Unlike Fig.~\ref{fig:phase_grad_pdf_lognormal}, these PDFs are stacked over the $N_L$ long-wavelength modes used for $\bar{R}(k)$. Each mode is first rotated to its own mean direction. The mean resultant length of the stacked PDF is then $\bar{R}(k)$. At $k \approx 0.18\,\mathrm{Mpc}^{-1}$ (top panel), the signal dominates. The PDF for $10\,000\,\mathrm{h}$ is nearly identical to the noiseless one, and the PDF for $1000\,\mathrm{h}$ remains more peaked than that of the GRF. Only the PDF for $100\,\mathrm{h}$ approaches that of the GRF. At $k \approx 0.45\,\mathrm{Mpc}^{-1}$ (bottom panel), the noise dominates. The noiseless PDF is still clearly peaked, but all the noisy PDFs lie close to the GRF curve, which is set by the GRF noise limit of $\bar{R}$ (Eq.~\ref{eq:grf_noise_limit}). Thermal noise has random phases and erases the phase coherence of the short-wavelength modes. At these scales, the signal can therefore no longer be distinguished from a GRF. The phase information of an SKA-Low observation therefore comes mainly from the largest scales.

\begin{figure*}
    \centering
    \includegraphics[width=0.498\linewidth]{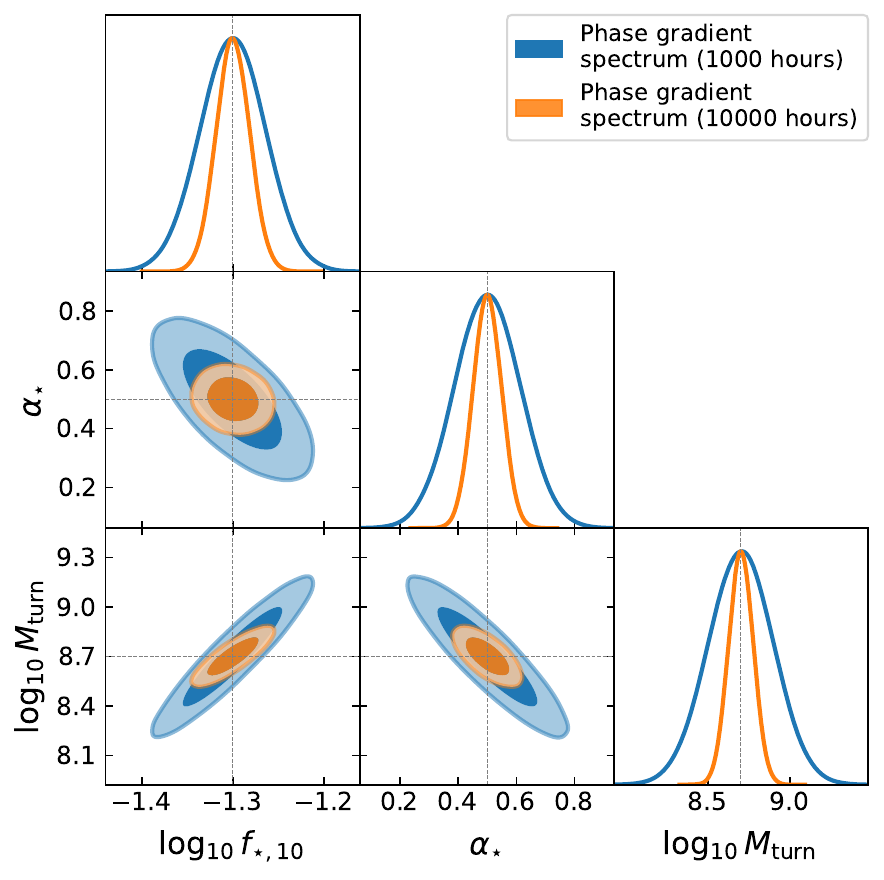}
    \includegraphics[width=0.498\linewidth]{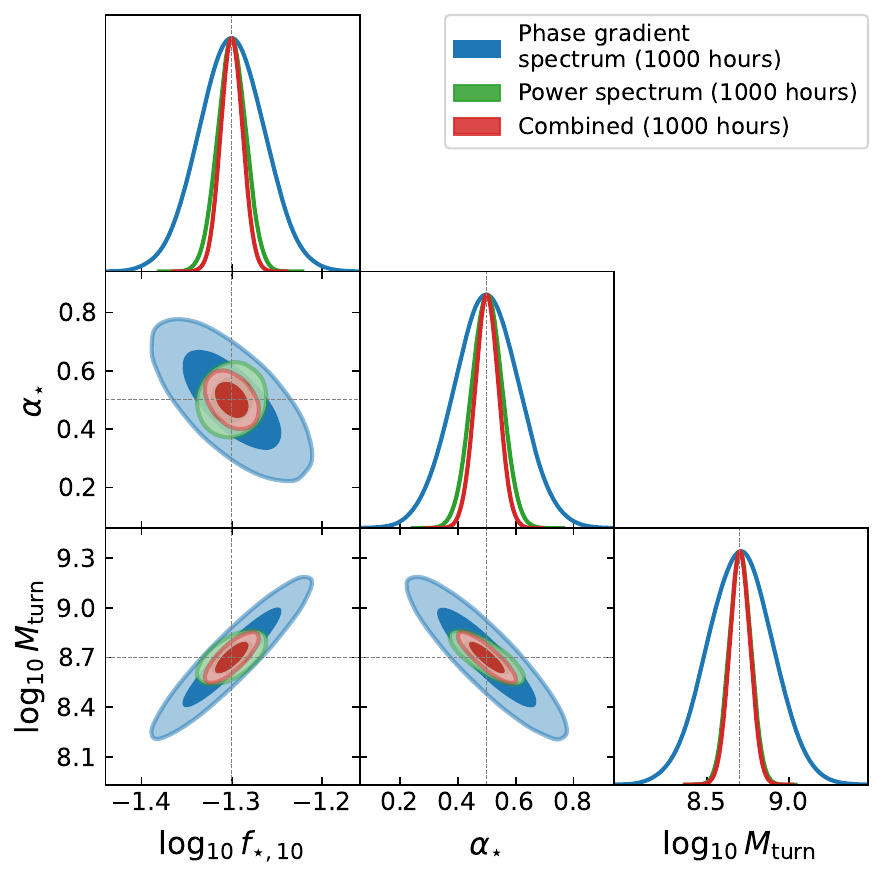}
    \caption{
    Fisher forecast constraints on the astrophysical parameters from SKA-Low observations, combining $z = 10$, 9, 8, 7 and 6.5. Contours show the 68 and 95 per cent confidence regions, and dotted lines mark the fiducial values. \textit{Left panel}: Phase gradient spectrum alone for $t_\mathrm{obs} = 1000\,\mathrm{h}$ (blue) and $10\,000\,\mathrm{h}$ (orange). \textit{Right panel}: Phase gradient spectrum alone (blue), power spectrum alone (green) and their combination (red) for $t_\mathrm{obs} = 1000\,\mathrm{h}$.
    }
    \label{fig:forecast_constraints}
\end{figure*}

The covariance matrix $\mathbf{C}$ of the summary statistics is estimated numerically as the sample covariance over independent realisations at the fiducial parameter values (see Appendix~\ref{app:covariance} for details). We generate $N_\mathrm{IC} = 200$ realisations with different cosmological initial conditions and $N_\mathrm{noise} = 100$ independent thermal noise realisations, so that $\mathbf{C}$ captures both cosmic variance and instrumental noise simultaneously. A Gaussian likelihood is assumed for both the power spectrum and the phase gradient spectrum. The Hartlap correction \citep{hartlap2007your} is applied to debias the inverted sample covariance. The resulting correlation matrix of the joint data vector shows weak cross-correlations between $P(k)$ and $\bar{R}(k)$ (Fig.~\ref{fig:covariance}).

\subsection{Constraints on the astrophysics}\label{sec:constraints}

We now assess the astrophysical constraints obtainable from the phase gradient spectrum in a mock SKA-Low observation of the fiducial \MassIndependent{} model with the Fisher matrix formalism. Fig.~\ref{fig:forecast_constraints} shows the forecast confidence contours on $(\log_{10} f_{\star,10},\, \alpha_\star,\, \log_{10}(M_\mathrm{turn}/M_\odot))$ with $z = 10$, 9, 8, 7 and 6.5 combined. These five redshifts span the early, middle and late stages of reionization and hence sample different points along the non-monotonic evolution of $\bar{R}(k)$ (Fig.~\ref{fig:power_phase_spect_21cmfast_fiducial}). The Fisher matrices use the $k$-bins below $0.5\,\mathrm{Mpc}^{-1}$. At smaller scales the thermal noise power exceeds the signal power by a factor of 20 or more at $t_\mathrm{obs} = 1000\,\mathrm{h}$. At $z = 9$, the noise already erases the phase coherence of the signal near this limit (Fig.~\ref{fig:phase_grad_spectrum_fiducial}).

The left panel of Fig.~\ref{fig:forecast_constraints} shows that the phase gradient spectrum alone constrains all three parameters. At $t_\mathrm{obs} = 1000\,\mathrm{h}$, the marginalised uncertainties are well within the plausible parameter ranges. These marginalised errors extend along the parameter degeneracies, where the model is not linear in the parameters, and are therefore approximate (Appendix~\ref{app:fisher_validity}). $\log_{10} f_{\star,10}$ and $\log_{10}(M_\mathrm{turn}/M_\odot)$ are positively correlated, and both are anti-correlated with $\alpha_\star$. These parameter combinations shift the ionizing photon budget between haloes of different mass while leaving the 21-cm signal nearly unchanged. This stand-alone constraining power relies on the average over the long-wavelength modes with $|\Delta\pmb{k}| \leq 3\,k_\mathrm{f}$, which reduces the sample variance of $\bar{R}(k)$ (Sec.~\ref{sec:lognormal_spectrum}). With only $|\Delta\pmb{k}| = k_\mathrm{f}$, the sample variance dominates and $\bar{R}(k)$ alone gives no useful constraint within the plausible parameter ranges. We compare the constraints for different sets of long-wavelength modes in Appendix~\ref{app:long_modes}. At $t_\mathrm{obs} = 10\,000\,\mathrm{h}$, the uncertainties shrink by factors of $\gtrsim 2$, as the lower thermal noise suppresses the phase coherence less (Fig.~\ref{fig:phase_grad_spectrum_fiducial}).

The right panel of Fig.~\ref{fig:forecast_constraints} compares the phase gradient spectrum with the power spectrum at $t_\mathrm{obs} = 1000\,\mathrm{h}$. On its own, $\bar{R}(k)$ gives constraints 2--3 times broader than the power spectrum, and the two statistics show similar degeneracy directions. Combining them tightens the marginalised $1\sigma$ uncertainties on $\log_{10} f_{\star,10}$ and $\alpha_\star$ by about 20 per cent and on $\log_{10}(M_\mathrm{turn}/M_\odot)$ by 7 per cent. This gain arises because $\bar{R}(k)$ responds to the ionization morphology (Sec.~\ref{sec:reion_models}) rather than to the amplitude of fluctuations, and its correlation with the power spectrum is weak (Fig.~\ref{fig:covariance}). The same pattern was found for the 21-cm bispectrum by \citet{watkinson2022}, where the bispectrum alone gives broader constraints than the power spectrum but their combination improves on the power spectrum alone. \citet{watkinson2022} also found that adding the bispectrum reduces the bias in the inferred parameters. A Fisher forecast assumes an unbiased estimator and cannot test this, which we leave for a full likelihood analysis in future work. 

A direct estimate of the normalised squeezed bispectrum uses the phase of each long-wavelength mode and keeps the sign of the response. With perfectly measured long-wavelength modes, it gives constraints tighter than those of $\bar{R}(k)$ by a factor of 1.3--1.6. Adding $\bar{R}(k)$ to this estimate and the power spectrum improves the constraints by only a further 3--5 per cent. The two statistics therefore largely carry the same information. The factor of 1.3--1.6 is an upper limit, as foreground residuals in the long-wavelength modes would reduce the gain of the direct estimate. We compare the constraints from the two statistics in Appendix~\ref{app:bispectrum_comparison}.

We emphasise that the constraints presented here represent the most optimistic scenario, as Fisher analysis provides a lower bound on the marginalised uncertainties. The degeneracy directions are sensitive to the choice of fiducial parameter values, and a full Bayesian inference pipeline, such as Markov chain Monte Carlo sampling or simulation-based inference, would explore other equally likely regions of parameter space and typically yield broader posteriors. The Fisher results should therefore be interpreted as indicative of the relative information content of different observables and observing strategies rather than as precise forecasts.

\section{Summary}\label{sec:summary}

The 21-cm signal from the epoch of reionization is highly non-Gaussian, and much of its information is carried by the Fourier phases. Radio interferometers measure these phases directly, but standard analyses reduce the data to the power spectrum and discard them. Every 21-cm dataset therefore already contains the phase information studied here, and exploiting it requires no change to the telescope or the observing strategy. We have introduced the phase gradient spectrum $\bar{R}(k)$, a statistic built from pairs of Fourier modes that measures how coherently the phases of short-wavelength modes shift relative to each other in the presence of a long-wavelength mode. Like the power spectrum, the phase gradient spectrum requires only a single fast Fourier transform. We showed analytically that $\bar{R}(k)$ measures the normalised squeezed bispectrum and can therefore be read as the response of small-scale structure to long-wavelength fluctuations. Tests on log-normal fields confirmed the analytic prediction across a wide range of non-Gaussianity.

Applied to simulations of reionization, the phase gradient spectrum reveals a distinctive signature. For the matter density field, the phase coherence grows steadily as structure forms. For the 21-cm signal, it rises, falls and rises again as ionized regions grow and merge. The trough marks the epoch at which the response of the small-scale 21-cm fluctuations to the long-wavelength modes changes sign, shortly after the 21-cm power spectrum peaks at the same scale. This behaviour is generic and appears for different prescriptions of the ionizing sources and for both an excursion-set and a radiative-transfer-based reionization simulation method, while the timing of the features still distinguishes the models.

The thermal noise of SKA-Low erases the phase coherence of the small scales, and the phase information of an observation therefore comes mainly from the largest scales. A Fisher forecast combining five redshifts across reionization shows that the phase gradient spectrum alone constrains the astrophysical parameters, although more weakly than the power spectrum, and that the constraints tighten by a factor of $\gtrsim 2$ when the observation time increases from 1000 to $10\,000\,\mathrm{h}$. Combining the two statistics reduces the marginalised uncertainties by up to $\sim 20$ per cent. The phase gradient spectrum is therefore best used together with the power spectrum, in line with earlier findings for the 21-cm bispectrum \citep{watkinson2022}. A direct estimate of the normalised squeezed bispectrum on the same long-wavelength modes gives somewhat tighter constraints, but requires the long-wavelength modes to be measured. The phase gradient spectrum does not need them, which is a practical advantage, as these modes are the most affected by foreground contamination.

The phase gradient spectrum uses the long-wavelength modes only as increments between short-wavelength modes and never measures their amplitudes or phases. Not measuring these modes has three costs. First, $\bar{R}(k)$ measures only the modulus of the squeezed bispectrum and discards its sign, and a sign change of the response appears as a trough rather than a zero crossing. Second, the signal for each increment scales with the amplitude of the corresponding long-wavelength mode in the observed volume, which varies strongly between realisations, and many increments must be averaged to suppress this scatter. Third, the modulus of a finite sum of random phase vectors is always positive, which sets a GRF noise limit below which no signal can be detected. The third cost is known analytically and can be subtracted, and our forecast shows that the first two reduce the constraining power only modestly relative to the direct bispectrum estimate. For an observation in which the long-wavelength modes are lost to foregrounds, this modest loss is outweighed by the ability to measure the squeezed bispectrum at all.

The phase gradient spectrum joins other statistics of the 21-cm signal that use the Fourier phases. The triangle correlation function of \citet{gorce2019studying} correlates the normalised phases of the field over triangles of all shapes and traces the typical size of the ionized regions. The closure phase measured on a triangle of baselines in radio interferometry is the phase of the bispectrum and is insensitive to antenna-based calibration errors \citep{thyagarajan2018detecting,thyagarajan2020detectionI}. Both are built from the phases of closed triangles. The phase gradient spectrum instead uses pairs of short-wavelength modes separated by a long-wavelength mode and requires no evaluation of triangles. $\bar{R}(k)$ measures the modulus of the angle-averaged normalised squeezed bispectrum, and hence the response of the small-scale 21-cm fluctuations to long-wavelength modes. The position-dependent power spectrum \citep{giri2019position} measures the same response in subvolumes of the survey, whereas the phase gradient spectrum uses the full volume. Both the closure phase of squeezed triangles and the position-dependent power spectrum keep the sign of the response, but, like the direct bispectrum estimate, they require the long-wavelength modes that foregrounds contaminate.

Two caveats apply to the present study. First, we have not modelled foregrounds, their mitigation or instrumental calibration errors, all of which alter the Fourier phases. Unlike the closure phase, the phase gradient is not immune to antenna-based phase errors. Testing how well the phase gradient spectrum survives calibration and foreground removal is a key step towards an observational application. Second, the forecast assumes a Gaussian likelihood and a model that is linear in the parameters. These assumptions hold approximately for the combination with the power spectrum, but not for the phase gradient spectrum alone. The likelihood of the 21-cm power spectrum is itself non-Gaussian \citep{prelogovic2024informative}, and a full likelihood analysis, for example with simulation-based inference, is needed to obtain robust constraints and to test whether the phase information also reduces biases in the inferred parameters \citep{watkinson2022}.

Several extensions are planned. Our simulations assume a fully heated IGM and neglect the lightcone effect and non-linear redshift-space distortions. Including spin temperature fluctuations and these line-of-sight effects will show how they alter the phase coherence of the 21-cm signal. Applying the statistic to more complete simulations, such as the radiation-hydrodynamics simulations of \citet{kannan2022introducing} and the radiative transfer simulations of \citet{giri2024end}, will test its robustness across a wider range of reionization scenarios. The phase gradient spectrum can also be applied to other epochs accessible to low-frequency radio observations, such as the dark ages, and to any cosmological field with measured Fourier phases, including the galaxy distribution and weak gravitational lensing maps.

\section*{Acknowledgements}

SKG acknowledges support from Olle Engkvist Stiftelse (grant no. 232-0238). PDM is partly supported by the Dutch Research Council (NWO) (grant no. \href{https://www.nwo.nl/projecten/iolfn50333}{OCENW.M.22.307}). LVEK acknowledges the financial support from the European Research Council (ERC) under the European Union's Horizon 2020 research and innovation programme (Grant agreement No. 884760, ``CoDEX'').
SKG also acknowledges support from the EDUCATE excellence centre funded by the Swedish Research Council through grant Dnr 2022-06627.
We thank the Center for Information Technology of the University of Groningen for their support and for providing access to the Hábrók high performance computing cluster.

\section*{Data Availability}
The simulations presented in this work can be reproduced with the publicly available \textsc{21cmFAST} \citep{murray202021cmfast} and \textsc{BEoRN} \citep{schaeffer2023beorn} codes. The phase gradient spectrum, the normalised squeezed bispectrum and the SKA-Low noise simulations are computed with the \textsc{Tools21cm} package \citep{giri2020tools21cm}, where the phase statistics will be made available upon publication of this article. The Fisher forecasts use the \textsc{psiphy} package (\url{https://github.com/sambit-giri/psiphy}), and the corner plots are made with the \textsc{GetDist} package \citep{lewis2025getdist}. The data underlying this article will be shared on reasonable request to the corresponding author.

\bibliographystyle{mnras}
\bibliography{references}

\appendix

\section{Phase gradient spectrum and the squeezed bispectrum}\label{app:bispectrum_connection}

In this appendix, we derive the analytic connection between the phase gradient spectrum $\bar{R}(k, \Delta\pmb{k}) = 1 - \sigma^2_\mathrm{circ}(k, \Delta\pmb{k})$ and the squeezed-limit bispectrum of the field. The single perturbative input is the first-order Edgeworth expansion of the joint distribution of Fourier modes \citep{Matsubara2003PhaseFields,matsubara2007statistics}. Every subsequent step is an exact manipulation of that expansion. The result is therefore valid in the weakly non-Gaussian regime, $|p^{(3)}| \ll 1$. While the late stages of reionization are strongly non-Gaussian, this framework provides a useful leading-order interpretation of the phase statistics.

\subsection{Conventions and the normalised bispectrum}\label{app:conventions}

Let $\delta(\pmb{x})$ be a real, statistically homogeneous and isotropic field in a periodic box of volume $V = L_B^3$, with Fourier coefficients $\delta_{\pmb{k}} = |\delta_{\pmb{k}}|\,e^{i\Phi_{\pmb{k}}}$ in the convention $\delta_{\pmb{k}} = \int_V d^3x\, \delta(\pmb{x})\, e^{-i\pmb{k}\cdot\pmb{x}}$. These are related to the coefficients of Sec.~\ref{sec:phases_log_normal} by $\delta_{\pmb{k}} = V\,\Tilde{\delta}(\pmb{k})$ and carry the same phases $\Phi_{\pmb{k}}$. The power spectrum and bispectrum are defined by
\begin{eqnarray}
    \langle \delta_{\pmb{k}_1}\delta_{\pmb{k}_2}\rangle &=& V P(k_1)\, \delta^K_{\pmb{k}_1+\pmb{k}_2,\,\pmb{0}} \, ,
    \label{eq:app_P_def}\\
    \langle \delta_{\pmb{k}_1}\delta_{\pmb{k}_2}\delta_{\pmb{k}_3}\rangle &=& V B(\pmb{k}_1,\pmb{k}_2,\pmb{k}_3)\, \delta^K_{\pmb{k}_1+\pmb{k}_2+\pmb{k}_3,\,\pmb{0}} \, ,
    \label{eq:app_B_def}
\end{eqnarray}
with $\delta^K$ the Kronecker delta. Throughout this appendix, bold symbols such as $\pmb{k}$ and $\Delta\pmb{k}$ denote the wavevectors of individual Fourier modes. Plain symbols such as $k$ and $\Delta k$ denote their magnitudes and are used for quantities averaged over directions or binned in wavenumber, such as $P(k)$ and $\bar{p}^{(3)}(k, \Delta k)$. By statistical isotropy, the ensemble mean of a quantity measured for a single increment, such as $\langle\bar{R}(k, \Delta\pmb{k})\rangle$, depends only on the magnitudes, and we write it as $\langle\bar{R}(k, \Delta k)\rangle$. The reality condition $\delta(\pmb{x}) \in \mathbb{R}$ imposes $\delta_{-\pmb{k}} = \delta_{\pmb{k}}^*$, i.e.\ $\Phi_{-\pmb{k}} = -\Phi_{\pmb{k}}$. The Kronecker deltas encode the structural fact underlying the whole derivation. By translation invariance, \textit{only closed polygons of wavevectors carry ensemble correlations}. The wavevectors $\pmb{k}+\Delta\pmb{k}$ and $-\pmb{k}$ of the two modes in the phase gradient $\mathcal{D}(\pmb{k},\Delta\pmb{k}) = \Phi_{\pmb{k}+\Delta\pmb{k}} - \Phi_{\pmb{k}}$ do not sum to zero, and their phases are therefore uncorrelated on their own. Together with the long-wavelength mode $-\Delta\pmb{k}$, they sum to zero and form a triangle. The phase gradient is therefore correlated with the phase of the long-wavelength mode, and the strength of this correlation is set by the bispectrum at leading order. Higher-order polyspectra contribute only subleading corrections.

We work with unit-normalised modes,
\begin{eqnarray}
    \alpha_{\pmb{k}} \equiv \frac{\delta_{\pmb{k}}}{\sqrt{\langle|\delta_{\pmb{k}}|^2\rangle}}
    = \frac{\delta_{\pmb{k}}}{\sqrt{V P(k)}}
    = |\alpha_{\pmb{k}}|\, e^{i\Phi_{\pmb{k}}} \, ,
    \quad \langle|\alpha_{\pmb{k}}|^2\rangle = 1 \, .
\end{eqnarray}
The third joint cumulant of three modes forming a closed triangle is then the \textit{normalised bispectrum} given as \citep{Matsubara2003PhaseFields},
\begin{eqnarray}
    p^{(3)}(\pmb{k}_1,\pmb{k}_2,\pmb{k}_3) \equiv \langle \alpha_{\pmb{k}_1}\alpha_{\pmb{k}_2}\alpha_{\pmb{k}_3}\rangle
    = \frac{B(\pmb{k}_1,\pmb{k}_2,\pmb{k}_3)}{\sqrt{V\, P_1 P_2 P_3}} \, .
    \label{eq:p3_def}
\end{eqnarray}
The factor $V^{-1/2}$ arises automatically once conventions are fixed and ensures that $p^{(3)}$ is dimensionless\footnote{Under the Fourier convention of Eqs.~\eqref{eq:app_P_def}--\eqref{eq:app_B_def}, the power spectrum scales as $[P] = \mathrm{Mpc}^3$ and the bispectrum as $[B] = \mathrm{Mpc}^6$. The denominator $\sqrt{V\,P_1 P_2 P_3}$ therefore carries the same dimensions as $B$, confirming that $p^{(3)}$ is dimensionless.}. This normalisation reflects a physical statement: any individual phase correlation vanishes as $V^{-1/2}$ in the infinite-volume limit, and non-Gaussian information survives only because the number of contributing modes grows as $V$ \citep{matsubara2007statistics}.

To derive the conditional distribution of phase gradients in Appendix~\ref{app:conditional}, we need an explicit form for the joint distribution of Fourier modes that captures leading-order non-Gaussianity. For a weakly non-Gaussian field, this is provided by the first-order Edgeworth (Gram--Charlier) expansion about the Gaussian given as \citep{Matsubara2003PhaseFields,matsubara2007statistics}
\begin{eqnarray}
    \mathcal{P}(\{\alpha_{\pmb{k}}\})
    &=& \left[\prod_{\pmb{k}} \frac{e^{-|\alpha_{\pmb{k}}|^2}}{\pi}\right] \nonumber \\
    && \times\,\Bigg[ 1 + \!\!\sum_{\pmb{k}_1+\pmb{k}_2+\pmb{k}_3=\pmb{0}}\!\! p^{(3)}\,
    \bar{\alpha}_{\pmb{k}_1}\bar{\alpha}_{\pmb{k}_2}\bar{\alpha}_{\pmb{k}_3} + \mathrm{c.c.} \Bigg] ,
    \label{eq:edgeworth}
\end{eqnarray}
where c.c.\ denotes the complex conjugate, the product and sum run over independent modes (one member of each conjugate pair) and, for three distinct modes, the multivariate complex Hermite polynomial reduces to $\bar{\alpha}_1\bar{\alpha}_2\bar{\alpha}_3$. The coefficient can be verified by computing $\langle\alpha_1\alpha_2\alpha_3\rangle$ directly from Eq.~\eqref{eq:edgeworth}. The Gaussian factor contributes zero (odd moment), and of the two correction terms only $p^{(3)}\bar{\alpha}_{\pmb{k}_1}\bar{\alpha}_{\pmb{k}_2}\bar{\alpha}_{\pmb{k}_3}$ gives a non-zero result, since $\langle\alpha^2\rangle = 0$ for a circularly symmetric Gaussian eliminates the c.c.\ term. The surviving term gives $p^{(3)}\langle|\alpha_1|^2\rangle\langle|\alpha_2|^2\rangle\langle|\alpha_3|^2\rangle = p^{(3)}$, consistent with Eq.~\eqref{eq:p3_def}. Higher-order corrections from $p^{(4)}$ and $(p^{(3)})^2$ are neglected throughout this appendix.

\subsection{Conditional distribution of phase gradients}\label{app:conditional}

Writing each normalised mode in polar form, $\alpha_{\pmb{k}} = A_{\pmb{k}}\,e^{i\Phi_{\pmb{k}}}$ with measure $d^2\alpha = A\,dA\,d\Phi$,\footnote{Unit-normalised means $\langle A^2\rangle = \langle|\alpha|^2\rangle = 1$. Individual amplitudes are still random variables, not fixed to unity.} the Gaussian factor in Eq.~\eqref{eq:edgeworth} makes each amplitude Rayleigh-distributed, $\mathcal{P}(A) = 2A e^{-A^2}$, with mean $\langle A\rangle = \sqrt{\pi}/2$, while each phase is uniform. For the triangle $(\Delta\pmb{k},\, \pmb{k},\, -\pmb{k}-\Delta\pmb{k})$, the reality condition gives the closure phase \citep{thyagarajan2018detecting,thyagarajan2020detectionI}
\begin{eqnarray}
    \psi \equiv \Phi_{\Delta\pmb{k}} + \Phi_{\pmb{k}} + \Phi_{-\pmb{k}-\Delta\pmb{k}}
    = \Phi_{\Delta\pmb{k}} - \mathcal{D} \, , \nonumber
    \\ %
    \cos\psi = \cos\!\left(\mathcal{D} - \Phi_{\Delta\pmb{k}}\right) .
    \label{eq:closure_phase}
\end{eqnarray}
The phase gradient $\mathcal{D} = \Phi_{\pmb{k}+\Delta\pmb{k}} - \Phi_{\pmb{k}}$ involves only the two short-wavelength phases. The phase of the long-wavelength mode, $\Phi_{\Delta\pmb{k}}$, enters only through the closure phase $\psi$.

We follow the derivation of the distribution of phase sums by \citet{Matsubara2003PhaseFields}, with the difference that we condition on the long-wavelength mode $\alpha_{\Delta\pmb{k}} = A_{\Delta\pmb{k}}\,e^{i\Phi_{\Delta\pmb{k}}}$, holding both its amplitude and its phase fixed. Of the terms in the sum of Eq.~\eqref{eq:edgeworth}, only the one for this triangle involves both short modes. All other terms contain at least one mode that appears once, and they vanish when that mode's uniform phase is integrated out. Writing $\pmb{k}' \equiv -\pmb{k}-\Delta\pmb{k}$ and taking $p^{(3)}$ real, as for a parity-invariant field, the surviving term in polar form is $p^{(3)}\bar{\alpha}_{\Delta\pmb{k}}\bar{\alpha}_{\pmb{k}}\bar{\alpha}_{\pmb{k}'} + \mathrm{c.c.} = 2p^{(3)} A_{\Delta\pmb{k}} A_{\pmb{k}} A_{\pmb{k}'}\cos\psi$. The joint distribution of the two short modes given the long-wavelength mode is therefore
\begin{eqnarray}
    \mathcal{P}(\alpha_{\pmb{k}}, \alpha_{\pmb{k}'} \,|\, \alpha_{\Delta\pmb{k}})\, d^2\alpha_{\pmb{k}}\, d^2\alpha_{\pmb{k}'}
    &=& \frac{2A_{\pmb{k}}e^{-A_{\pmb{k}}^2}\, 2A_{\pmb{k}'}e^{-A_{\pmb{k}'}^2}}{(2\pi)^2} \nonumber \\
    && \times\, \big[1 + 2p^{(3)} A_{\Delta\pmb{k}} A_{\pmb{k}} A_{\pmb{k}'}\cos\psi\big] \nonumber \\
    && \times\, dA_{\pmb{k}}\, dA_{\pmb{k}'}\, d\Phi_{\pmb{k}}\, d\Phi_{\pmb{k}'} \, .
\end{eqnarray}
We change the phase variables from $(\Phi_{\pmb{k}}, \Phi_{\pmb{k}'})$ to $(\Phi_{\pmb{k}}, \mathcal{D})$ with $\mathcal{D} = -\Phi_{\pmb{k}'} - \Phi_{\pmb{k}}$, which has unit Jacobian. The integrand depends on the phases only through $\psi = \Phi_{\Delta\pmb{k}} - \mathcal{D}$. Integrating over $\Phi_{\pmb{k}}$ therefore gives a factor $2\pi$, and integrating over each Rayleigh amplitude gives $\int_0^\infty A\, 2Ae^{-A^2}\, dA = \langle A\rangle = \sqrt{\pi}/2$ in the correction term and unity in the Gaussian term. This yields the conditional distribution of the phase gradient,
\begin{eqnarray}
    \mathcal{P}\!\left(\mathcal{D} \,\big|\, \alpha_{\Delta\pmb{k}}\right) = \frac{1}{2\pi}\left[1 + \frac{\pi}{2}\,|\alpha_{\Delta\pmb{k}}|\,\cos\!\left(\mathcal{D} - \Phi_{\Delta\pmb{k}}\right) p^{(3)}(\Delta\pmb{k}, \pmb{k})\right] ,
    \label{eq:conditional_phase_grad_dist}
\end{eqnarray}
where the prefactor decomposes as $\pi/2 = 2 \times (\sqrt{\pi}/2)^2$, i.e.\ the Edgeworth coefficient times the two Rayleigh amplitude means.
Eq.~\eqref{eq:conditional_phase_grad_dist} is a \textit{cardioid} distribution\footnote{The cardioid distribution is the simplest non-uniform distribution on the circle: $p(\theta;\rho,\mu) = (1/2\pi)(1 + \rho\cos(\theta-\mu))$, $|\rho|\leq 1$. It contains only the first circular harmonic and is the leading-order approximation to any unimodal circular distribution \citep[e.g.,][]{jammalamadaka2001topics}.}, which is uniform plus a single cosine harmonic.

The conditioning variable $\alpha_{\Delta\pmb{k}}$ carries both $|\alpha_{\Delta\pmb{k}}|$ and $\Phi_{\Delta\pmb{k}}$, and Eq.~\eqref{eq:conditional_phase_grad_dist} depends on both. The unconditional distribution of $\mathcal{D}$ follows by integrating over the long-wavelength mode,
\begin{eqnarray}
    \mathcal{P}(\mathcal{D}) = \int \mathcal{P}\!\left(\mathcal{D} \,\big|\, \alpha_{\Delta\pmb{k}}\right)\, \mathcal{P}(\alpha_{\Delta\pmb{k}})\, d^2\alpha_{\Delta\pmb{k}} = \frac{1}{2\pi} \, ,
    \label{eq:marginal_phase_grad_dist}
\end{eqnarray}
since $\Phi_{\Delta\pmb{k}}$ is uniform at leading order and the cosine term averages to zero. The marginal of $\mathcal{D}$ is therefore uniform at this order. Note that this is distinct from the exact theorem that one-point distributions of \textit{single} phases are uniform in homogeneous fields \citep{matsubara2007statistics}: $\mathcal{D}$ is a phase \textit{difference}, and its marginal acquires corrections at $\mathcal{O}(p^{(4)})$. The non-uniformity in Eq.~\eqref{eq:conditional_phase_grad_dist} is a realisation-dependent modulation: the long-wavelength mode coherently shifts the phase gradients of short-wavelength modes.

\subsection{Conditional mean resultant vector}\label{app:first_moment}

For a circular random variable $\mathcal{D}$, the first circular moment $\langle e^{i\mathcal{D}}\rangle$ is a vector in the complex plane, the mean resultant vector. Its length, $|\langle e^{i\mathcal{D}}\rangle|$, is the population mean resultant length, which is zero for a uniform distribution and unity for a distribution concentrated at a single angle \citep{jammalamadaka2001topics}. The statistic $\bar{R}$ of Sec.~\ref{sec:phases_phase_grad} is its sample counterpart, in which the ensemble average is replaced by the average over the modes in a $k$-bin (Appendix~\ref{app:estimator}). The conditional first circular moment follows by direct integration:
\begin{eqnarray}
\langle e^{i\mathcal{D}} | \alpha_{\Delta\pmb{k}} \rangle = \int_{-\pi}^{\pi} e^{i\mathcal{D}}\, \mathcal{P}(\mathcal{D}|\alpha_{\Delta\pmb{k}}) \, d\mathcal{D} \, .
\end{eqnarray}
The uniform part integrates to zero. Writing $\cos(\mathcal{D} - \Phi_{\Delta\pmb{k}}) = \frac{1}{2}\big[e^{i(\mathcal{D}-\Phi_{\Delta\pmb{k}})} + e^{-i(\mathcal{D}-\Phi_{\Delta\pmb{k}})}\big]$, the first exponential gives $\int e^{2i\mathcal{D}}\, d\mathcal{D} = 0$, and the second cancels the oscillation of $e^{i\mathcal{D}}$ and leaves a constant integrand. This yields
\begin{eqnarray}
\langle e^{i\mathcal{D}} | \alpha_{\Delta\pmb{k}} \rangle &=& \frac{1}{2\pi}\,\frac{\pi}{2}\,|\alpha_{\Delta\pmb{k}}|\, p^{(3)}(\Delta\pmb{k}, \pmb{k})\, \frac{e^{i\Phi_{\Delta\pmb{k}}}}{2} \int_{-\pi}^{\pi} d\mathcal{D} \nonumber \\
&=& \frac{\pi}{4}\, \alpha_{\Delta\pmb{k}} \, p^{(3)}(\Delta\pmb{k}, \pmb{k}) \, .
\label{eq:conditional_mean_resultant}
\end{eqnarray}
This is a special case of a general property of the cardioid distribution, $(1/2\pi)[1 + \rho\cos(\theta - \mu)]$ with mean direction $\mu$ and amplitude $\rho$, whose mean resultant vector is $(\rho/2)\,e^{i\mu}$ \citep{jammalamadaka2001topics}. Eq.~\eqref{eq:conditional_phase_grad_dist} has $\rho = \frac{\pi}{2}|\alpha_{\Delta\pmb{k}}|\,p^{(3)}(\Delta\pmb{k}, \pmb{k})$ and $\mu = \Phi_{\Delta\pmb{k}}$, which gives Eq.~\eqref{eq:conditional_mean_resultant} directly.

The conditional mean phase vector points along the \textit{phase} of the long-wavelength mode, with length proportional to its \textit{amplitude} times the normalised bispectrum. The long-wavelength mode acts as a realisation-specific reference direction that coherently drags all short-mode phase gradients. The magnitude of Eq.~\eqref{eq:conditional_mean_resultant}, $\tfrac{\pi}{4}|\alpha_{\Delta\pmb{k}}|\,|p^{(3)}(\Delta\pmb{k},\pmb{k})|$, is the conditional population mean resultant length, i.e.\ the phase coherence that $\bar{R}$ estimates for a given long-wavelength mode.

This result is consistent with the von Mises distributions used to describe the measured phase gradient PDFs in Sec.~\ref{sec:phases_phase_grad}. At small concentration, $e^{\kappa\cos x} \approx 1 + \kappa\cos x$, and the von Mises distribution reduces to a cardioid with $\rho = \kappa$. Its mean resultant length, $I_1(\kappa)/I_0(\kappa) \approx \kappa/2$, then matches the magnitude of Eq.~\eqref{eq:conditional_mean_resultant} for $\kappa = \tfrac{\pi}{2}|\alpha_{\Delta\pmb{k}}|\,|p^{(3)}(\Delta\pmb{k},\pmb{k})|$. As $p^{(3)}$ scales as $V^{-1/2}$, this small-concentration limit applies to the phase gradients in our simulation boxes, and the concentration of the phase gradient PDF for a given long-wavelength mode directly measures the normalised bispectrum, scaled by the amplitude of that mode.

\subsection{Mean resultant length and the squeezed bispectrum}\label{app:estimator}

In practice the ensemble average in Eq.~\eqref{eq:conditional_mean_resultant} is not available, and we replace it with the sample mean over all $N_k$ modes with $|\pmb{k}| \in \text{bin}$, all using the same $\Delta\pmb{k}$:
\begin{eqnarray}
    Z(k, \Delta\pmb{k}) \equiv \frac{1}{N_k}\sum_{|\pmb{k}|\in\mathrm{bin}} e^{i\mathcal{D}(\pmb{k},\Delta\pmb{k})} \, , \qquad \bar{R} = |Z| \, .
\end{eqnarray}
Every term is conditioned on the same $\alpha_{\Delta\pmb{k}}$, and to first order in $p^{(3)}$ different modes within the bin are conditionally independent given $\alpha_{\Delta\pmb{k}}$, with one exact exception. By the reality condition, the conjugate partner $\pmb{k}' = -\pmb{k}-\Delta\pmb{k}$ yields an \textit{identical} phase gradient, $\mathcal{D}(\pmb{k}',\Delta\pmb{k}) = \mathcal{D}(\pmb{k},\Delta\pmb{k})$. A mode whose partner lies in the same $k$-bin is therefore counted twice. If $n_p$ of the $N_k$ modes have their partner in the bin, the number of \textit{independent} phase vectors is $N_\mathrm{indep} = N_k^2/(N_k + n_p)$. This equals $N_k/2$ when every partner lies in the bin, as for $|\Delta\pmb{k}| \ll k$, and approaches $N_k$ in narrow bins at low $k$, where many partners $|\pmb{k}+\Delta\pmb{k}|$ fall outside the bin. Using Eq.~\eqref{eq:conditional_mean_resultant},
\begin{eqnarray}
    \langle Z \,|\, \alpha_{\Delta\pmb{k}} \rangle = \frac{\pi}{4}\,\alpha_{\Delta\pmb{k}} \cdot \bar{p}^{(3)}(k, \Delta k) \, ,
    \label{eq:Z_conditional_mean}
\end{eqnarray}
where
\begin{eqnarray}
    \bar{p}^{(3)}(k, \Delta k) &\equiv& \frac{1}{N_k}\sum_{|\pmb{k}|\in\mathrm{bin}} p^{(3)}(\Delta\pmb{k}, \pmb{k}) \nonumber\\
    &=& \int \frac{d\Omega_{\hat{\pmb{k}}}}{4\pi}\, \frac{B(\Delta\pmb{k},\, \pmb{k},\, -\pmb{k}-\Delta\pmb{k})}{\sqrt{V\,P(\Delta k)\,P(k)\,P(|\pmb{k}+\Delta\pmb{k}|)}} \qquad %
    \label{eq:p3bar_def}
\end{eqnarray}
is the angle-averaged normalised bispectrum at the triangle configuration $(\Delta\pmb{k}, \pmb{k}, -\pmb{k}-\Delta\pmb{k})$. When $\Delta k \equiv |\Delta\pmb{k}| \ll k$, the third side satisfies $|\pmb{k}+\Delta\pmb{k}| \approx k$ and this configuration approaches the \textit{squeezed limit}, giving
\begin{eqnarray}
    \bar{p}^{(3)}(k, \Delta k) \;\xrightarrow{\;\Delta k \ll k\;}\; \frac{B_\mathrm{sq}(\Delta k,\, k)}{\sqrt{V\,P(\Delta k)\,P(k)^2}} \, ,
    \label{eq:bispectrum_squeezed_limit}
\end{eqnarray}
where $B_\mathrm{sq}(\Delta k, k) \equiv B(\Delta k, k, k)$ is the squeezed bispectrum with one side of length $\Delta k$ and two sides of length $k$. The closing wavevector of every triangle is $\Delta\pmb{k}$, the increment used to compute the gradient. For $|\Delta\pmb{k}| \ll k$, the phase gradient statistic therefore probes the squeezed configuration. Larger increments probe less squeezed triangles, for which Eq.~\eqref{eq:bispectrum_squeezed_limit} no longer applies.

The phase $\psi$ of Eq.~\eqref{eq:closure_phase} is the closure phase of the triangle $(\Delta\pmb{k},\, \pmb{k},\, -\pmb{k}-\Delta\pmb{k})$, the phase of the triple product $\alpha_{\Delta\pmb{k}}\alpha_{\pmb{k}}\alpha_{\pmb{k}'}$ whose ensemble mean is the normalised bispectrum. The phase gradient differs from it by the phase of the long-wavelength mode, $\mathcal{D} = \Phi_{\Delta\pmb{k}} - \psi$. Averaged over the $k$-bin, the closure phases therefore give $Z_\psi = e^{i\Phi_{\Delta\pmb{k}}} Z^*$ (Eq.~\ref{eq:closure_phase_average}), and Eq.~\eqref{eq:Z_conditional_mean} gives
\begin{eqnarray}
    \langle Z_\psi \,|\, \alpha_{\Delta\pmb{k}} \rangle = \frac{\pi}{4}\,|\alpha_{\Delta\pmb{k}}|\, \bar{p}^{(3)}(k, \Delta k) \, .
    \label{eq:closure_phase_conditional_mean}
\end{eqnarray}
This conditional mean is real and carries the sign of $\bar{p}^{(3)}$. The noise part of $Z_\psi$ is circularly symmetric (Appendix~\ref{app:expected}) and averages to zero in its real part. $\mathrm{Re}\,Z_\psi$ is therefore an unbiased estimate of Eq.~\eqref{eq:closure_phase_conditional_mean}, without a GRF noise limit, but it requires the phase $\Phi_{\Delta\pmb{k}}$ of the long-wavelength mode. The modulus removes this phase, $|Z_\psi| = |Z| = \bar{R}(k, \Delta\pmb{k})$, at the cost of the sign and of the GRF noise limit. $\bar{R}$ is also computed from the three-dimensional Fourier modes of the image cube in the squeezed limit, rather than from triangles of baselines at each frequency. As $\mathcal{D}$ is a difference of two phases rather than a closed triangle, it does not inherit the insensitivity of the closure phase to antenna-based phase errors.

\subsection{Expected phase gradient spectrum}\label{app:expected}

We compute $\langle |Z|^2 \rangle$ by the law of total variance, $\langle |Z|^2 \rangle = \langle|\langle Z|\alpha_{\Delta\pmb{k}}\rangle|^2\rangle + \langle \mathrm{Var}(Z\,|\,\alpha_{\Delta\pmb{k}})\rangle$. The first (signal) term follows from Eq.~\eqref{eq:Z_conditional_mean} with $\langle|\alpha_{\Delta\pmb{k}}|^2\rangle = 1$. For the second (noise) term, each unit phase vector has conditional variance $1 - |\langle e^{i\mathcal{D}}\rangle|^2 \approx 1$, and $N_\mathrm{indep}$ independent phase vectors give $1/N_\mathrm{indep}$:
\begin{eqnarray}
    \langle |Z|^2 \rangle = \underbrace{\frac{1}{N_\mathrm{indep}}}_{\text{GRF noise}} + \underbrace{\frac{\pi^2}{16}\left[\bar{p}^{(3)}(k, \Delta k)\right]^2}_{\text{non-Gaussian signal}} \, .
    \label{eq:Z_variance}
\end{eqnarray}
The noise term is the random walk of $N_\mathrm{indep}$ independent unit phase vectors in the Gaussian limit ($p^{(3)} = 0$) and vanishes as the mode count grows. If this bias is subtracted (Sec.~\ref{sec:formalism}) from $\bar{R}$ averaged over the long-wavelength modes, the subtraction must be applied to $|Z|^2$ of each mode before averaging, since the square of the averaged $|Z|$ is not the average of the squares. We note that the signal term is $\mathcal{O}\big((p^{(3)})^2\big)$ while the Edgeworth expansion was truncated at $\mathcal{O}(p^{(3)})$. Trispectrum contributions and connected $(p^{(3)})^2$ cross-mode terms formally enter at the same order, but are incoherent with the long-wavelength mode and average down within the shell. We therefore neglect them.

To convert $\langle|Z|^2\rangle$ into $\langle|Z|\rangle$, note that $Z$ is approximately a zero-mean, circularly symmetric complex Gaussian. The noise part is Gaussian by the central limit theorem, which requires $N_\mathrm{indep} \gg 1$. The signal part, $\frac{\pi}{4}\alpha_{\Delta\pmb{k}}\bar{p}^{(3)}(k, \Delta k)$, is Gaussian because $\alpha_{\Delta\pmb{k}}$ is itself complex Gaussian with uniform phase at leading order in the Edgeworth expansion. The sum of the two independent parts is again a circularly symmetric complex Gaussian. The modulus $|Z|$ then follows a Rayleigh distribution, for which the mean and the second moment are related exactly by $\langle|Z|\rangle^2 = (\pi/4)\,\langle|Z|^2\rangle$. This gives
\begin{eqnarray}
    \langle \bar{R}(k, \Delta k) \rangle = \langle |Z| \rangle \approx \sqrt{\frac{\pi}{4}\langle|Z|^2\rangle} \, .
    \label{eq:rayleigh_mean}
\end{eqnarray}
The factor $\pi/4$ here is a property of the Rayleigh distribution and is unrelated to the factor $\pi/4$ in Eq.~\eqref{eq:Z_conditional_mean}, which arises from the Edgeworth coefficient and the mean short-mode amplitudes. The approximation in Eq.~\eqref{eq:rayleigh_mean} enters only through the Gaussianity of $Z$, i.e.\ through the assumptions of large $N_\mathrm{indep}$ and leading-order non-Gaussianity. Substituting Eq.~\eqref{eq:Z_variance} into Eq.~\eqref{eq:rayleigh_mean} gives the expected phase gradient spectrum quoted in the main text, Eq.~\eqref{eq:Rbar_full}.
In the Gaussian limit, Eqs.~\eqref{eq:Z_variance}--\eqref{eq:rayleigh_mean} give the mode-counting noise limit of the estimator,
\begin{eqnarray}
    \langle \bar{R} \rangle_\mathrm{GRF} = \sqrt{\frac{\pi}{4\,N_\mathrm{indep}}} \, ,
    \label{eq:noise_floor}
\end{eqnarray}
which is the null expectation of the Rayleigh test of circular uniformity \citep{mardia2009directional,jammalamadaka2001topics} and accurately reproduces the small non-zero values measured for Gaussian fields in Fig.~\ref{fig:phase_grad_spectrum_lognormal}. In the regime where the non-Gaussian signal dominates ($|\bar{p}^{(3)}| \gg N_\mathrm{indep}^{-1/2}$),
\begin{eqnarray}
    \langle \bar{R}(k, \Delta k) \rangle \approx \frac{\pi^{3/2}}{8}\left|\bar{p}^{(3)}(k, \Delta k)\right| \, .
    \label{eq:signal_dominated}
\end{eqnarray}
The coefficient can be cross-checked without the Rayleigh step: in the signal-dominated regime $|Z| = \frac{\pi}{4}|\alpha_{\Delta\pmb{k}}||\bar{p}^{(3)}|$ directly, and the Rayleigh amplitude mean $\langle|\alpha_{\Delta\pmb{k}}|\rangle = \sqrt{\pi}/2$ gives the same $\pi^{3/2}/8$. Equivalently, averaging Eq.~\eqref{eq:conditional_mean_resultant} over the long-mode amplitude recovers the known first-order three-point phase correlation $\langle e^{i(\Phi_{\pmb{k}+\Delta\pmb{k}} - \Phi_{\pmb{k}} - \Phi_{\Delta\pmb{k}})}\rangle = (\sqrt{\pi}/2)^3\, p^{(3)}$ of \citet{Matsubara2003PhaseFields}.

Using Eq.~\eqref{eq:bispectrum_squeezed_limit}, the phase gradient spectrum in the squeezed limit is therefore
\begin{eqnarray}
    \langle \bar{R}(k, \Delta k) \rangle \;\approx\; \frac{\pi^{3/2}}{8}\;\frac{\left|B_\mathrm{sq}(\Delta k,\, k)\right|}{\sqrt{V\;P(\Delta k)\;P(k)^2}} \, ,
    \label{eq:variance_spectrum_bispectrum}
\end{eqnarray}
valid to first order in the normalised bispectrum and in the limit $\Delta k \ll k$. This is the main result: \textit{the phase gradient spectrum directly measures the angle-averaged normalised squeezed bispectrum as a function of the short-wavelength mode $k$, with the long-wavelength mode set by $\Delta k$}. When $\Delta k = k_\mathrm{f}$ (the fundamental mode of the simulation box), the most squeezed configuration accessible in the volume is probed. In practice we average the resultant lengths of the $N_L = 61$ long-wavelength modes with $|\Delta\pmb{k}| \leq 3\,k_\mathrm{f}$ (Sec.~\ref{sec:phases_phase_grad}). Each is a statistically independent long-wavelength mode, which reduces the realisation scatter of the estimator without mixing the reference phases.

\subsection{Physical interpretation}\label{app:interpretation}

The above relation admits a simple physical interpretation in terms of the response of small-scale structure to long-wavelength perturbations. The long-wavelength mode is the Fourier coefficient $\delta_{\Delta\pmb{k}}$ at the increment $\Delta\pmb{k}$ used to define the phase gradient. With the Fourier convention of Appendix~\ref{app:conventions}, the inverse transform reads $\delta(\pmb{x}) = V^{-1}\sum_{\pmb{k}} \delta_{\pmb{k}}\, e^{i\pmb{k}\cdot\pmb{x}}$. In real space, this mode and its conjugate partner together contribute
\begin{eqnarray}
    \delta_L(\pmb{x}) = \frac{1}{V}\left[\delta_{\Delta\pmb{k}}\, e^{i\Delta\pmb{k}\cdot\pmb{x}} + \mathrm{c.c.}\right] ,
    \label{eq:long_mode}
\end{eqnarray}
which is a plane wave with a wavelength of $2\pi/\Delta k$, between one third of the box size and the box size. Different regions of the box therefore sit at different values of the large-scale overdensity $\delta_L$. If the short-wavelength modes respond to $\delta_L$, their local power spectrum varies across the box, and this modulation appears in Fourier space as the squeezed bispectrum. To first order in $\delta_L$, the local power spectrum is $P(k\,|\,\delta_L) \approx P(k)\left[1 + \mathcal{G}(k)\,\delta_L\right]$, where $P(k)$ is the power spectrum defined in Eq.~\eqref{eq:app_P_def} and
\begin{eqnarray}
\mathcal{G}(k) = \frac{\partial \ln P(k\,|\,\delta_L)}{\partial \delta_L}\bigg|_{\delta_L = 0}
\end{eqnarray}
is the response function. The squeezed bispectrum then takes the form \citep[see e.g.][]{bernardeau2002large,chiang2014position}
\begin{eqnarray}
    B(\Delta\pmb{k},\, \pmb{k},\, -\pmb{k}-\Delta\pmb{k}) \approx P(\Delta k)\,P(k)\,\mathcal{G}(k) \, ,
\end{eqnarray}
where $P(\Delta k)$ is the variance of the modulating long-wavelength mode.

Substituting into Eq.~\eqref{eq:variance_spectrum_bispectrum}, we obtain
\begin{eqnarray}
\langle \bar{R}(k, \Delta k) \rangle
\approx \frac{\pi^{3/2}}{8}\, \sqrt{\frac{P(\Delta k)}{V}}\; \left|\mathcal{G}(k)\right| \nonumber \\
= \frac{\pi^{3/2}}{8}\; \delta_\mathrm{rms}(\Delta k)\, \left|\mathcal{G}(k)\right| ,
\label{eq:response_form}
\end{eqnarray}
where $\delta_\mathrm{rms}(\Delta k) \equiv \sqrt{P(\Delta k)/V}$ is the rms amplitude of the long-wavelength mode $\Delta\pmb{k}$ in the volume. This follows from Eq.~\eqref{eq:long_mode}, as the real-space amplitude of the mode is $|\delta_{\Delta\pmb{k}}|/V$ and $\langle|\delta_{\Delta\pmb{k}}|^2\rangle = V P(\Delta k)$. For each long-wavelength mode, the statistic thus reads as the product of the typical amplitude of that mode and the response of small-scale structure to it. In reionization, this modulation is driven by ionized regions, which imprint large-scale structure that affects small-scale fluctuations \citep{giri2019position}. The phase gradient spectrum can therefore be viewed as a phase-based, computationally efficient probe of the response function.

For the 21-cm signal, both the long- and the short-wavelength modes in Eq.~\eqref{eq:response_form} are those of the 21-cm field, whereas the physical driver of the modulation is the long-wavelength matter density $\delta_{b,L}$. A linear bias model relates the two \citep{mcquinn2018observable,giri2019position}. For $T_\mathrm{s} \gg T_\mathrm{CMB}$, the brightness temperature scales as $\delta T_\mathrm{b} \propto x_\mathrm{HI}(1+\delta_b)$ (Eq.~\ref{eq:dTb}). On large scales, the neutral fraction is a biased tracer of the matter density, $x_\mathrm{HI} = \bar{x}_\mathrm{HI}(1 + b_1\delta_{b,L})$, with first-order bias $b_1$. To first order in $\delta_{b,L}$, the long-wavelength 21-cm fluctuation, in units of the brightness temperature of a fully neutral medium at mean density, is therefore
\begin{eqnarray}
    \delta_{21,L} = \bar{x}_\mathrm{HI}\,(1+b_1)\,\delta_{b,L} \, .
    \label{eq:linear_bias}
\end{eqnarray}
The long-wavelength 21-cm mode is thus a rescaled copy of the matter mode, with $P_{21}(\Delta k) = [\bar{x}_\mathrm{HI}(1+b_1)]^2 P_b(\Delta k)$. The response to it is $\mathcal{G}_{21}(k) = \mathcal{G}_b(k)/[\bar{x}_\mathrm{HI}(1+b_1)]$, where $\mathcal{G}_b(k) = \partial \ln P_{21}(k)/\partial \delta_{b,L}$ is the response of the 21-cm power spectrum to the long-wavelength matter density. The two responses have opposite signs once $b_1 < -1$. In Eq.~\eqref{eq:response_form}, the factor $\bar{x}_\mathrm{HI}(1+b_1)$ cancels between the amplitude and the response, and
\begin{eqnarray}
    \langle \bar{R}(k, \Delta k) \rangle \approx \frac{\pi^{3/2}}{8}\; \sqrt{\frac{P_b(\Delta k)}{V}}\, \left|\mathcal{G}_b(k)\right| \, .
    \label{eq:response_matter}
\end{eqnarray}
To this order, $\bar{R}(k)$ of the 21-cm field therefore measures the response of the small-scale 21-cm power to the long-wavelength matter density, weighted by the rms amplitude of the long-wavelength matter mode. The cancellation also keeps $\bar{R}$ finite when $b_1 \to -1$, where the long-wavelength 21-cm mode vanishes at linear order and $\mathcal{G}_{21}$ diverges. The model assumes a linear response of the large-scale neutral fraction to the density, which breaks down once the ionized regions become larger than the scales considered \citep{giri2019position}.

\section{Fisher forecast methodology}\label{app:fisher}

We forecast parameter constraints using the Fisher information matrix formalism \citep[see e.g.][]{tegmark1997karhunen}, as implemented in the \textsc{psiphy} package.\footnote{\url{https://github.com/sambit-giri/psiphy}} The method approximates the likelihood as Gaussian near its maximum, providing a lower bound on the marginalised uncertainties achievable by any unbiased estimator via the Cramér–Rao inequality.

\subsection{Data vector and likelihood}

At each redshift slice we construct a data vector $\pmb{d}$ that stacks the summary statistics evaluated in $N_\mathrm{b}$ wavenumber bins. For the phase gradient spectrum alone,
\begin{eqnarray}
    \pmb{d} = \left[\bar{R}(k_1),\, \ldots,\, \bar{R}(k_{N_\mathrm{b}})\right]^\mathsf{T} \; .
\end{eqnarray}
Since $\bar{R} = 1 - \sigma^2_\mathrm{circ}$ is a linear transformation, using $\sigma^2_\mathrm{circ}$ instead gives an identical Fisher matrix. When the power spectrum is included as a complementary statistic the data vector is extended by concatenation,
\begin{eqnarray}
    \pmb{d} = \left[P(k_1),\, \ldots,\, P(k_{N_\mathrm{b}}),\, \bar{R}(k_1),\, \ldots,\, \bar{R}(k_{N_\mathrm{b}})\right]^\mathsf{T} \; ,
\end{eqnarray}
so that the joint covariance matrix captures cross-correlations between the two statistics. Both statistics are measured from the same realisations, and the full covariance of the stacked data vector, including the cross-covariance between $P(k)$ and $\bar{R}(k)$, is estimated directly from them with Eq.~\eqref{eq:sample_covariance}. When several redshifts are combined, the data vectors of the individual redshifts are concatenated in the same way, and the covariance between redshifts is retained.

We assume a multivariate Gaussian likelihood for the summary statistics,
\begin{eqnarray}
    \ln \mathcal{L}(\pmb{\theta}) = -\frac{1}{2}\left[\pmb{d} - \pmb{\mu}(\pmb{\theta})\right]^\mathsf{T} \mathbf{C}^{-1} \left[\pmb{d} - \pmb{\mu}(\pmb{\theta})\right] + \mathrm{const.} \; ,
    \label{eq:gaussian_likelihood}
\end{eqnarray}
where $\pmb{\mu}(\pmb{\theta})$ is the model prediction at parameter point $\pmb{\theta} = (\log_{10}f_{\star,10},\, \alpha_\star,\, \log_{10}(M_\mathrm{turn}/M_\odot))$ and $\mathbf{C}$ is the covariance matrix of the data vector. For the phase gradient spectrum the Gaussian approximation is motivated by the central limit theorem, as $\bar{R}(k)$ is built from the mean of many unit phase vectors in each $k$-bin. The simulated distributions show mild skewness, and the likelihood is therefore only approximately Gaussian.

\subsection{Fisher matrix}

Under the assumption that the covariance $\mathbf{C}$ is evaluated at fixed fiducial parameters and does not depend appreciably on $\pmb{\theta}$, the Fisher matrix reduces to
\begin{eqnarray}
    F_{ij} = \frac{\partial \pmb{\mu}^\mathsf{T}}{\partial \theta_i} \, \mathbf{C}^{-1} \, \frac{\partial \pmb{\mu}}{\partial \theta_j} \; ,
    \label{eq:fisher_matrix}
\end{eqnarray}
where the derivatives are evaluated at the fiducial parameter values $\pmb{\theta}_\mathrm{fid}$. The forecasted $1\sigma$ marginalised uncertainty on parameter $\theta_i$ is given by the Cramér–Rao bound \citep[see e.g.][]{tegmark1997karhunen},
\begin{eqnarray}
    \sigma(\theta_i) \geq \sqrt{\left(\mathbf{F}^{-1}\right)_{ii}} \; .
    \label{eq:cramer_rao}
\end{eqnarray}
The unmarginalised (conditional) bound is $\sigma_\mathrm{cond}(\theta_i) \geq 1/\sqrt{F_{ii}}$, which assumes all other parameters are perfectly known.

\subsection{Covariance matrix estimation}\label{app:covariance}

\begin{figure*}
    \centering
    \text{\large ~~~$P(k)$ \qquad \qquad \qquad \qquad \qquad \qquad \qquad $\bar{R}(k)$ \qquad \qquad \qquad}\\
    \includegraphics[width=0.90\linewidth,trim={0 0 0 46},clip]{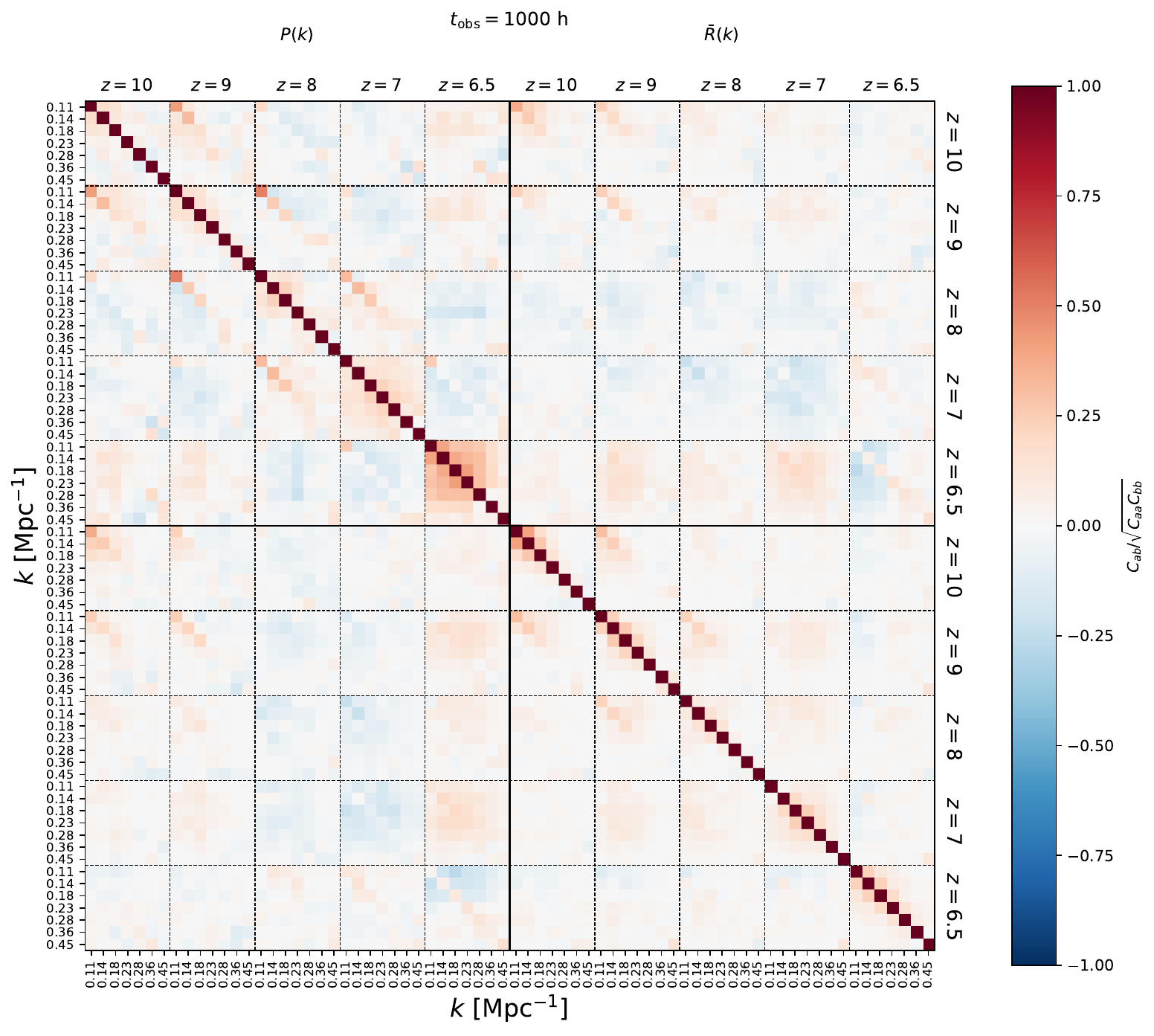}
    \caption{
    Correlation matrix $C_{ab}/\sqrt{C_{aa}C_{bb}}$ of the joint data vector of the power spectrum $P(k)$ and phase gradient spectrum $\bar{R}(k)$ at $z = 10$, 9, 8, 7 and 6.5 for a 1000-h SKA-Low observation. Solid lines separate the two statistics and dashed lines separate the redshifts.
    }
    \label{fig:covariance}
\end{figure*}

The covariance matrix is estimated from an ensemble of $N_\mathrm{sim}$ realisations of the data vector at the fiducial parameter values,
\begin{eqnarray}
    C_{ab} = \frac{1}{N_\mathrm{sim}-1}\sum_{s=1}^{N_\mathrm{sim}} \left(d_a^{(s)} - \bar{d}_a\right)\!\left(d_b^{(s)} - \bar{d}_b\right) \; ,
    \label{eq:sample_covariance}
\end{eqnarray}
where $\bar{d}_a = N_\mathrm{sim}^{-1}\sum_s d_a^{(s)}$ is the sample mean. We run $N_\mathrm{IC} = 200$ simulations with independent cosmological initial conditions and add each of $N_\mathrm{noise} = 100$ independent thermal noise realisations (generated as described in Sec.~\ref{sec:mock_obs}) to every simulation before computing the summary statistics. This gives $N_\mathrm{sim} = N_\mathrm{IC} \times N_\mathrm{noise} = 20\,000$ realisations, so that $\mathbf{C}$ accounts simultaneously for cosmic variance and thermal noise. The precision matrix $\mathbf{C}^{-1}$ is obtained by direct inversion. For $N_\mathrm{sim}$ realisations and a data vector of length $N_d$, the Hartlap correction factor \citep{hartlap2007your}
\begin{eqnarray}
    \mathbf{C}^{-1}_\mathrm{corrected} = \frac{N_\mathrm{sim} - N_d - 2}{N_\mathrm{sim} - 1}\,\mathbf{C}^{-1}
    \label{eq:hartlap}
\end{eqnarray}
is applied to debias the inverse sample covariance. For the joint data vector of $P(k)$ and $\bar{R}(k)$ in seven $k$-bins at five redshifts, $N_d = 70$ and the correction is below 1 per cent. The $N_\mathrm{sim}$ realisations are not fully independent, as the 100 noise realisations of each simulation share the same initial conditions. If the covariance rested only on the $N_\mathrm{IC}$ independent simulations, the bias of the inverse covariance would grow steeply for smaller $N_\mathrm{IC}$ and diverge for $N_\mathrm{IC} < N_d$. Instead, reducing $N_\mathrm{IC}$ from 200 to 50 changes the marginalised errors by less than 10 per cent. The noise realisations therefore add largely independent information, and the Hartlap correction based on $N_\mathrm{sim}$ is adequate. Increasing $N_\mathrm{IC}$ to 300 changes the marginalised errors from the phase gradient spectrum by less than 3 per cent.

Fig.~\ref{fig:covariance} shows the resulting correlation matrix $C_{ab}/\sqrt{C_{aa}C_{bb}}$ for $t_\mathrm{obs} = 1000\,\mathrm{h}$. The power spectrum block is close to diagonal, with two exceptions. The largest scale ($k \approx 0.11\,\mathrm{Mpc}^{-1}$) is positively correlated between neighbouring redshifts, with coefficients of up to 0.5, as cosmic variance of the few large-scale modes is shared across epochs. At $z = 6.5$, near the end of reionization, the $k$-bins are correlated with each other with coefficients of about 0.3, reflecting the strongly non-Gaussian signal of the last neutral islands. The $\bar{R}(k)$ block is also close to diagonal. Neighbouring $k$-bins are weakly correlated, with coefficients of about 0.2, because the pairs of modes $\pmb{k}$ and $\pmb{k}+\Delta\pmb{k}$ can straddle the bin edges. The lowest $k$-bin is correlated between neighbouring redshifts by up to 0.3, and the bins at higher $k$ are uncorrelated because thermal noise dominates there. The cross-correlations between $P(k)$ and $\bar{R}(k)$ are weak, with correlation coefficients of at most $\approx 0.37$ in magnitude. The largest value occurs in the lowest $k$-bin at $z = 10$, where both statistics are set by the same few large-scale modes. The phase gradient spectrum is therefore largely complementary to the power spectrum.

\subsection{Numerical derivatives}

The derivatives $\partial\pmb{\mu}/\partial\theta_i$ required by Eq.~\eqref{eq:fisher_matrix} are computed numerically using central-difference stencils \citep[see e.g.][]{press2007numerical}. For a step size $h_i = \delta_\mathrm{frac}\,|\theta_{i}^\mathrm{fid}|$, the general $n$-point stencil evaluates the simulator at offsets $s\,h_i$ for $s \in \{-(n/2),\ldots,-1,+1,\ldots,+(n/2)\}$ and combines the results with weights $w_s$ chosen to cancel successive Taylor coefficients,
\begin{eqnarray}
    \frac{\partial \pmb{\mu}}{\partial \theta_i} \approx \frac{1}{h_i}\sum_{s} w_s\,\pmb{\mu}\!\left(\pmb{\theta}_\mathrm{fid} + s\,h_i\,\hat{\pmb{e}}_i\right) + \mathcal{O}(h_i^n) \; .
    \label{eq:numerical_derivative}
\end{eqnarray}
We use a 4-point stencil ($n=4$), which evaluates $\pmb{\mu}$ at $\pmb{\theta}_\mathrm{fid} \pm h_i\,\hat{\pmb{e}}_i$ and $\pmb{\theta}_\mathrm{fid} \pm 2h_i\,\hat{\pmb{e}}_i$ with weights $(w_{-2},w_{-1},w_{+1},w_{+2}) = (1/12,\,-2/3,\,2/3,\,-1/12)$, giving
\begin{eqnarray}
    \frac{\partial \pmb{\mu}}{\partial \theta_i} \approx \frac{-\pmb{\mu}(+2h_i) + 8\,\pmb{\mu}(+h_i) - 8\,\pmb{\mu}(-h_i) + \pmb{\mu}(-2h_i)}{12\,h_i} + \mathcal{O}(h_i^4) \; ,
    \label{eq:4point_stencil}
\end{eqnarray}
where we write $\pmb{\mu}(\pm s\,h_i)$ as shorthand for $\pmb{\mu}(\pmb{\theta}_\mathrm{fid} \pm s\,h_i\,\hat{\pmb{e}}_i)$. The $\mathcal{O}(h_i^4)$ accuracy arises because all odd Taylor coefficients up to third order cancel in the weighted sum, making this stencil far less sensitive to the choice of $h_i$ than the simpler 2-point scheme $[\pmb{\mu}(+h_i)-\pmb{\mu}(-h_i)]/(2h_i) + \mathcal{O}(h_i^2)$.

We use a step size of $\delta_\mathrm{frac} = 0.02$ for all parameters. To further reduce the impact of cosmic variance on the derivative estimate, the same sequence of initial-condition random seeds is used across all stencil evaluations for a given parameter, so that differences in $\pmb{\mu}$ arise from parameter changes rather than stochastic fluctuations of the field. The derivatives are averaged over the first 40 of the 200 simulations. Thermal noise is additive and independent of $\pmb{\theta}$. It therefore does not change the derivative of the mean power spectrum, and we compute the power spectrum derivatives from noiseless fields. The phase gradient spectrum, by contrast, responds non-linearly to noise, which suppresses the phase coherence. Its derivatives are therefore computed from noisy fields and averaged over 30 noise realisations, using the same noise realisations at every stencil point. These are drawn in pairs of opposite sign, $\pm\pmb{n}$, which cancels the part of the noise response that is odd in $\pmb{n}$.

The estimated derivatives are averages over a finite number of realisations and carry residual noise, $\partial_i\hat{\pmb{\mu}} = \partial_i\pmb{\mu} + \pmb{\epsilon}_i$ with $\langle\pmb{\epsilon}_i\rangle = 0$. Inserting them into Eq.~\eqref{eq:fisher_matrix} biases the Fisher matrix high,
\begin{eqnarray}
    \langle \hat{F}_{ij} \rangle = F_{ij} + \mathrm{tr}\big[\mathbf{C}^{-1}\,\mathrm{Cov}(\pmb{\epsilon}_i,\, \pmb{\epsilon}_j)\big] \; ,
    \label{eq:fisher_derivative_bias}
\end{eqnarray}
since the cross terms between $\partial_i\pmb{\mu}$ and $\pmb{\epsilon}_j$ average to zero. We estimate $\mathrm{Cov}(\pmb{\epsilon}_i, \pmb{\epsilon}_j)$ as the sample covariance of the derivatives between initial conditions divided by their number, and add the same quantity computed between noise realisations. The sum slightly overestimates the bias, as their joint contribution is counted twice. We subtract the resulting term from every Fisher matrix that includes the phase gradient spectrum or the normalised squeezed bispectrum.

\subsection{Combining summary statistics}

When the power spectrum and the phase gradient spectrum are combined, the joint Fisher matrix is computed from the stacked data vector with its full covariance (Eq.~\ref{eq:fisher_matrix}). The cross-covariance between $P(k)$ and $\bar{R}(k)$ is therefore taken into account, and the joint Fisher matrix differs from the sum $\mathbf{F}^{P} + \mathbf{F}^{\bar{R}}$, which would double count the information shared by the two statistics.

\subsection{Validity of the Fisher approximation}\label{app:fisher_validity}

The Fisher matrix describes the likelihood only if the model $\pmb{\mu}(\pmb{\theta})$ is close to linear over the region the constraints occupy \citep{vallisneri2008use}. The constraints of Fig.~\ref{fig:forecast_constraints} lie well within the plausible ranges $\log_{10} f_{\star,10} \in [-3, 0]$, $\alpha_\star \in [-0.5, 1]$ and $\log_{10}(M_\mathrm{turn}/M_\odot) \in [8, 10]$, and we therefore do not include priors. Following \citet{vallisneri2008use}, we compare the exact and linearised models at a $1\sigma$ displacement of each parameter, using second derivatives from the same stencil evaluations, and require the log-likelihood ratio between them to satisfy $|\ln r| \lesssim 0.1$. For the conditional errors we find $|\ln r| \lesssim 0.01$ for the power spectrum, the phase gradient spectrum and their combination. The marginalised errors extend along the parameter degeneracies, where the linear approximation is less accurate. At a displacement of one marginalised $1\sigma$, $|\ln r| \lesssim 0.2$ for the power spectrum and the combination, but $|\ln r| > 1$ for the phase gradient spectrum alone, whose marginalised errors are the largest. The contours in Fig.~\ref{fig:forecast_constraints} are therefore best read as a measure of the relative information content of the two statistics.

With the average over the long-wavelength modes with $|\Delta\pmb{k}| \leq 3\,k_\mathrm{f}$, the Fisher matrix of the phase gradient spectrum alone is well conditioned, with a condition number of $\approx 1.7\times10^{3}$ at $t_\mathrm{obs} = 1000\,\mathrm{h}$. Its least constrained direction lies mainly along $\log_{10}(M_\mathrm{turn}/M_\odot)$ and is still constrained well within the plausible range. The derivatives of $\bar{R}(k)$ have a signal-to-noise ratio of 10--25 per parameter at $1000\,\mathrm{h}$, weighted by the inverse covariance. The derivative-noise correction changes the diagonal of the Fisher matrix by about 1 per cent, but increases the marginalised errors by 9--15 per cent, as the degenerate directions are sensitive to small changes in $\mathbf{F}$. Splitting the 15 pairs of noise realisations of opposite sign into two halves changes the marginalised errors by less than 3 per cent.

\begin{figure*}
    \centering
    \includegraphics[width=0.498\linewidth]{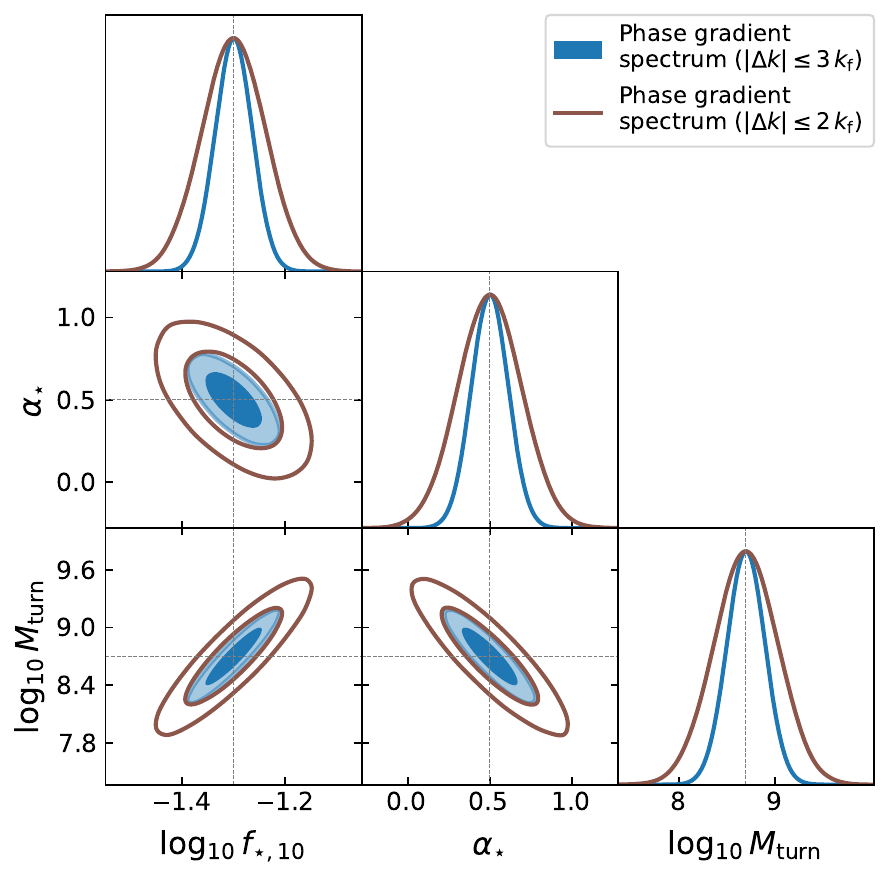}
    \includegraphics[width=0.498\linewidth]{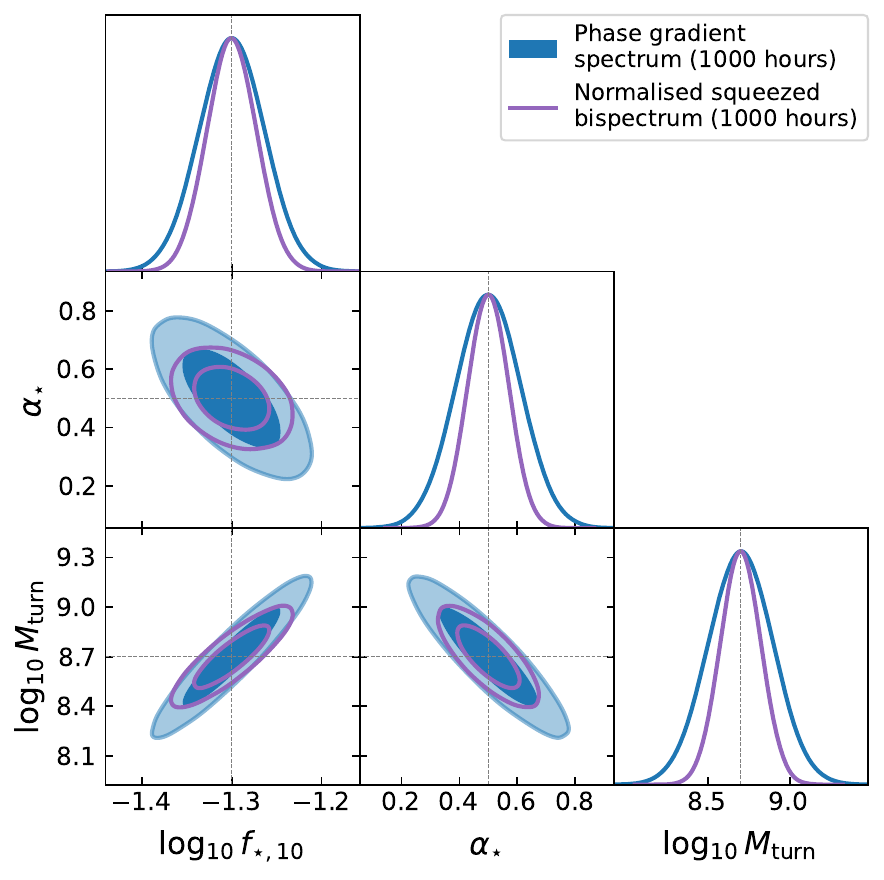}
    \caption{
    Fisher forecast constraints for $t_\mathrm{obs} = 1000\,\mathrm{h}$, combining observations at $z = 10$, 9, 8, 7 and 6.5. Contours show the 68 and 95 per cent confidence regions. \textit{Left panel}: Phase gradient spectrum averaged over the long-wavelength modes with $|\Delta\pmb{k}| \leq 3\,k_\mathrm{f}$ (blue) and $|\Delta\pmb{k}| \leq 2\,k_\mathrm{f}$ (brown). \textit{Right panel}: Phase gradient spectrum (blue) and normalised squeezed bispectrum (purple).
    }
    \label{fig:forecast_constraints_robustness}
\end{figure*}

\section{Information in the phase gradient spectrum}\label{app:phase_info_tests}

In this appendix, we test how the constraints from the phase gradient spectrum depend on the long-wavelength modes averaged over, and compare them with those from the normalised squeezed bispectrum. All forecasts use the setup of Sec.~\ref{sec:forecast_study} at $t_\mathrm{obs} = 1000\,\mathrm{h}$, with the same simulations, covariance estimation and derivatives (Appendix~\ref{app:fisher}).

\subsection{Long-wavelength modes}\label{app:long_modes}

The left panel of Fig.~\ref{fig:forecast_constraints_robustness} compares $\bar{R}(k)$ averaged over the long-wavelength modes with $|\Delta\pmb{k}| \leq 2\,k_\mathrm{f}$ and with $|\Delta\pmb{k}| \leq 3\,k_\mathrm{f}$, the case used in the main text. Both give the same degeneracy directions. The marginalised uncertainties for $|\Delta\pmb{k}| \leq 3\,k_\mathrm{f}$ are smaller by a factor of about 1.7, somewhat below the factor $\sqrt{61/16} \approx 2$ expected if the sample variance of the long-wavelength modes alone set the errors (Sec.~\ref{sec:lognormal_spectrum}). Thermal noise and the scatter of the short-wavelength modes also contribute. With only $|\Delta\pmb{k}| = k_\mathrm{f}$, the sample variance dominates and $\bar{R}(k)$ alone gives no useful constraint within the plausible parameter ranges, and we therefore do not show it. The gain from combining $\bar{R}(k)$ with the power spectrum grows in the same way. The marginalised uncertainties on $\log_{10} f_{\star,10}$ and $\alpha_\star$ shrink by 4--7 per cent for $|\Delta\pmb{k}| = k_\mathrm{f}$, 12 per cent for $|\Delta\pmb{k}| \leq 2\,k_\mathrm{f}$ and about 20 per cent for $|\Delta\pmb{k}| \leq 3\,k_\mathrm{f}$. Including longer increments would reduce the errors further, but at the cost of leaving the squeezed limit (Sec.~\ref{sec:lognormal_spectrum}).

\subsection{Comparison with the normalised squeezed bispectrum}\label{app:bispectrum_comparison}

Appendix~\ref{app:bispectrum_connection} shows that $\bar{R}(k)$ measures the normalised squeezed bispectrum $\bar{p}^{(3)}(k)$. We estimate $\bar{p}^{(3)}(k)$ directly by projecting the product $\delta_{\pmb{k}+\Delta\pmb{k}}\,\delta^*_{\pmb{k}}$ (Appendix~\ref{app:conventions}) onto the measured phase $\Phi_{\Delta\pmb{k}}$ of the long-wavelength mode and dividing by the power spectrum of the same realisation, averaged over the $k$-bin and the long-wavelength modes. We use the long-wavelength modes with $|\Delta\pmb{k}| \leq 3\,k_\mathrm{f}$, without those along the line of sight, which interferometers do not measure.

The right panel of Fig.~\ref{fig:forecast_constraints_robustness} shows that the normalised squeezed bispectrum gives tighter constraints than $\bar{R}(k)$, by a factor of 1.3--1.6, with the same degeneracy directions. Combined with the power spectrum, it tightens the constraints on $\log_{10} f_{\star,10}$ and $\alpha_\star$ by about 26 per cent and on $\log_{10}(M_\mathrm{turn}/M_\odot)$ by 14 per cent. Adding $\bar{R}(k)$ to this combination improves them by only a further 3--5 per cent. The two statistics therefore largely carry the same information. The bispectrum estimator gains for two reasons. It uses the phase of each long-wavelength mode and keeps only the part of the coupling in phase with it, whereas $\bar{R}(k)$ takes the modulus and keeps the noise in both components. It also weights each short-wavelength mode by its amplitude, which down-weights the noise-dominated modes. Normalising instead by the ensemble power spectra multiplies the squeezed bispectrum by fixed factors, which leaves the scatter of the long- and short-wavelength amplitudes between realisations in the data. The constraints are then broader than those of $\bar{R}(k)$, with marginalised $1\sigma$ uncertainties of 0.05, 0.17 and 0.33 on $\log_{10} f_{\star,10}$, $\alpha_\star$ and $\log_{10}(M_\mathrm{turn}/M_\odot)$.

The advantage of the bispectrum estimator relies on measuring the long-wavelength modes. With $|\Delta\pmb{k}| \leq 3\,k_\mathrm{f} \approx 0.06\,\mathrm{Mpc}^{-1}$, all of them have small line-of-sight wavenumbers, which are the modes most affected by foreground contamination. $\bar{R}(k)$ uses only the phase differences between short-wavelength modes and does not require the long-wavelength modes to be measured. The phase gradient spectrum therefore retains most of the information of the normalised squeezed bispectrum without requiring these modes.

\bsp	%
\label{lastpage}
\end{document}